\documentclass[11pt,a4paper]{article}
\pdfoutput=1
\usepackage{placeins,setup}
\usepackage{jheppub}
\usepackage{graphicx}
\usepackage{float}
\usepackage{afterpage}
\usepackage{epsfig}
\usepackage{amssymb}
\usepackage{amsmath}
\usepackage{bm}
\usepackage{multirow}
\usepackage{url}
\usepackage{xcolor}
\usepackage{float}
\usepackage{afterpage}
\usepackage{ulem}

\usepackage{url}

\usepackage{multirow,booktabs,multirow}

\newcommand{\be}{\begin{equation}}
\newcommand{\ee}{\end{equation}}
\newcommand{\bea}{\begin{eqnarray}}
\newcommand{\eea}{\end{eqnarray}}
\newcommand{\bi}{\begin{itemize}}
\newcommand{\ei}{\end{itemize}}
\newcommand{\ben}{\begin{enumerate}}
\newcommand{\een}{\end{enumerate}}

\newcommand{\lc}{\left[}
\newcommand{\rc}{\right]}
\newcommand{\lp}{\left(}
\newcommand{\rp}{\right)}

\def\frac#1#2{{{#1}\over {#2}}}
\def\gsim{\mathrel{\rlap{\lower4pt\hbox{\hskip1pt$\sim$}}
    \raise1pt\hbox{$>$}}}       
\def\lsim{\mathrel{\rlap{\lower4pt\hbox{\hskip1pt$\sim$}}
    \raise1pt\hbox{$<$}}}

\newcommand{\draft}[1]{}

\def\beq{\begin{equation}}
\def\eeq{\end{equation}}

\def\lapprox{\lower .7ex\hbox{$\;\stackrel{\textstyle <}{\sim}\;$}}
\def\gapprox{\lower .7ex\hbox{$\;\stackrel{\textstyle >}{\sim}\;$}}

\numberwithin{equation}{section}
\numberwithin{figure}{section}
\numberwithin{table}{section}

\usepackage{tabularx}
\newcolumntype{C}[1]{>{\centering\arraybackslash}p{#1}}

\subheader{\small Nikhef/2020-006}

\title{\boldmath nNNPDF2.0: Quark Flavor Separation in Nuclei from LHC Data}

\author[a,b]{Rabah Abdul Khalek,}
\author[a,b]{Jacob J. Ethier,}
\author[a,b]{Juan Rojo,}
\author[b]{and Gijs van Weelden}

\affiliation[a]{Department of Physics and Astronomy, VU Amsterdam, 1081HV Amsterdam, The Netherlands}
\affiliation[b]{Nikhef Theory Group, Science Park 105, 1098 XG Amsterdam, The Netherlands}

\emailAdd{rabah.khalek@gmail.com}
\emailAdd{j.j.ethier@vu.nl}
\emailAdd{j.rojo@vu.nl}
\emailAdd{gijsvanweelden@hotmail.com}

\abstract{We present a model-independent determination of the
nuclear parton distribution functions (nPDFs) 
using machine learning methods and Monte Carlo techniques 
based on the NNPDF framework.
The neutral-current deep-inelastic nuclear structure
functions used in our previous analysis, nNNPDF1.0, are complemented by
inclusive and charm-tagged cross-sections from charged-current scattering.
Furthermore, we include all available measurements of W and Z leptonic rapidity
distributions in proton-lead collisions
from ATLAS and CMS at $\sqrt{s}=5.02$ TeV and 8.16 TeV.
The resulting nPDF determination, nNNPDF2.0, achieves a good description of all datasets. 
In addition to quantifying
the nuclear modifications affecting individual quarks and antiquarks, 
we examine the implications for strangeness, assess the role that the
momentum and valence
sum rules play in nPDF extractions, and present predictions
for representative phenomenological applications.
Our results, made available via the {\tt LHAPDF} library, highlight the potential
of high-energy collider measurements to probe nuclear dynamics in a robust manner.
}

\begin{document}
\maketitle
\flushbottom

\section{Introduction}
\label{sec:introduction}

Decades of experimental investigations
have plainly revealed the inability to describe, within the framework
of perturbative QCD, high-energy scattering processes
involving heavy nuclei using a free-nucleon formalism. 
The parton distribution functions (PDFs)
of nucleons bound within nuclei, commonly known as
nuclear PDFs (nPDFs)~\cite{Rojo:2019uip,Ethier:2020way}, 
can therefore be significantly modified with respect to their
free-nucleon counterpart~\cite{Gao:2017yyd} as a result of non-perturbative dynamics. 
While a first-principles
understanding of the theoretical mechanisms that generate such QCD dynamics remains an
open challenge, phenomenological determinations of nPDFs have been able to
provide vital information about parton behavior in the cold nuclear medium. 

Precise extractions of nPDFs are not only crucial to study the strong interaction
in the high-density regime, but are also necessary to model
the initial state of heavy ion collisions which aim to characterize the Quark-Gluon
Plasma (QGP)~\cite{Abreu:2007kv,Adams:2005dq} using hard probes.
Furthermore, nPDFs also contribute to global QCD analyses
of the proton structure~\cite{Ball:2014uwa,Hou:2019efy,Harland-Lang:2014zoa,Alekhin:2017kpj}
via the inclusion of neutrino
structure function data collected in reactions involving heavy nuclear targets.
These measurements on nuclear targets
provide important information on the quark flavor
separation and strangeness in the proton~\cite{Ball:2018twp}.

Several groups have recently presented determinations
of the nuclear PDFs using
different input datasets, theoretical assumptions,
and methodological settings~\cite{Eskola:2016oht,Kovarik:2015cma,AbdulKhalek:2019mzd,Walt:2019slu,Khanpour:2016pph}.
While nPDF analyses are based on a significantly reduced dataset
compared to the free-nucleon case, the situation has improved
in recent years with the availability of hard-scattering 
cross-section data from proton-lead collisions at the LHC for processes such as jet,
W and Z, and heavy quark production~\cite{Adam:2015hoa,Adam:2016dau,Adam:2015xea,Aad:2016zif,Chatrchyan:2014hqa,1742-6596-612-1-012009,ATLAS-CONF-2015-056,Aad:2015gta,Khachatryan:2015pzs,Khachatryan:2015hha,CMS-PAS-HIN-15-012,Adam:2016mkz,Adam:2015qda,Abelev:2014hha,Khachatryan:2015sva,Khachatryan:2015uja,Aaij:2017gcy,Aaij:2019lkm}.
These collider measurements can clarify several
poorly understood aspects of nuclear PDFs, such as the quark 
flavor dependence of nuclear effects
and the nuclear modifications of the gluon distribution.
Several studies have indeed demonstrated the valuable constraints 
that can be provided for the nuclear PDFs from proton-lead 
collisions at the LHC, see {\it e.g.} Refs.~\cite{Eskola:2016oht,Kusina:2016fxy,
Kusina:2017gkz,Armesto:2015lrg,Eskola:2019dui}.

This work focuses on the determination of the quarks and anti-quark nuclear PDFs,
with emphasis on their flavor separation.
Since measurements of neutral-current (NC) deep-inelastic scattering (DIS) nuclear structure
functions on isoscalar targets
are only sensitive to a single quark PDF combination,
one needs to rely on the information provided by independent processes
to disentangle quark and antiquarks of different flavors.
The main options that are available to accomplish this are
neutrino-induced charged current (CC) DIS cross-sections on heavy nuclear targets,
sensitive to different quark combinations than the NC case, and
electroweak gauge boson production at the LHC.

From the methodological point of view, there exist two
primary limitations that affect the separation between  
quark and antiquark flavors in nPDF extractions.
The first one is the reliance on {\it ad-hoc} theoretical
assumptions required to model the dependence of the nuclear
modifications on both the parton momentum fraction $x$ and atomic mass
number $A$, where in some cases the expected behavior is hard-coded in
the nPDF parameterization.
The second is the lack of consistency between the nuclear PDF
determination and that of the corresponding proton baseline, to which the former should
reduce to in the $A\to 1$ limit in terms of central values and uncertainties. 
This consistency is particularly important given that the precision
LHC data impose stringent
constraints on the quark flavor separation for the proton PDFs,
for example via measurements 
of inclusive W and Z production characterized by per-mille level uncertainties.
Ensuring that the LHC constrains on the proton PDF baseline
are appropriately propagated to the nPDF determination for $A>2$ is 
therefore critical.

In this study we present a model-independent determination of nuclear PDFs
using machine learning methods and Monte Carlo techniques based on the NNPDF
framework~\cite{Forte:2002fg,DelDebbio:2004qj,DelDebbio:2007ee,Ball:2008by,Rojo:2008ke,Ball:2009qv,Ball:2010de,Ball:2011mu,Ball:2011uy,Ball:2012cx}.
We complement our previous nNNPDF1.0 analysis of NC DIS nuclear structure
functions with CC inclusive and charm-tagged measurements from fixed-target
neutrino experiments as well as with inclusive W
and Z production cross-sections in proton-lead collisions
from ATLAS and CMS at $\sqrt{s}=5.02$ TeV (Run I) and 8.16 TeV (Run II).
The $A=1$ proton PDF baseline used in the present analysis
is defined to be a variant of the NNPDF3.1 fit which excludes 
heavy nuclear target data.
This choice allows us to indirectly incorporate
the constraints on quark flavor separation provided
by the pp measurements from the LHC.

The nNNPDF2.0 results allow us to tackle several important issues
concerning nuclear effects among various quark flavors.
First, we assess the compatibility of the LHC W and Z leptonic rapidity distributions
from proton-lead collisions with the constraints coming from DIS structure functions,
and demonstrate that the former allow for a marked improvement in the quark
PDF uncertainties.
We also study the nuclear effects on the total strange content of heavy nuclei, highlighting the
interplay between the information provided by DIS and hadronic data.
This interplay is also interesting from the proton PDF point of view, where the pull
on strangeness provided by the ATLAS W, Z distributions~\cite{Aaboud:2016btc} is the opposite from
that of neutrino data and other LHC processes such as the W+c cross-sections.

We then analyze the impact that the momentum and total valence sum rule constraints
have in the global nPDF determination, and demonstrate that the corresponding
integrals agree with QCD predictions within uncertainties even
when the sum rules are not explicitly imposed.
We conclude the paper by providing theoretical
predictions based on nNNPDF2.0 for representative
processes of phenomenological interest in proton-ion collisions:
isolated photon production in the central and forward
rapidity regions and inclusive pion production.

This paper is organized in the following manner.
In Sect.~\ref{sec:expdata}, we provide
the input experimental observables used in this analysis
and detail the corresponding theoretical calculations.
We define a set of conventions and notation used in this work
and describe new aspects of our fitting methodology in Sect.~\ref{sec:fitting}.
The nNNPDF2.0 nuclear
parton distributions are then presented in Sect.~\ref{sec:results},
followed by a discussion of phenomenological implications
in Sect.~\ref{sec:phenomenology}.
Lastly, in Sect.~\ref{sec:summary} we conclude with a summary
and highlight future directions of study.

\section{Experimental data and theory calculations}
\label{sec:expdata}

In this section, we provide details of the experimental measurements 
used as input for the nNNPDF2.0 determination.
An emphasis is made in particular 
on the new datasets that are added with 
respect to those that were present in nNNPDF1.0.
We then discuss the theoretical calculations corresponding to these datasets 
and their numerical implementation in our fitting framework.

\subsection{Input dataset}
\label{sec:datasets}

Common to the previous nNNPDF1.0 analysis are the nuclear 
NC DIS measurements listed in Table~\ref{dataset}.
For each dataset, the target nuclei $A_1$ and $A_2$ used by each 
experiment are indicated together with their atomic mass numbers.
We also list the number of data points after the DIS kinematical cuts 
and provide the corresponding publication reference.
The DIS kinematic cuts are the same as in our previous study, 
i.e. $Q^2 = 3.5$ GeV$^2$ and $W^2 = 12.5$ GeV$^2$,
consistent with the NNPDF3.1 proton PDF baseline 
used to satisfy our boundary condition.

\begin{table}[htbp]
  \centering
  \small
   \renewcommand{\arraystretch}{1.09}
\begin{tabular}{c c c c}
Experiment & ${\rm A}_1/{\rm A}_2$ & ${\rm N}_{\rm dat}$ & Reference\\
\toprule
  SLAC E-139 & $^4$He/$^2$D & 3 & \cite{Gomez:1993ri} \\
  NMC 95, re. & $^4$He/$^2$D & 13 & \cite{Amaudruz:1995tq}\\
\midrule
  NMC 95 & $^6$Li/$^2$D & 12 & \cite{Arneodo:1995cs}\\
\midrule
  SLAC E-139 & $^9$Be/$^2$D & 3 & \cite{Gomez:1993ri}\\
  NMC 96 & $^9$Be/$^{12}$C & 14 & \cite{Arneodo:1996rv}\\
\midrule
  EMC 88, EMC 90 & $^{12}$C/$^2$D & 12 & \cite{Ashman:1988bf,Arneodo:1989sy}\\
  SLAC E-139 & $^{12}$C/$^2$D & 2 & \cite{Gomez:1993ri}\\
  NMC 95, NMC 95, re.  & $^{12}$C/$^2$D & 26 & \cite{Arneodo:1995cs,Amaudruz:1995tq}\\
  FNAL E665 & $^{12}$C/$^2$D & 3 & \cite{Adams:1995is}\\
  NMC 95, re. & $^{12}$C/$^6$Li & 9 & \cite{Amaudruz:1995tq}\\
\midrule
  BCDMS 85 & $^{14}$N/$^2$D & 9 & \cite{Alde:1990im}\\
\midrule
  SLAC E-139 & $^{27}$Al/$^2$D & 3 & \cite{Gomez:1993ri}\\
  NMC 96 & $^{27}$Al/$^{12}$C & 14 & \cite{Arneodo:1996rv}\\
\midrule
  SLAC E-139 & $^{40}$Ca/$^2$D & 2 & \cite{Gomez:1993ri}\\
  NMC 95, re. & $^{40}$Ca/$^2$D & 12 & \cite{Amaudruz:1995tq}\\
  EMC 90 & $^{40}$Ca/$^2$D & 3 & \cite{Arneodo:1989sy}\\
  FNAL E665 & $^{40}$Ca/$^2$D & 3 & \cite{Adams:1995is}\\
  NMC 95, re. & $^{40}$Ca/$^6$Li & 9 & \cite{Amaudruz:1995tq}\\
  NMC 96 & $^{40}$Ca/$^{12}$C & 23 & \cite{Arneodo:1996rv}\\
\midrule
  EMC 87 & $^{56}$Fe/$^2$D & 58 & \cite{Aubert:1987da}\\
  SLAC E-139 & $^{56}$Fe/$^2$D & 8 & \cite{Gomez:1993ri}\\
  NMC 96 & $^{56}$Fe/$^{12}$C & 14 & \cite{Arneodo:1996rv}\\
  BCDMS 85, BCDMS 87 & $^{56}$Fe/$^2$D & 16 & \cite{Alde:1990im,Benvenuti:1987az}\\
\midrule
  EMC 88, EMC 93 & $^{64}$Cu/$^2$D & 27 & \cite{Ashman:1988bf,Ashman:1992kv}\\
\midrule
  SLAC E-139 & $^{108}$Ag/$^2$D & 2 & \cite{Gomez:1993ri}\\
\midrule
 EMC 88 & $^{119}$Sn/$^2$D & 8 & \cite{Ashman:1988bf}\\
 NMC 96, $Q^2$ dependence  & $^{119}$Sn/$^{12}$C & 119 & \cite{Arneodo:1996ru}\\
\midrule
 FNAL E665 & $^{131}$Xe/$^2$D & 4 & \cite{Adams:1992vm}\\
\midrule
  SLAC E-139 & $^{197}$Au/$^2$D & 3 & \cite{Gomez:1993ri}\\
\midrule
 FNAL E665 & $^{208}$Pb/$^2$D & 3 & \cite{Adams:1995is}\\
 NMC 96 & $^{208}$Pb/$^{12}$C & 14 & \cite{Arneodo:1996rv}\\
 \midrule
 \midrule
 {\bf Total NC DIS} & & {\bf 451} & \\
\bottomrule
\end{tabular}
\vspace{4mm}
\caption{\small The
neutral-current nuclear deep-inelastic
  input datasets included in nNNPDF2.0.
  For each dataset, we indicate the nuclei $A_1$ and $A_2$
  involved, the number of data points that
  satisfy the baseline kinematical cuts,
  and the publication reference.
}
\label{dataset}
\end{table}

Note that all NC DIS measurements listed in Table~\ref{dataset} 
are provided in terms of ratios of structure functions between two 
different nuclei.
In most cases the denominator is given by the deuterium structure 
function, but ratios to carbon and lithium are also provided.
As we will discuss in Sect.~\ref{sec:fitting}, our fitting approach 
parameterizes the PDFs entering the absolute structure functions 
for each value of $A$, after which their ratios are constructed.

The remaining input data which is newly added to our nNNPDF2.0 
analysis is presented in terms of absolute cross-sections, without 
normalizing to any baseline nucleus.
We list these data in Table~\ref{datasetnew}, divided into two 
categories: CC neutrino DIS reduced cross-sections on 
nuclear targets and leptonic rapidity distributions in electroweak gauge boson production 
from proton-lead collisions at the LHC.
The neutrino and anti-neutrino reduced cross-sections are further 
separated into inclusive cross-sections from 
CHORUS~\cite{Onengut:2005kv} and charm-tagged cross-sections 
from NuTeV~\cite{Goncharov:2001qe}.
The LHC measurements are divided into data from ATLAS and from 
CMS from the Run I and Run II data-taking periods.
In this table we also indicate the total number of data points included 
in the fit, combining the NC and CC cross-sections measurements 
with the LHC data. In total, the nNNPDF2.0 global fit contains 
$n_{\rm dat}=1467$ data points.

\begin{table}[t]
  \centering
  \small
   \renewcommand{\arraystretch}{1.45}
\begin{tabular}{c c c c}
Experiment & ${\rm A}$ & ${\rm N}_{\rm dat}$ & Reference\\
\toprule
CHORUS $\nu$  & 208  &  423  & \cite{Onengut:2005kv}  \\
CHORUS $\bar{\nu}$  & 208  &  423  &  \cite{Onengut:2005kv} \\
NuTeV $\nu$  & 56  &  39  &   \cite{Goncharov:2001qe}\\
NuTeV $\bar{\nu}$  & 56  &  37  & \cite{Goncharov:2001qe}   \\
{\bf Total CC DIS} & & {\bf 922} & \\
\midrule
CMS $W^{\pm}$ $\sqrt{s}=8.16$ TeV  &  208  & 48     & \cite{Sirunyan:2019dox}   \\
CMS $W^{\pm}$ $\sqrt{s}=5.02$ TeV  &  208  & 20     & \cite{Khachatryan:2015hha}   \\
CMS $Z$ $\sqrt{s}=5.02$ TeV  &  208  &  12    & \cite{Khachatryan:2015pzs}   \\
ATLAS $Z$ $\sqrt{s}=5.02$ TeV  &  208  & 14     & \cite{Aad:2015gta}   \\
{\bf Total LHC} & & {\bf 94} & \\
\midrule
{\bf Total } & & {\bf 1467} & \\
\bottomrule
\end{tabular}
\vspace{4mm}
\caption{\small \label{datasetnew} Same as Table~\ref{dataset}
  for the new datasets that have been added to nNNPDF2.0.
  As opposed to the NC structure function measurements,
  these datasets are presented as absolute distributions
  rather than as as cross-sections ratios.
  We also indicate the total number of data points in the fit,
  combining the NC and CC structure functions with the LHC data. 
}
\end{table}


Starting with the CC measurements from CHORUS,
we fit the inclusive neutrino and anti-neutrino 
double-differential cross-sections, $d^2\sigma^{\nu N}/dxdQ^2$.
After imposing kinematic cuts, the dataset consists of 
$n_{\rm dat}=846$ data points equally distributed between neutrino 
and anti-neutrino beams.
The fitted cross-sections are not corrected for non-isoscalarity 
of the lead target, and therefore the corresponding theory 
calculations take into account effects related to the 
difference between $Z=82$ and $Z=A/2=104$.
The situation is therefore different from the treatment 
of NC nuclear structure functions, where the experimental results
are presented with non-isoscalar
effects already subtracted, as discussed in~\cite{AbdulKhalek:2019mzd}.

In addition to the CHORUS reduced cross-sections, nNNPDF2.0 also 
includes the NuTeV di-muon cross-sections from 
neutrino-iron scattering.
Dimuon events in neutrino DIS are associated with the 
W$^\pm + s~(d) \to c$ scattering process, where the charm 
quark hadronizes into a charmed meson and then decays 
into a final state containing a muon.
This process is dominated by the strange-initiated contributions 
since other initial states are CKM-suppressed, thus providing 
direct sensitivity to the strange quark nuclear PDF.
In fact, the NuTeV di-muon data are known to play an important 
role in studies of proton strangeness in global QCD analyses.
After kinematic cuts, we end up with $n_{\rm dat}=39$ and 37 
data points for the neutrino and the anti-neutrino 
cross-sections, respectively.
Together with the CHORUS cross section data, the 
CC measurements comprise a majority of the
input dataset with a total of $n_{\rm dat}=922$ data points.

Moving now to the LHC electroweak gauge boson cross-sections,
we consider in this work the four datasets that are listed in 
Table~\ref{datasetnew}.
Three of the datasets come from the Run I data-taking period,
corresponding to a per-nucleon center-of-mass energy 
of $\sqrt{s}=5.02$ TeV.
These are the ATLAS Z rapidity distributions~\cite{Aad:2015gta} 
and the CMS W$^\pm$~\cite{Khachatryan:2015hha} and 
Z~\cite{Khachatryan:2015pzs} rapidity distributions,
which contain $n_{\rm dat}=14$, 20, and 12 data points respectively.
Note that ATLAS does not have a published measurement of the 
W$^\pm$ rapidity distributions from Run I and that only preliminary 
results have been presented~\cite{ATLAS-CONF-2015-056}.

In the same way as the CC reduced cross-sections, the LHC measurements 
of electroweak gauge boson production are provided as absolute 
distributions.
In this case, however,
it is possible to construct new observables with the LHC W$^\pm$ and 
Z production data that might be beneficial for nPDF determinations.
For example, the EPPS16 analysis composed and analyzed the 
forward-to-backward ratio, where cross sections at 
positive lepton rapidities are divided by the ones at negative rapidities.
Nevertheless, in this work we choose only to work with the absolute 
rapidity distributions presented in the experimental publications.

In addition to the three Run I results, we add also for the first 
time in an nPDF analysis measurements from Run II corresponding 
to a per-nucleon center-of-mass energy of $\sqrt{s}=8.16$ TeV.
More specifically, the measurements correspond to W$^+$ and 
W$^-$ leptonic rapidity distributions~\cite{Sirunyan:2019dox} from CMS, which 
provide an additional $n_{\rm dat}=48$ data points.
The fact that the amount of data is more than doubled compared 
to the corresponding Run I measurements is a consequence of the 
increase in the CoM energy as well as the higher integrated luminosity. 
In particular, the Run II measurements are based on $\mathcal{L}=173$ 
nb$^{-1}$ compared to $\mathcal{L}=34.6$ nb$^{-1}$ available from Run I.
The CMS Run II data are therefore expected to provide 
important constraints on the quark flavor separation of the nuclear PDFs.

As opposed to the situation in proton-proton collisions, the LHC gauge 
production measurements do not provide information on the correlation 
between experimental systematic uncertainties.
For this reason, we construct the total experimental error by adding the 
various sources of uncertainty in quadrature.
The only source of systematic error which is kept as fully correlated 
among all the data bins of a given dataset is the overall normalization uncertainty.
Note that this normalization uncertainty is correlated within a single 
experiment and LHC data-taking period, but elsewhere is uncorrelated 
between different experiments.

\begin{figure}[t]
\begin{center}
  \includegraphics[width=0.8\textwidth]{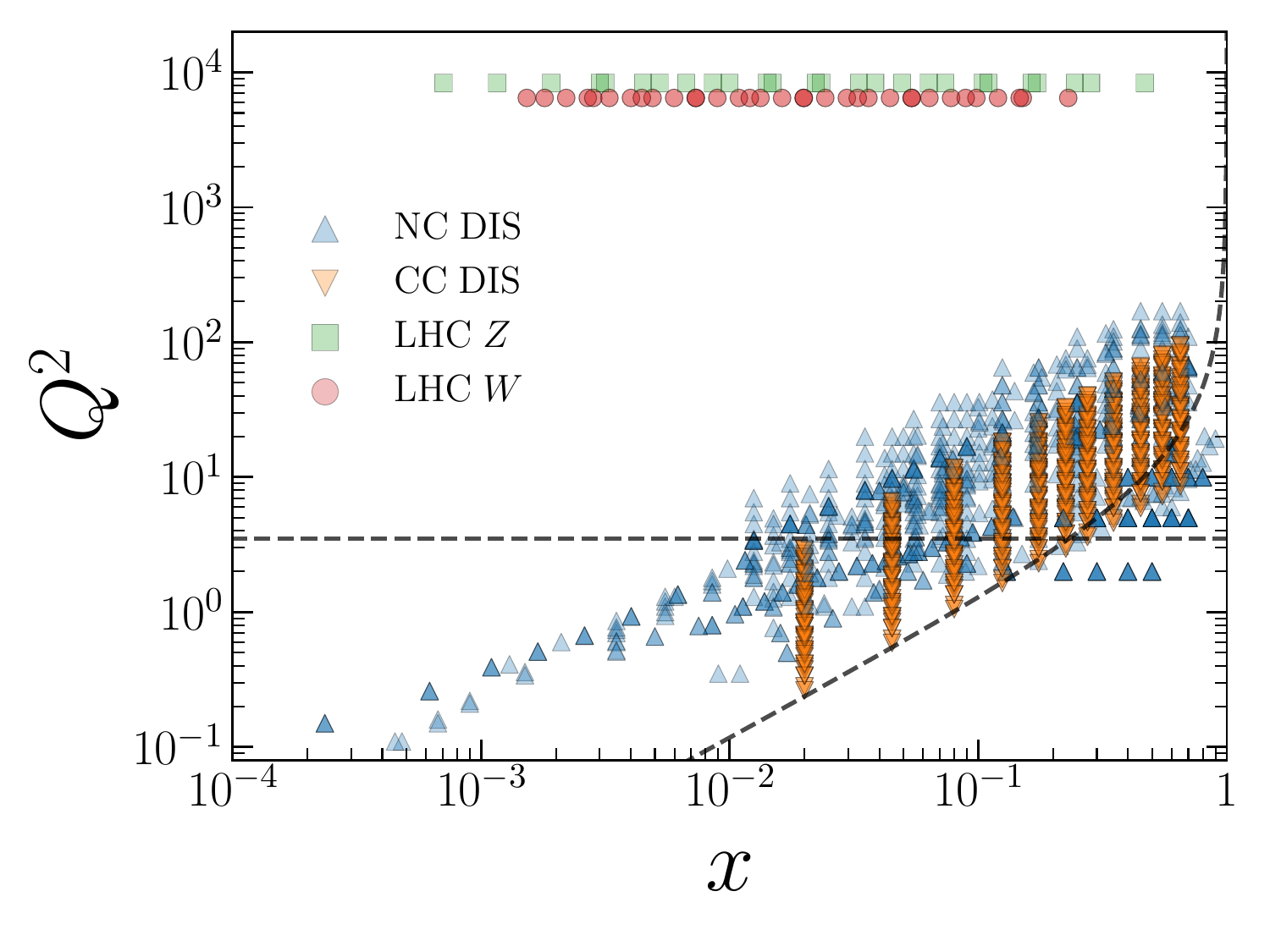}
 \end{center}
\vspace{-0.9cm}
\caption{\small Kinematical coverage in $x$ and $Q^2$ of the data points 
included in the nNNPDF2.0 determination.
  The horizontal dashed and curved dashed lines correspond to 
  $Q^2 = 3.5$ GeV$^2$ and $W^2 = 12.5$
  GeV$^2$, respectively, which are the kinematic cuts imposed in this 
  analysis.
  For each LHC measurement, there are two values of $x$ associated 
  with leading order kinematics of proton-lead scattering being displayed.
  \label{figkinplot}
}
\end{figure}

In order to illustrate the coverage of the experimental data that is 
included in nNNPDF2.0 and summarized in Tables~\ref{dataset} 
and~\ref{datasetnew}, we display in Fig.~\ref{figkinplot} their kinematical 
range in the $(x,Q^2)$ plane.
Here the horizontal dashed and curved dashed lines correspond to 
the kinematic cuts of $Q^2 = 3.5$ GeV$^2$ and 
$W^2 = 12.5$ GeV$^2$, respectively.
In addition, we show for each hadronic data point the 
two values of $x$ corresponding to the 
parton momentum fractions of the incoming 
proton and lead beams, computed to leading order.

There are several interesting observations that one can make regarding 
Fig.~\ref{figkinplot}.
First of all, the LHC proton-lead measurements significantly extend the kinematic 
coverage of the fixed-target
DIS reduced cross-sections, both in terms of $x$ and $Q^2$. 
In particular, the LHC data reside at $Q^2$ values that are orders of magnitude 
larger while the coverage in partonic
momentum fraction is extended down to $x \simeq 10^{-3}$.  
Secondly, the CC reduced cross-sections have a similar coverage compared 
to the NC ones, providing sensitivity to different quark and antiquark 
combinations across the shared medium- to large-$x$ region.
Finally, the kinematics of the LHC W and Z measurements largely 
overlap. The ability to describe them simultaneously can therefore 
demonstrate the compatibility between the experimental data and theoretical 
calculations.

\subsection{Theoretical calculations}
\label{sec:theorycalculations}

\paragraph{DIS structure functions.}
For the NC DIS structure functions we use the same 
theoretical settings as in nNNPDF1.0, i.e. the structure functions 
are evaluated at NLO using {\tt APFEL}~\cite{Bertone:2013vaa} 
in the FONLL-B general-mass variable flavor number 
scheme~\cite{Forte:2010ta}.
The value of the strong coupling constant 
is taken to be the same as in the NNPDF3.1 proton PDF fit, 
$\alpha_S(m_Z)=0.118$, as well as the charm and bottom 
mass thresholds $m_c=1.51$ GeV and $m_b=4.92$ GeV, respectively.
The charm PDF is generated perturbatively by the DGLAP evolution equations
and is thus absent from the $n_f=3$ scheme.
Lastly, the structure functions are processed by the 
{\tt APFELgrid}~\cite{Bertone:2016lga} fast interpolation tables
which allow for efficient evaluations during the PDF fit.

Concerning the CC neutrino reduced cross-sections, 
most of the theory settings are shared with their NC
counterparts.
The main difference is that the heavy quark contributions in the
CC predictions at NLO are accounted for in the FONLL-A scheme 
instead to maintain consistency with the proton baseline. 
Massive $\mathcal{O}\lp \alpha_s^2\rp$ corrections to charm 
production in CC DIS have been presented in 
Ref.~\cite{Berger:2016inr}, and subsequently used to study their 
impact in the determination of the strange content of the nucleon 
in Ref.~\cite{Gao:2017kkx}.
Further details about the implementation of heavy quark mass 
corrections in the NNPDF framework for charged-current scattering can be found in 
Ref.~\cite{Ball:2011uy}.

\begin{figure}[t]
\begin{center}
  \includegraphics[width=0.75\textwidth,angle=0]{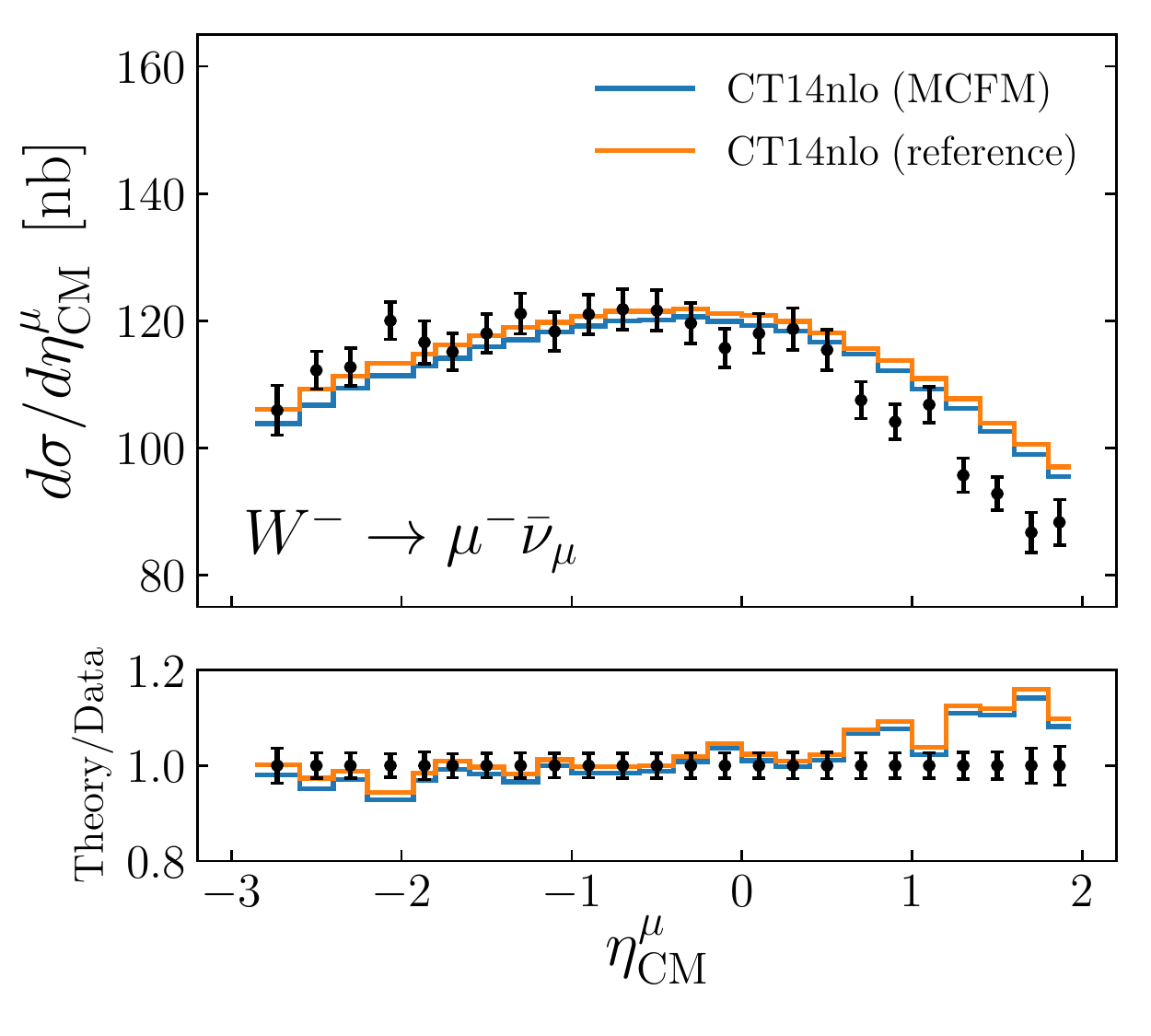}
 \end{center}
\vspace{-0.75cm}
\caption{\small The leptonic rapidity distributions for W$^-$ boson production
  at $\sqrt{s}=8.16$ TeV in the center-of-mass frame.
  Our {\tt MCFM}-based  calculation is compared
  with the theory predictions
  provided in~\cite{Sirunyan:2019dox}
  both as absolute cross-sections (upper) and as
  ratios to the experimental data (lower panel).
  In both cases the CT14 NLO proton PDF set is adopted and
  nuclear corrections are neglected.
  \label{fig:benchmarking}
}
\end{figure}

\paragraph{Hadronic cross-sections.}
The rapidity distributions from W and Z boson production 
in proton-lead collisions are evaluated at NLO using 
{\tt MCFM}~\cite{Boughezal:2016wmq} v6.8 interfaced with
{\tt APPLgrid}~\cite{Carli:2010rw}.
We have ensured that the numerical integration uncertainties 
in the {\tt MCFM} calculations are always much smaller than 
the corresponding experimental errors.
Furthermore, our calculations are benchmarked with the 
reference theoretical values whenever provided by the 
corresponding experimental publications.
To illustrate this benchmarking, we display in 
Fig.~\ref{fig:benchmarking} the muon rapidity distributions for 
W$^-$ boson production at $\sqrt{s}=8.16$ TeV in the 
center-of-mass frame.
Here we compare our {\tt MCFM}-based calculation with the theory 
predictions provided in Ref.~\cite{Sirunyan:2019dox} at the level 
of absolute cross-sections (upper panel) and also as ratios to the 
central experimental values (lower panel).
In both cases, the CT14 NLO proton PDF set is adopted as input and nuclear corrections 
are neglected.
As can been seen by the figure, there is good agreement at the 
$\sim$1\% level between our calculations and the reference results 
provided in the CMS paper.
Similar agreement is obtained for the rest of hadronic datasets 
included in the present analysis.

%
Since the fast interpolation grids are computed in the 
center-of-mass frame of the proton-lead collision,
rapidity bins that are given in the laboratory frame $\eta_{\rm lab}$
are shifted to the center-of-mass frame $\eta_{\rm CM}$ when
required. 
This shift is given by $\eta_{\rm lab} = \eta_{\rm CM}+0.465$
both at $\sqrt{s}=5.02$ and 8.16 center-of-mass energies.
Lastly, we note that the same theoretical settings were used for the 
evaluation of W and Z production in proton-proton collisions
for the baseline NNPDF3.1 fit.

\section{Fitting methodology}
\label{sec:fitting}

In this section, we describe the fitting methodology that was adopted 
for the nNNPDF2.0 determination,
focusing in particular on the differences and improvements 
with respect to the nNNPDF1.0 analysis.
We begin by establishing the PDF notation and conventions
that will be used throughout this work.
We then detail our strategy to parameterize the nuclear parton
distributions, including the treatment of 
sum rules, preprocessing factors, and the 
proton boundary condition.
Lastly, we outline the implementation of the cross-section
positivity constraint.

\subsection{Notation and conventions}
\label{sec:defs}

Parton distributions can be parametrized in 
a number of different bases, all of which are
related by linear transformations.
Two popular ones are the flavor basis, corresponding
to the individual quark and anti-quark PDFs, and the evolution basis,
given by the eigenvectors of the DGLAP
evolution equations~\cite{Ball:2014uwa}.
For illustrative purposes we consider here
three active quarks,
a vanishing strangeness asymmetry, and heavy quark 
PDFs that are generated via perturbative evolution.
At the parameterization scale $Q_0 < m_c$, 
the flavor basis is composed of the
$u,\, \bar{u},\, d,\, \bar{d},\, s,\,$ and $g$ PDFs, 
with $s=\bar{s}$, while the corresponding evolution basis is given by
$\Sigma,\, T_3,\, T_8,\, V,\, V_3,$ and $g$.
Expressed in terms of the elements of the flavor basis, the evolution basis
distributions are given by
\bea
\Sigma(x,Q_0) &=& \lp u^+ + d^+ + s^+ \rp (x,Q_0)\, ,  \nonumber\\
T_3(x,Q_0) &=& \lp u^+ - d^+  \rp (x,Q_0)\, ,\nonumber \\
T_{8}(x,Q_0) &=& \lp u^+ + d^+ - 2s^+\rp (x,Q_0)\, , \label{eq:evolbasis}\\
V(x,Q_0) &=& \lp u^- + d^-  \rp (x,Q_0)\, ,\nonumber\\
V_3(x,Q_0) &=& \lp u^- - d^-  \rp (x,Q_0)\, , \nonumber
\eea
where $q^\pm = q \pm \bar{q}$.
Although the results of an nPDF analysis should be
independent of the basis choice for the parameterization, 
some bases offer practical advantages, for example in 
the implementation of the sum rules which are discussed later.

In this work, we define $f^{(N/A)}$ to be the PDF for the 
flavor $f$ associated to the average nucleon $N$ bound in
a nucleus with atomic number $Z$ and mass number $A$.
This object can be written as
\be
\label{eq:qNAdefinition}
f^{(N/A)}(x,Q_0) = \frac{Z}{A} f^{(p/A)}(x,Q_0)  + \lp 1-\frac{Z}{A} \rp f^{(n/A)}(x,Q_0) \, ,
\ee
where $f^{(p/A)}$ and $f^{(n/A)}$ represent the PDFs of a proton and a neutron,
respectively, bound in the same nucleus of mass number $A$.
Assuming isospin symmetry, the PDFs of the neutron in Eq.~(\ref{eq:qNAdefinition})
can be expressed in terms of the proton PDFs via
\bea
u^{(n/A)}(x,Q_0) = d^{(p/A)}(x,Q_0) \, , \quad && \nonumber
\bar{u}^{(n/A)}(x,Q_0) = \bar{d}^{(p/A)}(x,Q_0) \\
d^{(n/A)}(x,Q_0) = u^{(p/A)}(x,Q_0)\, , \quad \label{eq:isospin}  &&
\bar{d}^{(n/A)}(x,Q_0) = \bar{u}^{(p/A)}(x,Q_0) \\
s^{(n/A)}(x,Q_0) = s^{(p/A)}(x,Q_0)\, , \quad  &&  \nonumber
g^{(n/A)}(x,Q_0) = g^{(p/A)}(x,Q_0) \, .
\eea
Using the relations above, the strange and gluon distributions
of the average bound nucleon ($f^{(N/A)}$)
and bound proton ($f^{(p/A)}$) become equivalent, while 
the up and down flavored distributions of the average 
bound nucleon are instead linear combinations 
of the bound proton PDFs with coefficients depending on the values of $A$ and $Z$.

\subsection{nPDF parameterization}
\label{sec:parametrisation}

The cross-sections for hard scattering processes
involving heavy nuclei can be expressed 
either in terms of $f^{(N/A)}$ or $f^{(p/A)}$.
The two options are fully equivalent, as is highlighted
by the LO expressions of the observables collected
in Appendix~\ref{sec:LOxsecs}.
One can therefore
choose to parameterize either the PDFs of
the average bound nucleon or those 
of the bound proton in a global nPDF 
analysis.
In this study we choose the latter option for two main reasons.
First, the connection with the free-proton boundary condition 
is more straightforward.
In addition, the valence sum rules 
for non-isoscalar nuclei are independent of $A$ and $Z$.
If instead $f^{(N/A)}$ is parameterized, one of the valence sum rules
would depend on the value of the $Z/A$ ratio and thus be different for each nuclei, 
making it inconvenient from the parameterization point of view.

The relation between $f^{(N/A)}$ and $f^{(p/A)}$ is trivial also for
PDF combinations that comprise the evolution basis.
Consider for example the total quark singlet, where the flavor combination 
is the same in the proton and in the neutron, i.e. $\Sigma^{(p/A)}=\Sigma^{(n/A)}$.
From Eq.~(\ref{eq:qNAdefinition}), it simply follows that $\Sigma^{(N/A)}=\Sigma^{(p/A)}$.
The same equivalence holds also for $V$ and $T_8$.
However, the distinction is important for $T_3$ and $V_3$, for which we have
\be
V_3^{(N/A)} = \frac{Z}{A} V_3^{(p/A)}   + \lp 1-\frac{Z}{A}\rp   V_3^{(n/A)}
= \lp  \frac{2Z}{A} -1\rp  V_3^{(p/A)}  \, ,
\ee
\be
T_3^{(N/A)} = \frac{Z}{A} T_3^{(p/A)}   + \lp 1-\frac{Z}{A}\rp  T_3^{(n/A)}
= \lp \frac{2Z}{A} -1\rp  T_3^{(p/A)}  \, ,
\ee
so there is an overall rescaling factor of $(2Z/A-1)$ between $f^{(N/A)}$ and $f^{(p/A)}$.
The main consequence of this relation is highlighted by 
assuming an isoscalar nucleus, with $Z=A/2$.
In this case, $V_3^{(N/A)}=T_3^{(N/A)}=0$ 
while their bound proton counterparts are different from zero.
Unless otherwise indicated, the nPDFs discussed in this section will
always correspond to those of the bound proton.

\paragraph{Fitting basis and functional form.}
In our previous nNNPDF1.0 analysis, we parameterized only 
three independent evolution basis distributions at the initial scale $Q_0$,
namely the total quark singlet $\Sigma(x,Q_0)$,
the quark octet $T_{8}(x,Q_0)$, and the gluon nPDF $g(x,Q_0)$.
From the LO expression of  Eq.~(\ref{eq:F2NC}), it is clear that
NC structure functions are  sensitive only to a 
specific combination of $\Sigma$ and $T_8$ for isoscalar nuclei, 
in particular $\Sigma+T_8/4$, while the contribution
proportional to $T_3$ vanishes. 
In other words,  $\Sigma$ and
$T_8$ are strongly anti-correlated and only the combination 
$\Sigma + T_8/4$ can be meaningfully determined from the data.

The picture is quite different in the present study,
where the availability of charged current DIS data and electroweak gauge
boson production cross-sections in proton-lead collisions allow additional 
elements of the evolution PDF basis to be parameterized (see App.~\ref{sec:LOxsecs}).
If non-isoscalar effects are neglected,
there is only a single distribution to be added
to our evolution basis choice, namely the total valence quark
combination $V =  u^- + d^- $.
However, non-isoscalar corrections are necessary for 
the targets considered in this analysis, particularly for lead.
In this case, the quark triplet
$T_3 = u^+ - d^+  $ and the valence triplet $V_3= u^- - d^- $ 
must also be parameterized.
Note that since $T_3$ and $V_3$ correspond to bound protons, they will be different
from zero even for isoscalar nuclei. However, in such cases their contribution
to the scattering cross-section vanishes and therefore the data provides no constraint
on these combinations.

Putting together these considerations, in this work we parameterize six independent
PDF combinations in the evolution basis as follows
\bea
x\Sigma^{(p/A)}(x,Q_0) &=&x^{\alpha_\Sigma} (1-x)^{\beta_\Sigma} {\rm NN}_\Sigma(x,A) \, , \nonumber \\
xT_3^{(p/A)}(x,Q_0) &=&x^{\alpha_{T_3}} (1-x)^{\beta_{T_3}} {\rm NN}_{T_3}(x,A) \, , \nonumber \\
xT_8^{(p/A)}(x,Q_0) &=&x^{\alpha_{T_8}} (1-x)^{\beta_{T_8}} {\rm NN}_{T_8}(x,A) \, , \label{eq:param2} \\
xV^{(p/A)}(x,Q_0) &=&B_{V}x^{\alpha_V} (1-x)^{\beta_V} {\rm NN}_V(x,A) \, , \nonumber \\
xV_3^{(p/A)}(x,Q_0) &=&B_{V_3}x^{\alpha_{V_3}} (1-x)^{\beta_{V_3}} {\rm NN}_{V_3}(x,A) \nonumber\, ,  \\
xg^{(p/A)}(x,Q_0) &=&B_gx^{\alpha_g} (1-x)^{\beta_g} {\rm NN}_g(x,A) \, . \nonumber
\eea
In these expressions, ${\rm NN}_f(x,A)$ stands for the value of the
neuron in the output layer of the neural network associated to each specific
distribution.
As was done in nNNPDF1.0, we use a single artificial neural network
consisting of an input layer, one hidden layer with sigmoid activation
function, and an output layer with linear activation function.
The input layer contains three neurons that take as input the values of the
momentum fraction $x$, $\ln(1/x)$, and atomic mass number $A$,
respectively.
Since the hidden layer contains 25 neurons,
there are a total of $N_{\rm par}=256$ free parameters
(weights and thresholds) in the neural network used to model our nPDFs.

The neural-net parameterization in Eq.~(\ref{eq:param2}) is then
complemented by three normalization coefficients 
$B_g$, $B_V$, and $B_{V_3}$ which are fixed by the sum rules,
and by twelve preprocessing exponents $\alpha_f$ and $\beta_f$
which are fitted simultaneously with the network parameters.
Since our proton baseline is a variant of the NNPDF3.1 global
NLO fit~\cite{Ball:2017nwa}
with perturbative charm, we adopt for consistency the
same parameterization scale of $Q_0=1$ GeV.

It is important to emphasize here that the parameterization
in Eq.~(\ref{eq:param2}) is valid from $A=1$ (free-proton)
up to $A=208$ (lead).
As a result, the nNNPDF2.0 analysis incorporates an independent
determination of the free-proton PDFs, where agreement
with the proton PDF baseline is enforced by means of a boundary 
condition as explained below.
This is a relevant distinction, implying that the $A=1$
PDF can by construction differ slightly from 
our proton baseline, for example as a result of positivity 
constraints that are more general in the former case, or
by new information contained in the LHC proton-lead cross-sections.

\paragraph{Sum rules.}
For every nuclei, we assume that the fitted 
nuclear PDFs satisfy the same valence and momentum sum 
rules as in the proton case.
The sum rules are implemented via an overall normalization
factor in the PDF parameterization, 
which are adjusted each time the neural network parameters are 
modified in order to ensure that the sum rules remain satisfied.
Note that these sum rules need only to be applied at the input scale 
$Q_0$, since the properties of DGLAP perturbative evolutions guarantee
that they will also be satisfied for other $Q > Q_0$.
First, energy conservation leads to the momentum sum rule constraint,
\be
\label{eq:MSR}
\int_0^1 dx \,x \left(\Sigma^{(p/A)}(x,Q_0) + g^{(p/A)}(x,Q_0)\right) = 1 \, , \quad \forall \, A \, ,
\ee
which is enforced by fixing the gluon normalization to be
\be
\label{eq:NormG}
B_g(A) = \frac{1 - \int_0^1 dx\, x\Sigma^{(p/A)}(x,Q_0)}{\int_0^1 dx\, xg^{(p/A)}(x,Q_0)} \, ,
\ee
where the denominator of Eq.~(\ref{eq:NormG}) is evaluated using 
Eq.~(\ref{eq:param2}) and setting $B_g=1$.
Our nuclear PDFs are also constructed to comply with the three valence sum 
rules that follow from the valence quark quantum numbers of the proton:
\be
\label{eq:valencesr1}
\int_0^1 dx~ \left(u^{(p/A)}(x,Q_0) - \bar{u}^{(p/A)}(x,Q_0)\right) = 2 \, ,\quad \forall \, A \, ,
\ee
\be
\label{eq:valencesr2}
\int_0^1 dx~ \left(d^{(p/A)}(x,Q_0) - \bar{d}^{(p/A)}(x,Q_0)\right) = 1 \, ,\quad \forall \, A \, ,
\ee
\be
\label{eq:valencesr3}
\int_0^1 dx~ \left(s^{(p/A)}(x,Q_0) - \bar{s}^{(p/A)}(x,Q_0)\right) = 0 \, ,\quad \forall \, A \, ,
\ee
where the final relation is trivially satisfied due to our inherent flavor assumption
of $s=\bar{s}$.

To implement the former two valence sum rules in our analysis,
we first must derive the corresponding constraints in the evolution basis. 
Adding Eqns.~(\ref{eq:valencesr1}) and~(\ref{eq:valencesr2}) results in
\begin{align}
\label{eq:valencesr4}
\int_0^1 dx~ &\left(u^{(p/A)}(x,Q_0) - \bar{u}^{(p/A)}(x,Q_0)+ d^{(p/A)}(x,Q_0) - \bar{d}^{(p/A)}(x,Q_0)\right) = \nonumber \\
&\int_0^1 dx~ V^{(p/A)}(x,Q_0) = 3\, , \quad \forall \, A \, .
\end{align}
This condition can then be implemented in the same way as the 
momentum sum rule, namely by setting the overall normalization 
factor of $V$ as
\be
\label{eq:NormV}
B_V(A) = \frac{3}{\int_0^1 dx\, V^{(p/A)}(x,Q_0,A)}\, ,
\ee
where the denominator of Eq.~(\ref{eq:NormV}) is evaluated using 
Eq.~(\ref{eq:param2}) and setting $B_V=1$.

The second valence sum rule in the evolution basis 
is the one related to the quark valence triplet $V_3$.
Subtracting Eq.~(\ref{eq:valencesr2}) from~(\ref{eq:valencesr1})
gives
\begin{align}
\label{eq:valencesr5}
\int_0^1 dx~ &\left(u^{(p/A)}(x,Q_0,A) - \bar{u}^{(p/A)}(x,Q_0,A) - d^{(p/A)}(x,Q_0,A) + \bar{d}^{(p/A)}(x,Q_0,A)\right) = \nonumber\\ 
&\int_0^1 dx~ V_3^{(p/A)}(x,Q_0,A) = 1 \, , \quad \forall \, A \, .
\end{align}
which again is imposed by setting
\be
\label{eq:NormV3}
B_{V_3}(A) = \frac{1}{\int_0^1 dx\, V_3^{(p/A)}(x,Q_0,A)}\, ,
\ee
where the denominator of Eq.~(\ref{eq:NormV3}) is evaluated using 
Eq.~(\ref{eq:param2}) with $B_{V_3}=1$.

In this analysis, 
the normalization pre-factors $B_g(A)$, $B_V(A)$, and $B_{V_3}(A)$ 
are computed using the trapezoidal rule for
numerical integration between
$x_{\rm min}=10^{-9}$ and $x_{\rm max}=1$
each time the fit parameters are 
updated by the minimization procedure.
With a suitable choice of the ranges for the preprocessing
exponents (see discussion below), we guarantee that each quark
combination satisfies the corresponding physical integrability conditions.
Lastly, we have confirmed that individual replicas
satisfy the sum rules with a precision
of a few per-mille or better.

An interesting question in the context of nuclear
global QCD analyses is the extent to which
theoretical constraints such as the sum rules are satisfied by the
experimental data when not explicitly imposed.
In fact, it was shown in Ref.~\cite{Ball:2011uy} that the momentum
sum for the free proton agrees with the QCD expectation 
within $\sim 1\%$ in this scenario.
Here we will revisit this analysis for the nuclear case,
and will present in Sect.~\ref{sec:SR} variants of
the nNNPDF2.0 fit where either the momentum
sum rule, Eq.~(\ref{eq:MSR}), or the valence sum rule,
Eq.~(\ref{eq:valencesr4}), is not enforced.
Interestingly, we will find that the experimental data
is in agreement with sum rule expectations, 
albeit within larger uncertainties, 
demonstrating the remarkable consistency of the 
nuclear global QCD analysis.

\vspace{-0.1cm}
\paragraph{Preprocessing exponents.}
The $x^{\alpha_f}(1-x)^{\beta_f}$ polynomial pre-factors
appearing in Eq.~(\ref{eq:param2}) are included to increase
the efficiency of the parameter optimization, since they
approximate the general PDF behavior 
in the small- and large-$x$ asymptotic limits~\cite{Ball:2016spl}.
Note that the exponents $\alpha_f$ and $\beta_f$
are $A$-independent, implying that $A$ dependence of the
nPDFs will arise completely from the output of the neural network.
As in the case of the nNNPDF1.0 analysis, the values of $\alpha_f$
and $\beta_f$ are fitted for each Monte Carlo replica
on the same footing as the weights and thresholds of the
neural network.

The ranges of the preprocessing parameters are determined
both by physical considerations and by empirical observations.
First of all, the lower limit of the small-$x$ parameter is
set so that each PDF combination satisfies various
integrability conditions.
In particular, the non-singlet combinations 
$xV$, $xV_3$, $xT_3$, and $xT_8$
must tend to zero at small-$x$, else the valence sum rules
and other relations such as the Gottfried sum 
rule~\cite{Forte:1992df,Abbate:2005ct} 
would be ill-defined.
Moreover, the singlet combinations $x\Sigma$ and $xg$ must be exhibit 
finite integrable behavior as $x\rightarrow0$, 
otherwise the momentum integral cannot be computed.
Concerning the large-$x$ exponents $\beta_f$, the lower limits 
of their ranges ensure that PDFs vanish in the elastic limit, 
while the upper limit is determined largely from general arguments 
related to sum rule expectations. 
In general, however, the upper values of both 
$\alpha_f$ and $\beta_f$ are chosen to be sufficiently 
large to allow flexibility 
in exploring the parameter space while keeping 
fit efficiency optimal.

Under these considerations,
we restrict the parameter values for the pre-processing factors
during the fit to the following intervals,
\begin{align}
\label{eq:preprocessing}
\alpha_\Sigma & \in [-1,5]~~([-1,1]) \, , & \beta_\Sigma &\in [1,10]~~([1,5]) \, ,\nonumber \\
\alpha_g & \in [-1,5]~~([-1,1]) \, ,      & \beta_g &\in [1,10]~~([1,5]) \, ,\nonumber\\
\alpha_V & \in [0,5]~~([1,2]) \, ,        & \beta_V &\in [1,10]~~([1,5]) \, ,\\
\alpha_{T_8} & \in [-1,5]~~([-1,1]) \, ,  & \beta_{T_8} &\in [1,10]~~([1,5]) \, ,\nonumber\\
\alpha_{V_3} & \in [0,5]~~([1,2]) \, ,    & \beta_{V_3} &\in [1,10]~~([1,5]) \, , \nonumber\\
\alpha_{T_3} & \in [-1,5]~~([-1,1]) \, ,  & \beta_{T_3} &\in [1,10]~~([1,5]) \, ,\nonumber
\end{align}
where the ranges in parentheses are those used to randomly select the initial values
of $\alpha_f$ and $\beta_f$ at the start of the minimization.
We do not impose any specific relation between the 
small- or large-$x$ exponents of the different quark combinations, 
so that each are fitted independently.
It is also worth emphasizing here that
the neural network has the ability to compensate 
for any deviations in the shape of the preprocessing function,
so the dependence on $x$ and $A$ of the nPDFs in the data region 
will be dominated by the neural network output.
This implies that the preprocessing exponents will primarily
affect the results in the extrapolation regions.

\begin{figure}[t]
\begin{center}
  \includegraphics[width=0.95\textwidth]{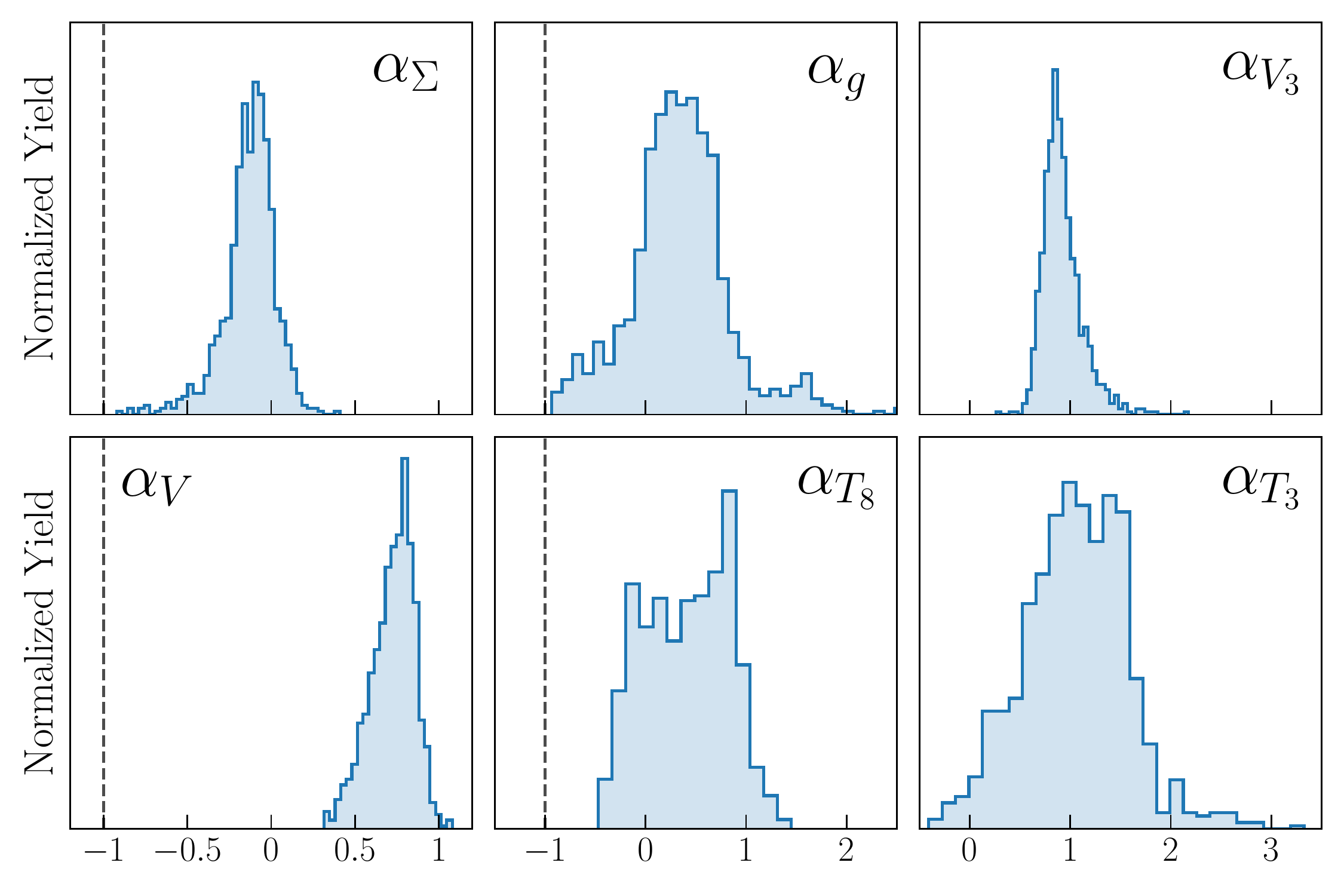}
 \end{center}
\vspace{-0.5cm}
\caption{\small Same as Fig.~\ref{fig:beta_exp} for
   the fitted 
   small-$x$ preprocessing 
  exponents $\alpha_f$ 
}
\label{fig:alpha_exp}
\end{figure}

\begin{figure}[t]
\begin{center}
  \includegraphics[width=0.95\textwidth]{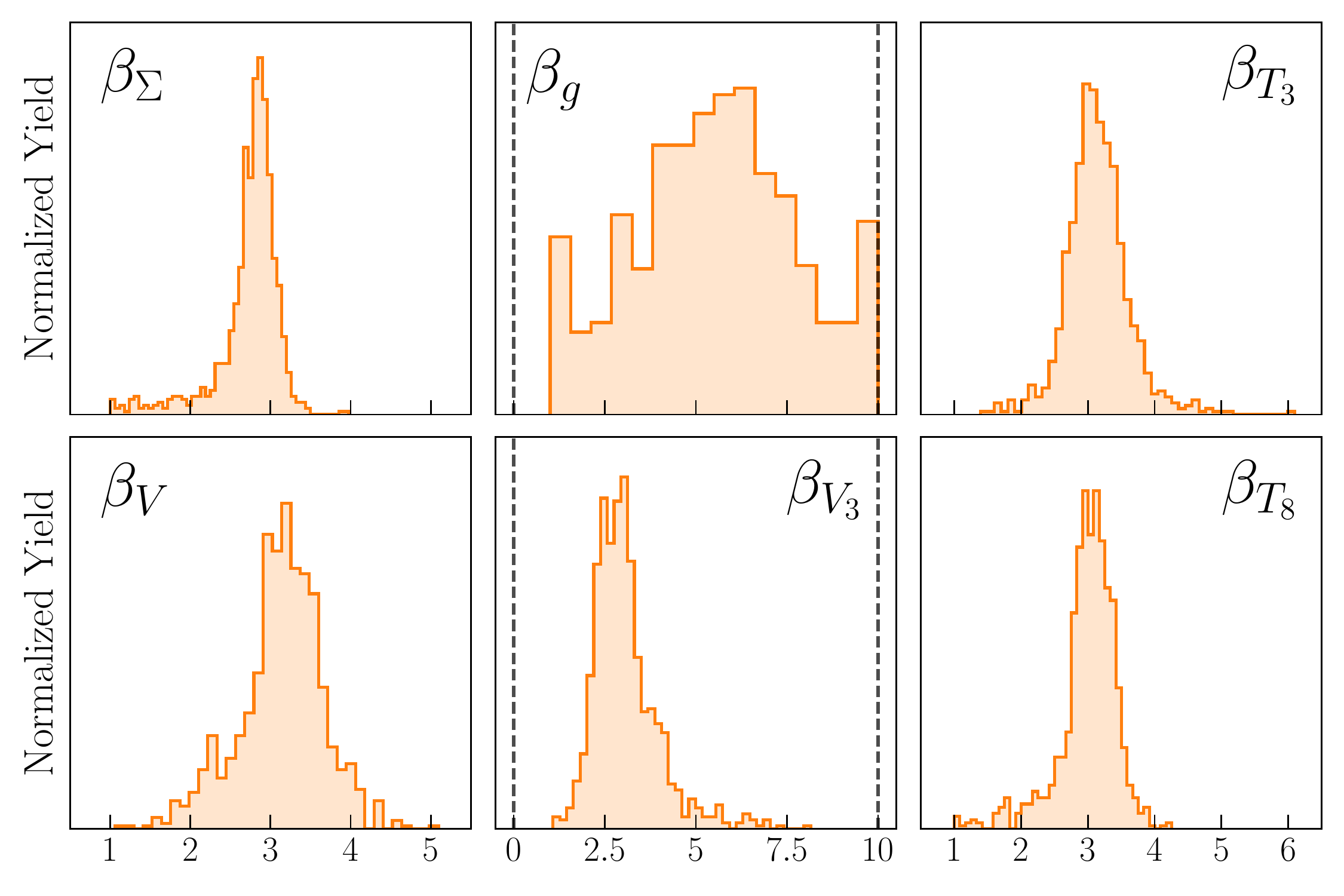}
 \end{center}
\vspace{-0.5cm}
\caption{\small The probability distribution associated to the fitted 
   large-$x$ preprocessing 
  exponents $\beta_f$ computed with the $N_{\rm rep}=1000$ replicas of the nNNPDF2.0
  NLO set.
  The ranges for which these exponents
  are allowed to vary, Eq.~(\ref{eq:preprocessing}),
  are indicated by horizontal dashed lines.
}
\label{fig:beta_exp}
\end{figure}

To illustrate the values of the small and
large-$x$ preprocessing exponents preferred by the experimental
data, we display in Figs.~\ref{fig:alpha_exp} and~\ref{fig:beta_exp}
the probability distributions associated with the $\alpha_f$ and $\beta_f$
exponents, respectively, computed using the $N_{\rm rep}=1000$ replicas of the nNNPDF2.0
analysis.
Note how these exponents are restricted to lie in the interval given by
Eq.~(\ref{eq:preprocessing}).
For $T_3$ and $T_8$, we can see that despite
not imposing the strict integrability requirement that
$\alpha_f > 0$, it is still being satisfied for a large 
majority of the replicas, especially for $T_3$.
Interestingly, the gluon seems to prefer a valence-like
behavior at small-$x$.
However, such behaviour is only observed 
at the parameterization scale and as soon as 
$Q > Q_0$, DGLAP evolution drives it to its
expected singlet-like behavior.

Concerning the behavior of the large-$x$ exponents $\beta_f$, 
we find that they are reasonably well
constrained for the quark distributions,
where the best-fit values are located in a 
region close to $\beta_f \simeq 3$.
The fact that they share similar $\beta_f$ exponents
can be explained by the fact that in the large-$x$ region 
the quark combinations are dominated by valence 
components.
Interestingly, a best-fit value of $\beta_f \simeq 3$ for the 
valence quarks is consistent with the expectations from 
the QCD counting rules, as discussed in~\cite{Ball:2016spl}.
Furthermore, the best-fit value for $\beta_g$
is also consistent with the QCD counting rules 
prediction of $\beta_g \simeq 5$, although with
significant uncertainties.
The fact that $\beta_g$ is found to vary in a wide range
is due to the lack of direct constraints on the 
large-$x$ nuclear gluon PDF in the present analysis.

\paragraph{The free-proton boundary condition.}
\label{sec:freeproton}
As was done in our previous study, we again implement
the condition that the proton PDF baseline, obtained with consistent theoretical
and methodological choices, is reproduced when $A\rightarrow1$.
This condition should be constructed to 
match the free-proton distributions
not only in terms of central values
but also at the level of PDF uncertainties.
In other words, it should allow a full propagation of the information 
contained in the proton baseline,
which is particularly important to constrain the 
nPDFs of relatively light nuclei.
Note, however, that 
 for the reasons explained above,
 the nNNPDF2.0 $A=1$ set will in general not be strictly identical
 to the corresponding proton baseline.

The proton boundary condition constraint is implemented by adding a quadratic 
penalty term to the $\chi^2$ of the form
\be
\label{eq:chi2}
\chi^2 \to \chi^2 +
\lambda_{\rm BC} \sum_{f}\sum_{i=1}^{N_x} \lp q_f^{(p/A)}(x_i,Q_0,A=1)
- q_f^{(p)}(x_i,Q_0) \rp^2 \, ,
\ee
where the sum over flavors $f$ runs over the six independent elements
in the PDF evolution basis.
Since the properties 
 of DGLAP evolution ensure that the distributions for 
 $Q > Q_0$ also satisfy the condition, only the PDFs at
 the parameterization scale $Q_0$ enter the penalty term.
 In Eq.~(\ref{eq:chi2}), we use a grid with $N_x=60$ points of which
 10 are distributed logarithmically from $x_{\rm min}=10^{-3}$ to $x_{\rm mid}=0.1$ and
 the remaining 50 points are linearly distributed from $x_{\rm mid}=0.1$ to $x_{\rm max}=0.7$.
 The value of the boundary condition
 hyper-parameter is fixed to be $\lambda_{\rm BC}=100$
 as was done in the previous nNNPDF1.0 analysis.
 For such a value, we find that the
 central values and uncertainties of the proton baseline
 are reasonably well described (see Sect.~\ref{sec:results}).

In this analysis the proton baseline, $f^{(p)}(x,Q_0)$, is taken to be a variant of the
NNPDF3.1 NLO fit~\cite{Ball:2017nwa} with perturbative charm, 
where the neutrino DIS cross-sections
from NuTeV and CHORUS are removed along with the di-muon production
measurements in proton-copper collisions from the E605 experiment~\cite{Moreno:1990sf}.
As such, the proton baseline not only avoids double counting of the CC 
DIS data but also excludes constraints from heavy nuclear target
data where nuclear effects are neglected.
This choice is different to that used for nNNPDF1.0, where 
the global NNPDF3.1 fit was used and double-counting of 
experimental data was not an issue.
%

\begin{figure}[t]
\begin{center}
  \includegraphics[width=0.99\textwidth]{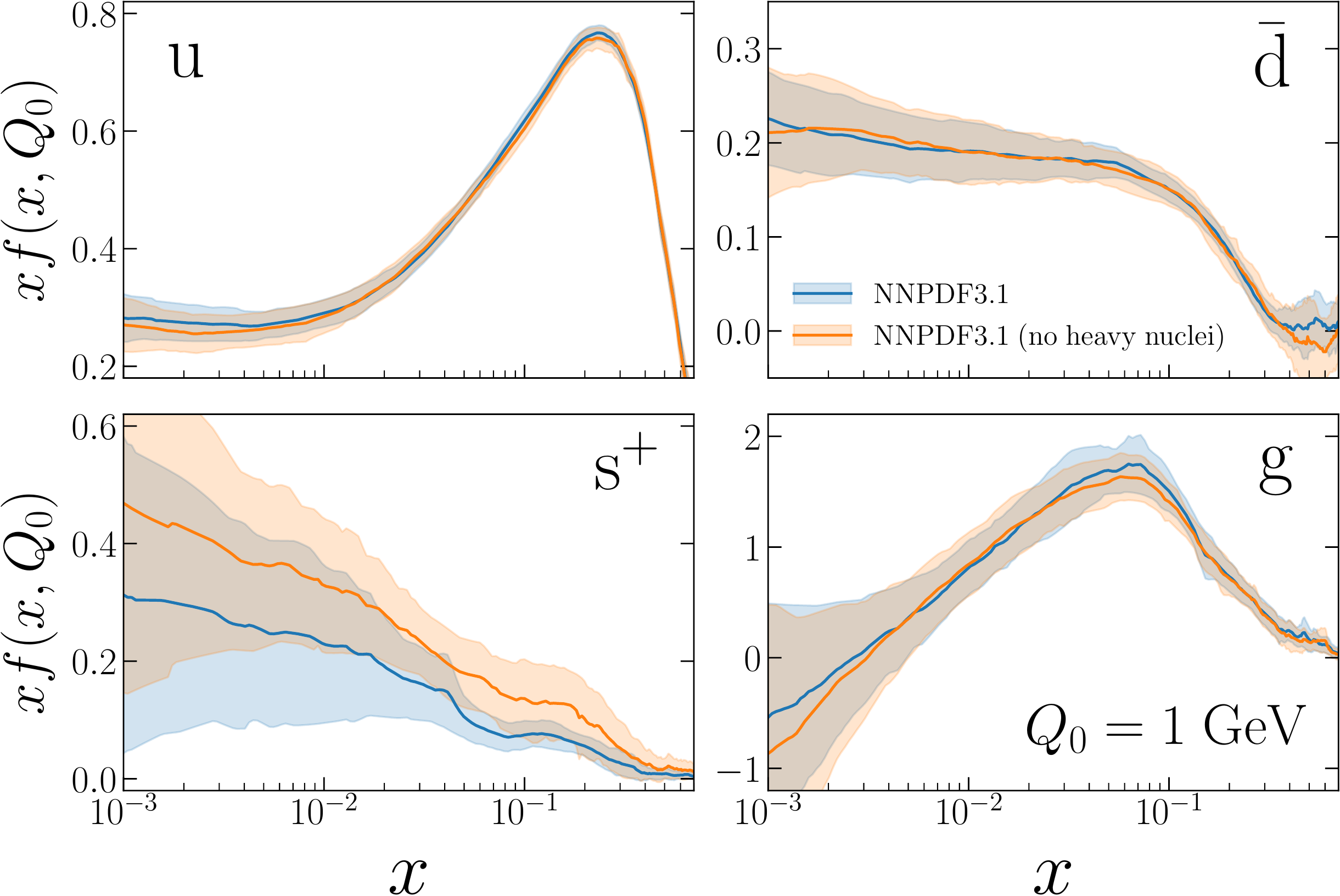}
 \end{center}
\vspace{-0.5cm}
\caption{\small A comparison between the global NNPDF3.1 free-proton analysis
  with its variant with heavy nuclear data excluded.
  We show results for the up quark, down antiquark, total strangeness, and the gluon at $Q_0=1$ GeV.
  The comparison is presented for
  the range of $x$ for which the proton boundary condition is implemented
  in nNNPDF2.0 using Eq.~(\ref{eq:chi2}).
  The PDF uncertainty bands correspond to 90\% CL intervals.
}
\label{fig:NNPDF31_BC_comp}
\end{figure}

In order to illustrate the differences between the free-proton
boundary condition used in nNNPDF1.0 and that employed in the present
analysis, we compare in Fig.~\ref{fig:NNPDF31_BC_comp} the NNPDF3.1 NLO global
and no heavy nuclear fits at the initial parameterization scale of $Q_0=1$ GeV.
Displayed are the gluon, up quark, down sea quark,
and total strange PDFs in the
range of $x$ with which the proton boundary condition 
is constrained by Eq.~(\ref{eq:chi2}).
Here one can see that removing the heavy nuclear data 
from NNPDF3.1 results in a moderate
increase of the PDF uncertainties associated to the quarks
as well as an upward shift of the central value of
the total strange distribution.
The former effect is primarily a consequence of 
information loss on quark flavor separation with
the removal of neutrino scattering measurements.
The strangeness feature, on the other hand, 
arises due to the absence of sensitivity 
from the NuTeV neutrino dimuon cross-sections, 
resulting in an upward pull by the ATLAS $W,Z$ 2011 rapidity distributions 
which are known to produce an enhanced strange
with respect to the up and down quark sea. 
The results of Fig.~\ref{fig:NNPDF31_BC_comp} highlight the importance of a 
consistent choice of the free-proton baseline in order to draw solid conclusions 
on the nuclear modifications, for example those associated to the nucleon's 
strange content.

In order to ensure that all central values
and PDF uncertainties are reproduced,
we select a different replica from the NNPDF3.1 
proton baseline when constructing Eq.~(\ref{eq:chi2}) 
for each replica of nNNPDF2.0. 
Since we perform a large $N_{\rm rep}$ number of fits to 
estimate the uncertainties in nNNPDF2.0, 
we are able to propagate the necessary information 
contained in NNPDF3.1 to the resulting nPDFs in a robust manner.
Lastly, we note that Eq.~(\ref{eq:chi2}) is the only place in the analysis
where the free-proton NNPDF3.1 baseline is inserted.
In other parts of the fit where a free-nucleon PDF is required, for example
in the theoretical predictions of the proton-lead scattering cross-sections,
the nNNPDF2.0 set with $A=1$ is used instead.

\subsection{Cross-section positivity}
\label{sec:positivity}

While parton distributions are scheme-dependent and thus not necessarily positive-definite
beyond leading order in perturbative QCD, physical cross-sections constructed from 
them are scheme independent and should be positive-definite in the region of validity
of the perturbative expansion.\footnote{A recent study~\cite{Candido:2020yat} suggests, however,
that from a practical point of view  PDFs in the $\overline{\rm MS}$-scheme
should also satisfy positivity beyond the LO approximation in the perturbative region.}
In the NNPDF family of proton PDF fits, the requirement that cross-sections 
remain positive is implemented by adding to the $\chi^2$ a penalty term 
in the presence of negative cross-section values~\cite{Ball:2014uwa}.
The cross-sections that enter this penalty term
correspond to theoretical predictions based on 
pseudo-data generated for representative 
processes that are directly sensitive
to a sufficient number of PDF combinations.

In the nNNPDF1.0 analysis, cross-section positivity was not imposed
and led to some observables, such as the longitudinal structure
function $F_L(x,Q^2)$, becoming negative at small-$x$ values
outside the data region.
To bypass this problem, and also to improve the methodological consistency
with the free-proton baseline, in nNNPDF2.0 we impose the positivity of physical
cross-sections for all nuclei used in the fit 
by adding a suitable penalty to the figure of merit.
In this case, the positivity penalty is expressed as
\be
\label{eq:positivity}
\chi^2 \to \chi^2 + \lambda_{\rm pos} \sum_{l=1}^{N_{\rm pos}}\sum_{j=1}^{N_A} \sum_{i_l=1}^{N_{\rm dat}^{(l)}} {\rm max}\lp 0, -\mathcal{F}_{i_l}^{(l)}(A_j) \rp \, ,
\ee
for $N_{\rm pos}$ positivity observables $\mathcal{F}^{(l)}$. 
Each of the observables
contain $N_{\rm dat}^{(l)}$ kinematic points that are chosen to
cover an adequately large region of phase space
relevant to various PDF combinations, 
as we discuss in more detail below.
The computed observables are summed over
all $N_A$ nuclei for which we have experimental data, as
listed in Tables~\ref{dataset} and~\ref{datasetnew}, 
as well as for the free-proton at $A=1$.
Finally, the value of the hyper-parameter $\lambda_{\rm pos} = 1000$ is 
determined by manual inspection of the optimization process and is 
chosen so that positivity is satisfied without distorting the training 
on the real experimental data.\footnote{In future work
it might be advantageous to determine dynamically the fit hyper-parameters 
such as $\lambda_{\rm BC}$ and $\lambda_{\rm pos}$ using 
the hyper-optimization method presented in Ref.~\cite{Carrazza:2019mzf}.}

\begin{table}[htbp]
  \footnotesize
  \begin{center}
    \renewcommand{\arraystretch}{1.55}
    \begin{tabular}{|c| c| c|c |}
      \toprule
      Observable  &  LO expression  &  $N_{\rm dat}$   &  Kinematic coverage \\
      \midrule
      \multirow{2}{*}{ $F_2^{u}(x,Q^2,A)$}  &  $\propto \lp u^{N/A}+\bar{u}^{N/A} \rp$  &   \multirow{2}{*}{20}  &
      $Q^2=5$ GeV\\
      \multirow{2}{*}{ }  &
      $\propto  \lc (Z/A)\lp u^{p/A}+\bar{u}^{p/A} \rp+ (1-Z/A)\lp d^{p/A}+\bar{d}^{p/A}\rp\rc $  & & $5\times 10^{-7}\le x \le 0.9$ \\
       \midrule
      \multirow{2}{*}{ $F_2^{d}(x,Q^2,A)$}  &  $\propto \lp d^{N/A}+\bar{d}^{N/A} \rp$  &   \multirow{2}{*}{20}  &
      $Q^2=5$ GeV\\
      \multirow{2}{*}{ }  &
      $\propto  \lc (Z/A)\lp d^{p/A}+\bar{d}^{p/A} \rp+ (1-Z/A)\lp u^{p/A}+\bar{u}^{p/A}\rp\rc $  & & $5\times10^{-7}\le x \le 0.9$ \\
       \midrule
      \multirow{2}{*}{ $F_2^{s}(x,Q^2,A)$}  &  $\propto \lp s^{N/A}+\bar{s}^{N/A} \rp$  &   \multirow{2}{*}{20}  &
      $Q^2=5$ GeV\\
      \multirow{2}{*}{ }  &
      $\propto  \lp s^{p/A}+\bar{s}^{p/A} \rp$   & & $5\times 10^{-7}\le x \le 0.7$ \\
       \midrule
      \multirow{2}{*}{ $F_L(x,Q^2,A)$}  & \multirow{2}{*}{sensitive to $xg(x,Q^2)$ (see text)}  &   \multirow{2}{*}{20}  &
      $Q^2=5$ GeV\\
      \multirow{2}{*}{ }  &
       & & $5\times 10^{-7}\le x \le 0.9$ \\
        \midrule
        \multirow{2}{*}{ $\sigma_{u\bar{u}}^{DY}(y,M^2,A)$}  &
        $\propto \lp  u^{p}(x_1)\times \bar{u}^{N/A}(x_2) \rp $  &   \multirow{2}{*}{20}  &
      $Q^2=5$ GeV\\
      \multirow{2}{*}{ }  &  $\propto \lp  u^{p}(x_1)\times \lp Z \bar{u}^{p/A}(x_2) + (A-Z)\bar{d}^{p/A}(x_2) \rp \rp $
      & & $10^{-2}\le x \le 0.9$ \\
        \midrule
        \multirow{2}{*}{ $\sigma_{d\bar{d}}^{DY}(y,M^2,A)$}  &
        $\propto \lp  d^{p}(x_1)\times \bar{d}^{N/A}(x_2) \rp $  &   \multirow{2}{*}{20}  &
      $Q^2=5$ GeV\\
      \multirow{2}{*}{ }  &  $\propto \lp  d^{p}(x_1)\times \lp Z \bar{d}^{p/A}(x_2) + (A-Z)\bar{u}^{p/A}(x_2) \rp \rp $
      & & $10^{-2}\le x \le 0.9$ \\
          \midrule
        \multirow{2}{*}{ $\sigma_{s\bar{s}}^{DY}(y,M^2,A)$}  &
        $\propto \lp  s^{p}(x_1)\times \bar{s}^{N/A}(x_2) \rp $  &   \multirow{2}{*}{20}  &
      $Q^2=5$ GeV\\
      \multirow{2}{*}{}  &$\propto \lp  s^{p}(x_1)\times \bar{s}^{p/A}(x_2) \rp $   
         & & $10^{-2}\le x \le 0.9$ \\
          \midrule
        \multirow{2}{*}{ $\sigma_{\bar{u}d}^{DY}(y,M^2,A)$}  &
        $\propto \lp  \bar{u}^{p}(x_1)\times d^{N/A}(x_2) \rp $  &   \multirow{2}{*}{20}  &
      $Q^2=5$ GeV\\
      \multirow{2}{*}{}  &   $\propto \lp  \bar{u}^{p}(x_1)\times \lp Z d^{p/A}(x_2) + (A-Z)u^{p/A}(x_2) \rp \rp $
      & & $10^{-2}\le x \le 0.9$ \\
       \midrule
        \multirow{2}{*}{ $\sigma_{\bar{d}u}^{DY}(y,M^2,A)$}  &
        $\propto \lp  \bar{d}^{p}(x_1)\times u^{N/A}(x_2) \rp $  &   \multirow{2}{*}{20}  &
      $Q^2=5$ GeV\\
      \multirow{2}{*}{}  &   $\propto \lp  \bar{d}^{p}(x_1)\times \lp Z u^{p/A}(x_2) + (A-Z)d^{p/A}(x_2) \rp \rp $
      & & $10^{-2}\le x \le 0.9$ \\
       \midrule
        \multirow{2}{*}{ $\sigma_{u\bar{s}}^{DY}(y,M^2,A)$}  &
        $\propto \lp  u^{p}(x_1)\times \bar{s}^{N/A}(x_2) \rp $  &   \multirow{2}{*}{20}  &
      $Q^2=5$ GeV\\
      \multirow{2}{*}{}  &   $\propto \lp  u^{p}(x_1)\times\bar{s}^{p/A}(x_2)    \rp $
      & & $10^{-2}\le x \le 0.9$ \\
       \midrule
        \multirow{2}{*}{ $\sigma_{\bar{u}s}^{DY}(y,M^2,A)$}  &
        $\propto \lp  \bar{u}^{p}(x_1)\times s^{N/A}(x_2) \rp $  &   \multirow{2}{*}{20}  &
      $Q^2=5$ GeV\\
      \multirow{2}{*}{}  &   $\propto \lp  \bar{u}^{p}(x_1)\times s^{p/A}(x_2)    \rp $ & & $10^{-2}\le x \le 0.9$ \\
      \bottomrule
    \end{tabular}
  \end{center}
  \vspace{-0.3cm}
  \caption{\label{tab:positivity}
    The processes used
    to impose the positivity of physical cross-sections by means
    of the constraint of Eq.~(\ref{eq:positivity}).
    For each process we indicate the corresponding LO expressions,
    the number of data points $N_{\rm dat}$, and the kinematic coverage
    spanned by the pseudo-data.
  }
\end{table}

In Table~\ref{tab:positivity} we list
the $\mathcal{F}^{(l)}$ processes 
used in this analysis for which the positivity 
of physical cross-sections is imposed using Eq.~(\ref{eq:positivity}).
For each observable, the LO expressions in terms of the average bound nucleon PDFs and
bound proton distributions are given together with the number of pseudo-data points 
$N_{\rm dat}$ and the corresponding kinematic coverage.
Note that the LO expressions in Table~\ref{tab:positivity} are shown
for illustration purposes only, and in our analysis
these observables are computed using the full NLO formalism.

Here we consider two types of positivity observables.
The first type are the DIS structure functions $F_2^u$, $F_2^d$, 
$F_2^s$, and $F_L$.
The former three quantities, which contain only $u$, $d$, and $s$ 
contributions, respectively, are constructed to be positive-definite since there
exists consistent physical theories where the photon couples only to up-, down-,
or strange-type quarks.
The longitudinal structure function $F_L$, on the other hand,
largely impacts the nuclear gluon PDF since $F_L$ enters 
only at NLO and is dominated by the gluon contribution.
Lastly, we evaluate each of these structure functions on a grid of $N_{\rm}=20$ 
pseudo-data points between $x=10^{-7}$ and $x=0.9$
at $Q=\sqrt{5}$ GeV, which is slightly above the input 
parameterization scale $Q_0=1$ GeV to ensure
perturbative stability.

The second type of observable for which the
cross-section positivity is imposed
is the double-differential Drell-Yan cross-section.
In particular, we enforce the positivity of both neutral- and charged-current
Drell-Yan cross-sections in pA scattering for specific combinations of 
quark-antiquark annihilation listed in Table~\ref{tab:positivity}.
At leading order, the Drell-Yan cross-section
can be written schematically as
\be
\label{eq:DYpos}
\frac{d^2\sigma_{q_{f_1}\bar{q}_{f_2}}^{DY}}{dydQ^2} \propto \lp 
f_1^{(p)}(x_1,Q^2)  \bar{f_2}^{(p/A)}(x_2,Q^2) \rp \, ,
\ee
where the momentum fractions $x_1$
and $x_2$ are related to the rapidity $y$ and invariant mass of the
final state $Q$ at at leading order by $x_{1,2} = \lp Q/\sqrt{s}\rp e^{\pm y}$.
Here we set $Q^2=5$ GeV$^2$ and
adjust the rapidity range and center-of-mass energy $\sqrt{s}$
so that the LO kinematic range for $x_1$ and $x_2$ correspond 
to $10^{-2} \le x_{1,2} \le 0.9$.

Note that positivity of Eq.~(\ref{eq:DYpos}) will affect also the 
fitted nNNPDF2.0 $A=1$ distribution which enters as the free-proton PDF. 
While most of the positivity observables coincide with 
those included in the free-proton baseline from
which the $A=1$ distribution is derived, 
the $u\bar{d}$, $\bar{u}d$, $u\bar{s}$, and $\bar{u}s$
combinations of Table~\ref{tab:positivity} are
new in the nNNPDF2.0 determination.
Consequently, we impose these new observables 
only for proton-iron and proton-lead collisions, 
the two nuclei for which experimental
data from charged-current DIS and Drell-Yan reactions
are analyzed to study quark flavor separation.

In Sect.~\ref{sec:posresults} we will demonstrate the
positivity of cross-sections for all relevant processes
in the entire kinematical range.
We have verified that, in the absence of these constraints, 
the DIS structure functions
and the DY cross-section will in general not satisfy positivity.

\section{Results}
\label{sec:results}

In this section we present the main results of this work, namely the
nNNPDF2.0 determination of nuclear PDFs.
We first study the features of the nNNPDF2.0 fit by assessing the
quality of its agreement with experimental data, focusing largely on the
LHC weak boson production cross-sections, and by studying the behavior
of nuclear modification ratios across different nuclei.
Subsequently, we contrast this new nPDF determination with its
predecessor, nNNPDF1.0, and trace the origin of observed differences via
a series of fits with systematic changes.
We then study the role that the valence and momentum sum rules play in
the global nPDF determination by presenting two variants of the
nNNPDF2.0 fit in which one of the two sum rules is not imposed.
Finally, we demonstrate that the nNNPDF2.0 fit satisfies the positivity
of physical cross-sections in the kinematic range where experimental
data is available.

\subsection{The nNNPDF2.0 determination}

We begin by discussing the fit quality which is assessed across the
various datasets and quantified by the $\chi^2$ figure of merit.
A comparison is then made using the nNNPDF2.0 predictions with the LHC
weak gauge boson production measurements.
Following this, we take a closer look on the nNNPDF2.0 parton
distributions and the corresponding ratios to the free-nucleon case.
Lastly, we investigate the sensitivity of the nuclear modification
factors with respect to the atomic mass $A$, in particular on the sea
quarks and strangeness, and compare our results with those from the
EPPS16 analysis.

\paragraph{Fit quality.}
In Tables~\ref{tab:chi2} and~\ref{tab:chi2_2} we collect the values of
the $\chi^2$ per data point for all the datasets included in the
nNNPDF2.0 analysis, i.e. the neutral and charged current DIS structure
functions as well as gauge boson production measurements at the LHC.
We compare the nNNPDF2.0 results with a variant fit where we exclude all
LHC datasets (DIS only) and with EPPS16.\footnote{For the EPPS16
calculation we use CT14nlo as the free-proton PDF set for consistency.}
 Values in italics indicate predictions for datasets
 not included in the corresponding fits.
The numbers presented in Tables~\ref{tab:chi2} and~\ref{tab:chi2_2}
contain only the contribution to the $\chi^2$ associated with the
experimental data, and do not include penalty from the proton boundary
condition or positivity constraint (the latter of which vanishes for all
final nNNPDF2.0 replicas anyway).
Moreover, while we use the $t_0$ prescription~\cite{Ball:2009qv} during
the optimization to avoid the D'Agostini bias, the quoted numbers
correspond to the experimental definition of the $\chi^2$
instead~\cite{Ball:2012wy}, in which the central experimental value is
used to compute the correlated multiplicative uncertainties.

\begin{table}[htp]
  \centering
  \footnotesize
    \renewcommand{\arraystretch}{1.25}
\begin{tabular}{|l|c|C{2.7cm}|C{2.5cm}|C{2.5cm}|}
\toprule
 \multicolumn{2}{|c|}{} & nNNPDF2.0 (DIS) & nNNPDF2.0 & EPPS16nlo \\ \midrule
Dataset & $n_{\rm dat}$ &$\chi^ 2/n_{\rm dat}$ &$\chi^ 2/n_{\rm dat}$ &$\chi^ 2/n_{\rm dat}$ \\ \hline
${\rm NMC \,\, (He/D)}$ & 13 & 1.11 & 1.129 & 0.829 \\ \hline
${\rm SLAC \,\, (He/D)}$ & 3 & 0.623 & 0.638 & 0.152 \\ \hline
${\rm NMC \,\, (Li/D)}$ & 12 & 1.083 & 1.166 & 0.74 \\ \hline
${\rm SLAC \,\, (Be/D)}$ & 3 & 1.579 & 1.719 & 0.098 \\ \hline
${\rm EMC \,\, (C/D)}$ & 12 & 1.292 & 1.321 & 1.174 \\ \hline
${\rm FNAL \,\, (C/D)}$ & 3 & 0.932 & 0.838 & 0.985 \\ \hline
${\rm NMC \,\, (C/D)}$ & 26 & 2.002 & 2.171 & 0.872 \\ \hline
${\rm SLAC \,\, (C/D)}$ & 2 & 0.286 & 0.251 & 1.075 \\ \hline
${\rm BCDMS \,\, (N/D)}$ & 9 & 2.439 & 2.635 & n/a \\ \hline
${\rm SLAC \,\, (Al/D)}$ & 3 & 0.606 & 0.864 & 0.326 \\ \hline
${\rm EMC \,\, (Ca/D)}$ & 3 & 1.72 & 1.722 & 1.82 \\ \hline
${\rm FNAL \,\, (Ca/D)}$ & 3 & 1.253 & 1.194 & 1.354 \\ \hline
${\rm NMC \,\, (Ca/D)}$ & 12 & 1.503 & 1.747 & 1.772 \\ \hline
${\rm SLAC \,\, (Ca/D)}$ & 2 & 0.82 & 0.771 & 1.642 \\ \hline
${\rm BCDMS \,\, (Fe/D)}$ & 16 & 2.244 & 2.743 & 0.765 \\ \hline
${\rm EMC \,\, (Fe/D)}$ & 58 & 0.827 & 0.875 & 0.445 \\ \hline
${\rm SLAC \,\, (Fe/D)}$ & 8 & 2.171 & 2.455 & 1.06 \\ \hline
${\rm EMC \,\, (Cu/D)}$ & 27 & 0.523 & 0.572 & 0.714 \\ \hline
${\rm SLAC \,\, (Ag/D)}$ & 2 & 0.667 & 0.691 & 1.595 \\ \hline
${\rm EMC \,\, (Sn/D)}$ & 8 & 2.197 & 2.248 & 2.265 \\ \hline
${\rm FNAL \,\, (Xe/D)}$ & 4 & 0.414 & 0.384 & n/a \\ \hline
${\rm SLAC \,\, (Au/D)}$ & 3 & 1.216 & 1.353 & 1.916 \\ \hline
${\rm FNAL \,\, (Pb/D)}$ & 3 & 2.243 & 2.168 & 2.044 \\ \hline 
${\rm NMC \,\, (Be/C)}$ & 14 & 0.268 & 0.269 & 0.27 \\ \hline
${\rm NMC \,\, (C/Li)}$ & 9 & 1.063 & 1.117 & 0.9 \\ \hline
${\rm NMC \,\, (Al/C)}$ & 14 & 0.345 & 0.354 & 0.396 \\ \hline
${\rm NMC \,\, (Ca/C)}$ & 23 & 0.468 & 0.44 & 0.564 \\ \hline
${\rm NMC \,\, (Fe/C)}$ & 14 & 0.663 & 0.667 & 0.751 \\ \hline
${\rm NMC \,\, (Sn/C)}$ & 119 & 0.607 & 0.638 & 0.626 \\ \hline
${\rm NMC \,\, (Ca/Li)}$ & 9 & 0.259 & 0.276 & 0.15 \\ 
\bottomrule
\end{tabular}
\vspace{0.3cm}
\caption{The values of the $\chi^2$ per data point for the
DIS neutral current structure function
  datasets included in  nNNPDF2.0.
  We compare the nNNPDF2.0 global and DIS-only
  results with those obtained using EPPS16 as input for the theory predictions.
  \label{tab:chi2}
}
\end{table}

\begin{table}[htp]
  \centering
  \small
  \renewcommand{\arraystretch}{1.30}
  \begin{tabular}{|l|c|C{2.9cm}|C{2.5cm}|C{2.5cm}|}
    \toprule
  \multicolumn{2}{|c|}{} & nNNPDF2.0 (DIS) & nNNPDF2.0 & EPPS16nlo \\ \midrule
Dataset & $n_{\rm dat}$ &$\chi^ 2/n_{\rm dat}$ &$\chi^ 2/n_{\rm dat}$ &$\chi^ 2/n_{\rm dat}$ \\ \hline
${\rm NuTeV \,\, (\bar{\nu} Fe)}$ & 37 & 0.946 & 1.094 & \textit{0.639} \\ \hline
${\rm NuTeV \,\, (\nu Fe)}$ & 39 & 0.287 & 0.264 & \textit{0.381} \\ \hline
${\rm CHORUS \,\, (\bar{\nu} Pb)}$ & 423 & 0.938 & 0.97 & 1.107 \\ \hline
${\rm CHORUS \,\, (\nu Pb)}$ & 423 & 1.007 & 1.015 & 1.024 \\ \hline \hline
${\rm ATLAS^{5TEV} \,\, Z}$ & 14 & \textit{1.469} & 1.134 & 1.12 \\ \hline
${\rm CMS^{5TeV} \,\, W^{-}}$ & 10 & \textit{1.688} & 1.078 & 0.857 \\ \hline
${\rm CMS^{8TeV} \,\, W^{-}}$ & 24 & \textit{1.453} & 0.72 & \textit{0.825} \\ \hline
${\rm CMS^{5TeV} \,\, W^{+}}$ & 10 & \textit{2.32} & 1.125 & 1.211 \\ \hline
${\rm CMS^{8TeV} \,\, W^{+}}$ & 24 & \textit{3.622} & 0.772 & \textit{0.951} \\ \hline
${\rm CMS^{5TeV} \,\, Z}$ & 12 & \textit{0.58} & 0.52 & 0.639 \\ \hline \hline
{\bf Total} & {\bf 1467} & {\bf 1.013} & {\bf 0.976} & {\bf 0.896} \\ \bottomrule
\end{tabular}
\vspace{0.3cm}
\caption{Same as Table~\ref{tab:chi2} now
  for the new datasets
  included in nNNPDF2.0:
  charged current DIS structure functions
  and gauge boson production at the LHC.
  We also provide the values of $\chi^2/n_{\rm dat}$
  for the total dataset.
  Values in italics indicate predictions for datasets
  not included in the corresponding fit.
  \label{tab:chi2_2}
}
\end{table}


From the results of Table~\ref{tab:chi2} and~\ref{tab:chi2_2}, one can
see that the nNNPDF2.0 determination achieves a satisfactory description
of all datasets included in this analysis.
A good $\chi^2$ is obtained in particular for the charged-current DIS
cross-sections and LHC gauge boson production distributions.
For instance, for the precise W boson rapidity distributions at
$\sqrt{s}=8.16$ TeV from CMS one finds $\chi^2/n_{\rm dat} = 0.74$ for
$n_{\rm dat}=48$ data points.
The corresponding predictions using EPPS16  (which do not include this dataset)
also lead to a good agreement with
$\chi^2/n_{\rm dat} = 0.88$.
The description of most neutral current DIS datasets is comparable
to that of nNNPDF1.0.
Some datasets, such as the SLAC iron structure functions,
are somewhat deteriorated with respect to nNNPDF1.0,
possibly due to some mild tension with the CC cross-sections.
Further, we find that our resulting fit quality to the CC
deep-inelastic structure functions is similar to that obtained in the
corresponding proton PDF analysis~\cite{Ball:2017nwa}.

Overall, the resulting $\chi_{\rm tot}^2/n_{\rm dat}= 0.976$ for the $n_{\rm
dat}=1467$ data points included in the fit highlights the remarkable
consistency of the experimental data on nuclear targets and the
corresponding theory predictions based on the QCD factorization
framework.
A similar total $\chi^2$ is obtained for the theoretical predictions
computed using EPPS16 as input.
As will be shown below, the fact that both global fits lead to comparable
$\chi^2$ values can be explained by the corresponding similarities at the nPDF level.

\paragraph{Comparison with experimental data.}
To facilitate our discussion regarding the comparison between
data and theory calculations, we first introduce here the conventions that we use to
define the nuclear modification factors.
Following the notations of Sect.~\ref{sec:defs}, the Drell-Yan rapidity
  distributions in proton-nucleus collision can be expressed as
  \be
  \label{eq:DYdecomposition}
  \frac{d\sigma_{\rm DY}(y)}{dy} \equiv A\frac{d\sigma^{(N/A)}_{\rm DY}(y)}{dy} = Z\frac{d\sigma^{(p/A)}_{\rm DY}(y)}{dy}
  + \lp A-Z\rp\frac{d\sigma^{(n/A)}_{\rm DY}(y)}{dy} \, ,
  \ee
  where the superindices $N/A$, $p/A$, and $n/A$ indicate respectively the collision
  between a proton with an average nucleon, a proton, or a neutron bound
  within a nucleus of atomic number $Z$ and mass number $A$.
  As in the case of the PDFs, the bound proton and nucleon
  cross-sections $\sigma_{\rm DY}^{(p/A)}$ and $\sigma_{\rm DY}^{(n/A)}$
  are related to each other via isospin symmetry.
  
  The expression in Eq.~(\ref{eq:DYdecomposition}) helps in emphasizing
  the two reasons why the Drell-Yan cross-sections will be different
  between pp and pA collisions.
  The first is due to the modifications of the bound proton PDFs in
  nuclei, namely the fact that $\sigma_{\rm DY}^{(p/A)} \ne \sigma_{\rm
  DY}^{(p)}$.
  Secondly, the predictions of pA collisions using non-isoscalar nuclei
  would still differ from those of pp reactions in the absence of
  nuclear modifications, i.e. $\sigma_{\rm DY}^{(p/A)}= \sigma_{\rm
  DY}^{(p)}$, as a consequence of the unequal amount of protons and
  neutrons in the target, resulting in $\sigma_{\rm DY}^{(N/A)} \ne \sigma_{\rm DY}^{(p)}$.
  
  Taking these considerations into account, one should define the
  nuclear modification factor in Drell-Yan proton-nuclear collisions as
   \be
   \label{eq:RADY}
   R_A^{\rm DY}(y) \equiv \lp \frac{d\sigma^{(p/A)}_{\rm DY}(y)}{dy} +
   \lp \frac{A}{Z}-1\rp  \frac{d\sigma^{(n/A)}_{\rm DY}(y)}{dy} \rp \Bigg/ \lp 
   \frac{d\sigma^{(p)}_{\rm DY}(y)}{dy} + \lp \frac{A}{Z}-1\rp \frac{d\sigma^{(n)}_{\rm
   DY}(y)}{dy} \rp \, .
   \ee
   With the above definition, one should find that $R_A^{\rm DY}(y) \ne
   1$ only in the presence of genuine nuclear corrections, that is, when
   $\sigma_{\rm DY}^{(p/A)} \ne \sigma_{\rm DY}^{(p)}$.
   The definition of Eq.~(\ref{eq:RADY}) differs from that of an
   observable frequently measured in proton-lead collisions, where the
   proton-nucleus cross-section is normalized to a proton-proton
   baseline,
    \bea
   \label{eq:RADYexp}
   R_{A, \rm exp}^{\rm DY}(y) &\equiv&
   \frac{d\sigma^{(N/A)}_{\rm DY}(y)}{dy}
    \Bigg/ \frac{d\sigma^{(p)}_{\rm DY}(y)}{dy} \\ &=&  \nonumber \lp  \frac{Z}{A}
      \frac{d\sigma^{(p/A)}_{\rm DY}(y)}{dy} + \lp 1- \frac{Z}{A}\rp
      \frac{d\sigma^{(n/A)}_{\rm DY}(y)}{dy} \rp \Bigg/ \lp 
      \frac{d\sigma^{(p)}_{\rm DY}(y)}{dy} \rp \, .
   \eea
   As  explained above, in proton-nuclear collisions one will find
   $R_{A, \rm exp}^{\rm DY}(y) \ne 1$ for non-isoscalar targets
   even when $\sigma_{\rm
     DY}^{(p/A)}= \sigma_{\rm DY}^{(p)}$ due to the imbalance between
   the number of protons and neutrons.
   In this section, we will exclusively use the definition of
   Eq.~(\ref{eq:RADY}) when evaluating the theoretical predictions of
   nuclear modification ratios in Drell-Yan distributions.
   
\begin{figure}[t]
\begin{center}
  \includegraphics[width=0.99\textwidth]{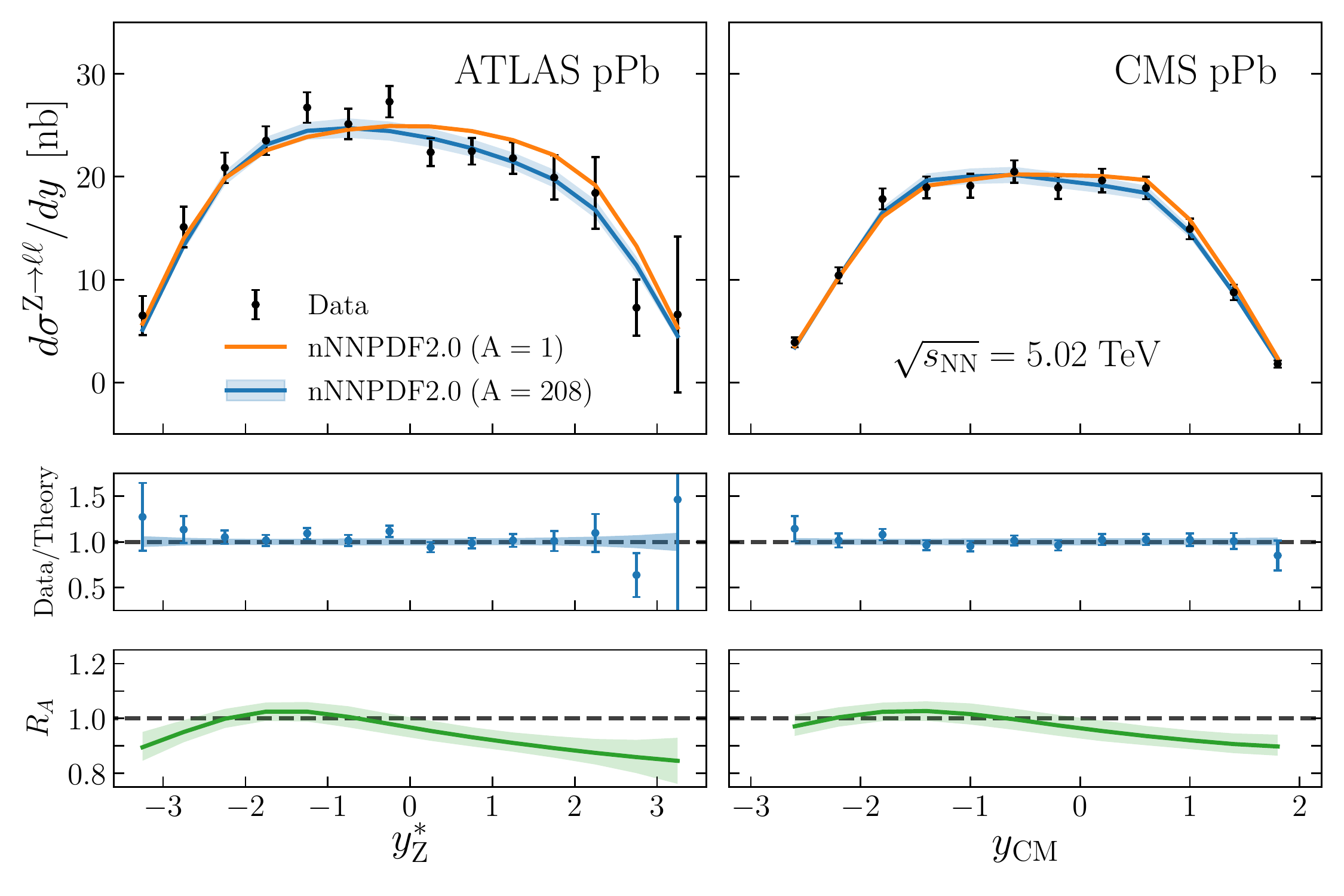}
 \end{center}
\vspace{-0.8cm}
\caption{ Comparison between the ATLAS (left) and CMS (right panel)
  measurements of Z boson production in proton-lead collisions at
  $\sqrt{s}=5.02$ TeV with the theoretical predictions using nNNPDF2.0
  as input.
  We show the predictions obtained both for $A=1$ and $A=208$, in the
  later case including the 90\% confidence level band.
  From top to bottom, the three panels display the absolute
  cross-sections, the ratio between data and theory, and the nuclear
  modification $R_A(y)$.
  \label{nNNPDF20comp_data_1}
}
\end{figure}

In Fig.~\ref{nNNPDF20comp_data_1} we display the comparison between the
ATLAS and CMS measurements of Z boson production in proton-lead collisions at
$\sqrt{s}=5.02$ TeV with the theoretical predictions using nNNPDF2.0.
  The theory calculations were computed using
  Eq.~(\ref{eq:DYdecomposition}) with the nNNPDF2.0 $A=1$ and $A=208$
  distributions, in the latter case also including the corresponding
  90\% CL uncertainty band.
  Note that for the theory cross-section obtained with the $A=1$ PDFs,
  nuclear effects are absent since it corresponds to the free proton
  distributions.
  From top to bottom, the three panels display the absolute
  cross-sections as a function of the dilepton rapidity $y$, the ratio
  between data and theory, and the nuclear modification factor ratio
  $R_A(y)$ defined by Eq.~(\ref{eq:RADY}).
  
  Here the ATLAS and CMS measurements of the dilepton rapidity
  distributions are both provided in the Z-boson center-of-mass
  reference frame.
  The CMS absolute cross-sections are lower than ATLAS due to the
  different kinematical selection cuts.
  We also note that for these datasets, as well as for the rest of LHC
  measurements, forward rapidities correspond to the direction of the
  incoming lead nuclei.
  From the comparisons in Fig.~\ref{nNNPDF20comp_data_1} we can see that
  the theory predictions are in good agreement with the experimental
  data.
Interestingly, the $R_A(y)$ ratios exhibit a strong preference for
nuclear modifications, especially at forward rapidities which correspond
to small values in $x$ for the bound nucleons.
As will be discussed below, this behavior can be explained at the level
of the nuclear PDFs by a notable shadowing effect at small-$x$ for up
and down quarks and antiquarks.

\begin{figure}[t]
\begin{center}
  \includegraphics[width=0.99\textwidth]{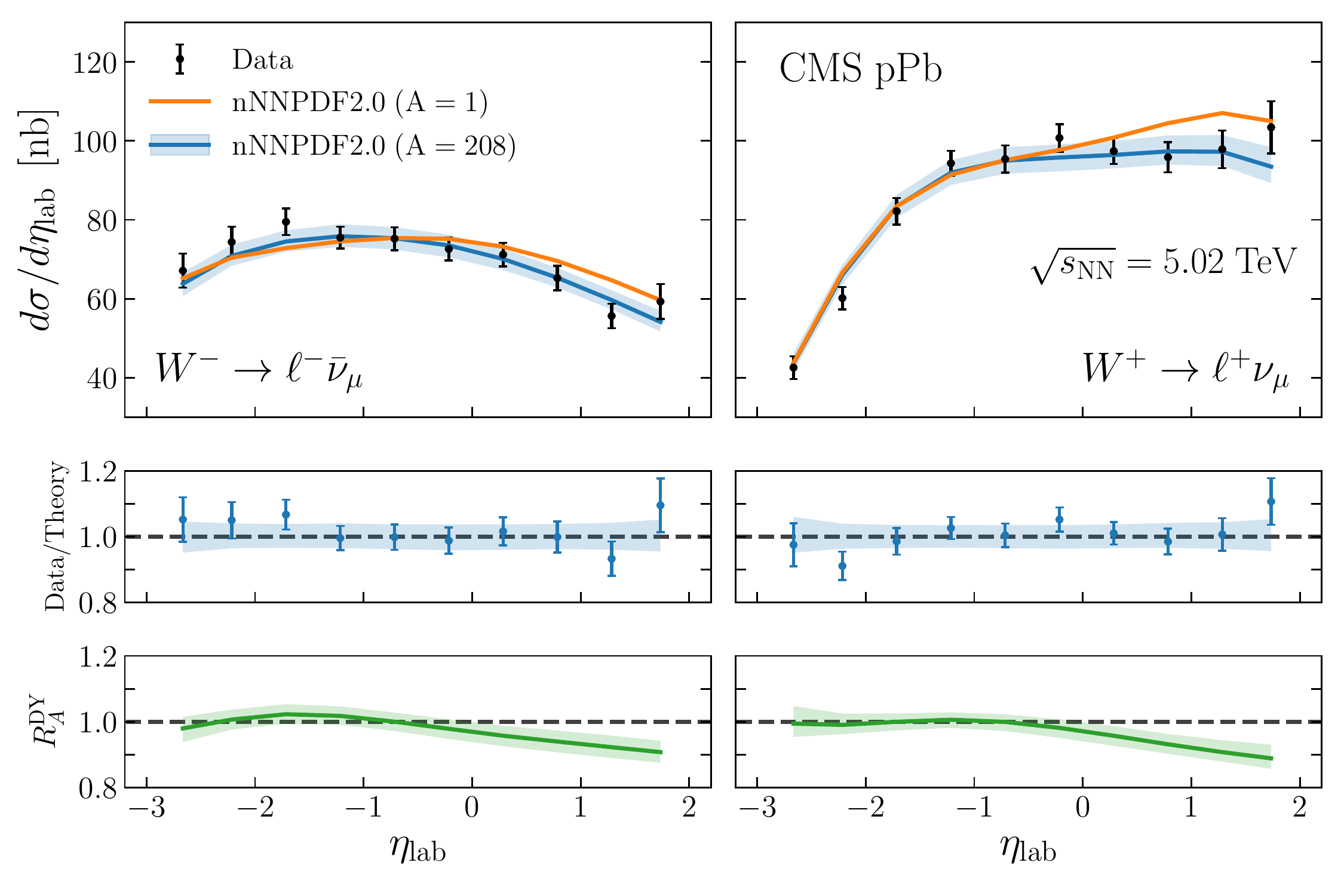}
 \end{center}
\vspace{-0.8cm}
\caption{Same as Fig.~\ref{nNNPDF20comp_data_1} now for the CMS muon
  rapidity distributions in W$^-$ and W$^+$ production at
  $\sqrt{s}=5.02$ TeV.
  \label{nNNPDF20comp_data_2}
}
\end{figure}

\begin{figure}[t]
\begin{center}
  \includegraphics[width=0.99\textwidth]{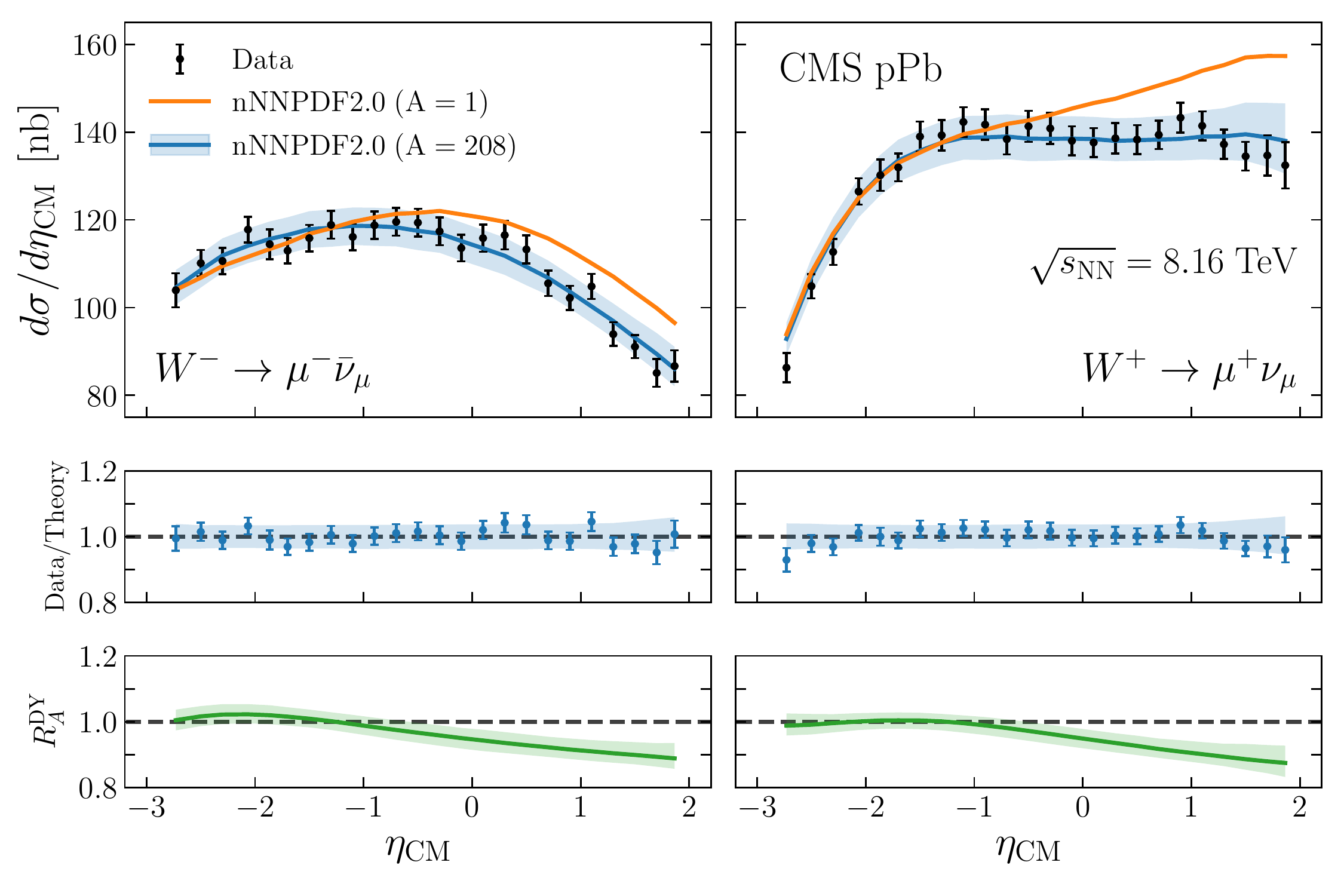}
 \end{center}
\vspace{-0.8cm}
\caption{Same as Fig.~\ref{nNNPDF20comp_data_1} now for the CMS muon
  rapidity distributions in W$^-$ and W$^+$ production at
  $\sqrt{s}=8.16$ TeV.
  \label{nNNPDF20comp_data_3}
}
\end{figure}

The corresponding comparisons for the CMS muon rapidity distributions in
W$^-$ and W$^+$ production at $\sqrt{s}=5.02$ TeV and 8.16 TeV are
displayed in Figs.~\ref{nNNPDF20comp_data_2}
and~\ref{nNNPDF20comp_data_3} respectively.
The results are presented as functions of the rapidity of the charged
lepton from the W boson decay in the laboratory frame.
In all cases the theoretical predictions based on nNNPDF2.0 provide a
satisfactory description of the experimental data.
It is interesting to note that for the high-statistics CMS measurement at
8 TeV, the $A=208$ prediction is remarkably better than the free-proton
one, particularly at forward rapidities where one is sensitive to the
small-$x$ region of the bound nucleons.
This feature highlights how the CMS 8 TeV W production data provides
direct evidence for the nuclear modifications of valence and sea quark
distributions.


\begin{figure}[t]
\begin{center}
  \includegraphics[width=0.99\textwidth]{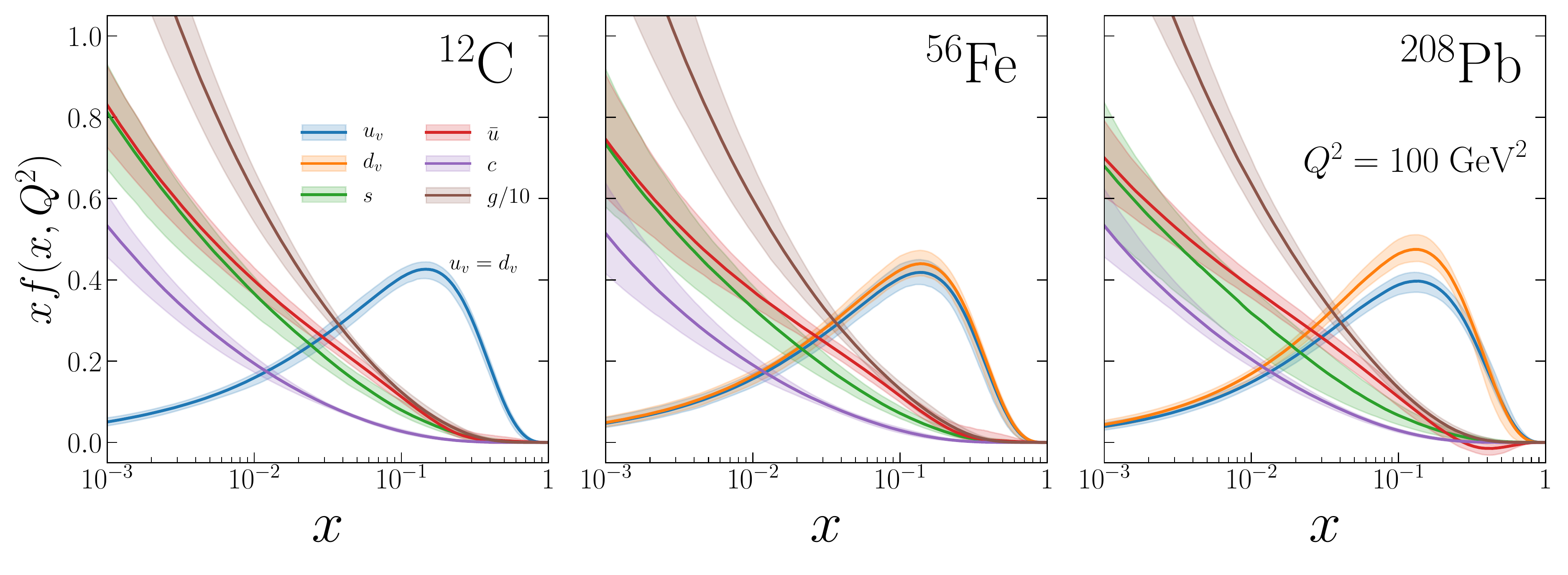}
 \end{center}
\vspace{-0.6cm}
\caption{The nNNPDF2.0  set of nuclear PDFs for $^{12}$C, $^{56}$Fe, and
  $^{208}$Pb at the scale $Q=10$ GeV.
  We display the gluon, the up and down valence quarks (which coincide
  for isoscalar nuclei), as well as the down, strange, and charm sea
  quark distributions.
  The bands indicate the 90\% CL uncertainty range.
  \label{fig:nNNPDF20_PDFs}
}
\end{figure}

\paragraph{Nuclear parton distributions.}
In Fig.~\ref{fig:nNNPDF20_PDFs} we display the nNNPDF2.0 set of nuclear
PDFs for three different nuclei, $^{12}$C, $^{56}$Fe, and $^{208}$Pb,
constructed using Eq.~(\ref{eq:qNAdefinition}) at a scale of $Q=10$ GeV.
Specifically, we display the gluon, the up and down valence quarks, and
the down, strange, and charm sea quark distributions.
For isoscalar nuclei such as $^{12}$C, the up and down valence
distributions are equivalent, $u_v^{(N/A)}=d_v^{(N/A)}$, as well as the
up and down sea PDFs, $\bar{u}^{(N/A)}=\bar{d}^{(N/A)}$, as a result of
Eq.~(\ref{eq:qNAdefinition}).

From the comparisons in Fig.~\ref{fig:nNNPDF20_PDFs}, we can see that
the nuclear PDFs exhibit a moderate dependence on the atomic number $A$.
The resulting pattern of PDF uncertainties can partly be
explained by the input data.
For example, nPDF uncertainties on strangeness are smaller in $^{12}$C
and $^{56}$Fe compared to $^{208}$Pb, due to the impact of the proton
boundary condition and the NuTeV dimuon data, respectively.
We also observe that the PDF uncertainties on the gluon (and
correspondingly on the dynamically generated charm PDF) at medium and
small-$x$ are larger in iron than in carbon and lead.
While the gluon uncertainties for carbon are largely determined by the impact of
the free-proton boundary condition, those on lead nuclei can
likely be attributed to the LHC measurements of W and Z production
and the large amount of charged-current DIS data, which indirectly
provide constraints via DGLAP evolution.

Fig.~\ref{fig:nNNPDF20_PDFs} also shows that in the case of $^{208}$Pb,
there is a clear difference between the $d_v$ and $u_v$ distributions
due to the isoscalar nature of the nucleus, where $d_v$ is larger due to
the significant neutron excess in lead.
The fact that $u_v$ and $d_v$ do not overlap within the 90\% CL bands in
a wide range of $x$ highlights how a careful treatment of the quark and
antiquark flavor separation is essential in order to describe the
precise data available on lead targets, especially the weak boson
production measurements in proton-lead collisions at the LHC.

To further illustrate the features of the nNNPDF2.0 determination, it is useful to study
them in terms of ratios with respect to the corresponding free-nucleon
baseline.
In the following we will define the nuclear modification ratios of PDFs
for a nucleus with mass number $A$ as,
\be
\label{eq:nuclearratioRv3}
R_f^{A}(x,Q^2) \equiv \frac{Z  q^{(p/A)}_f(x,Q^2) + (A-Z)
  q^{(n/A)}_f(x,Q^2) }{Z q^{(p)}_f(x,Q^2) + (A-Z) q^{(n)}_f(x,Q^2)} \, .
\ee
When evaluating Eq.~(\ref{eq:nuclearratioRv3}), it is important to
account both for the uncertainties associated to the nuclear and free-proton
PDFs.
In the case of a Monte Carlo set such as nNNPDF2.0, this entails
evaluating $R_f^{A}$ for each of the $N_{\rm rep}$ replicas and then
determining the resulting median and 90\% CL interval.

\begin{figure}[t]
\begin{center}
  \includegraphics[width=0.93\textwidth]{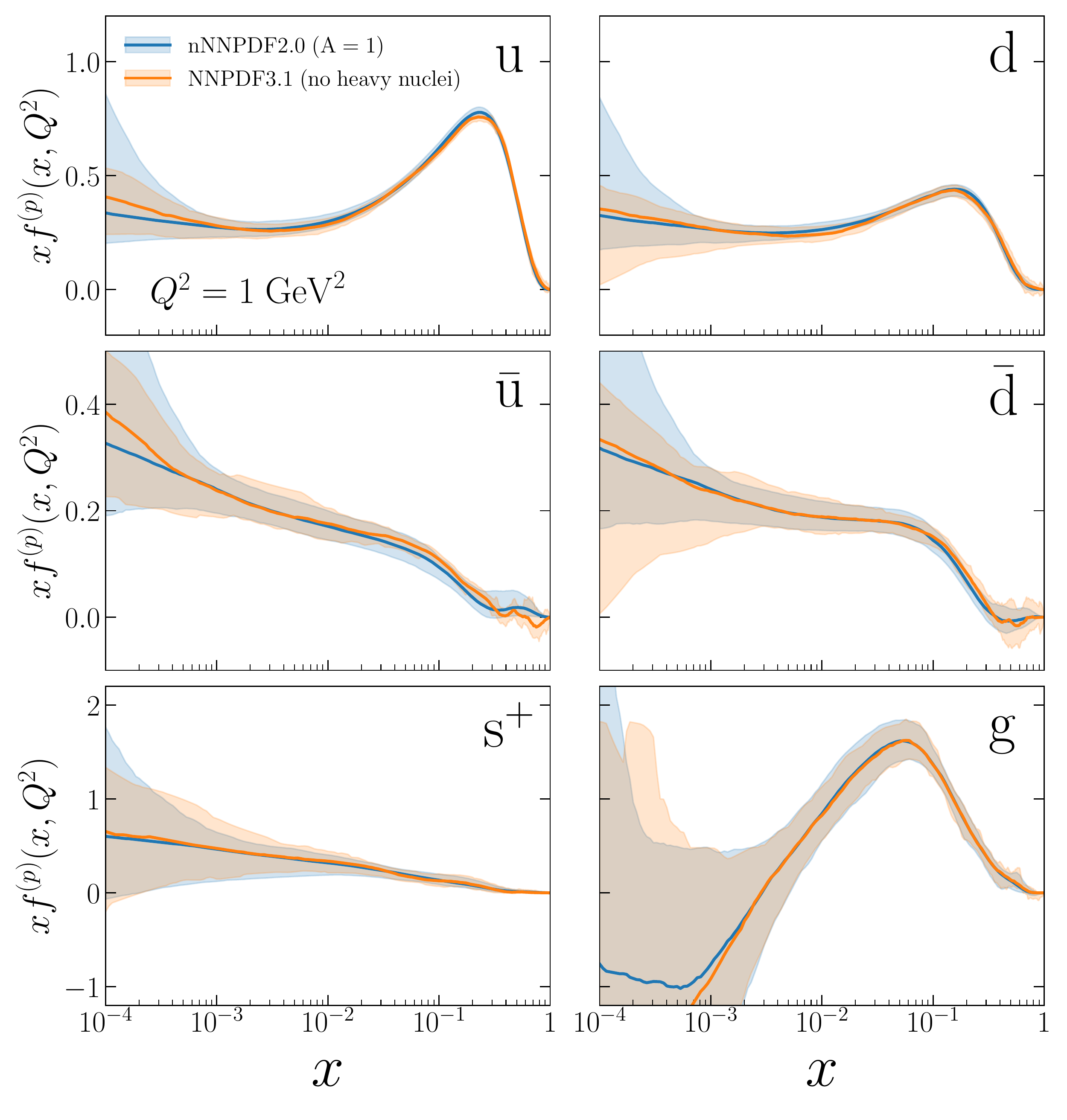}
 \end{center}
\vspace{-0.8cm}
\caption{Comparison of the nNNPDF2.0  parton distributions for $A=1$
  with the NNPDF3.1
  baseline used for the boundary condition in Eq.~(\ref{eq:chi2})
  at the parametrisation scale  $Q_0=1$ GeV.
  \label{fig:nNNPDF20_vs_BC}
}
\end{figure}

In Fig.~\ref{fig:nNNPDF20_vs_BC} we show the nNNPDF2.0 
distributions  for $A=1$ that enter the denominator of Eq.~(\ref{eq:nuclearratioRv3}).
They  are compared with the NNPDF3.1
  proton baseline used for implementation the boundary condition via Eq.~(\ref{eq:chi2})
  at the input parametrisation scale  $Q_0=1$ GeV.
Overall, there is very good agreement between our $A=1$ result
and the proton boundary condition, particularly in the region
of $x$ where the constraint is being imposed, $10^{-3} < x < 0.7$. 
It is important to emphasize that due to the nNNPDF2.0 $A=1$ set being 
determined not only by the boundary condition but also by the positivity constraints and  the LHC
cross-sections, one expects some moderate differences with the NNPDF3.1 proton baseline.
The most notable differences indeed are found in the $\bar{u}$
and $\bar{d}$ PDFs at medium to large $x$, where
the newly added DY positivity observables for 
$\bar{u}d$ and $u\bar{d}$ quark combinations, 
as well as the LHC proton-lead data, play a significant role. 
Nevertheless, the level of agreement
reported in Fig.~\ref{fig:nNNPDF20_vs_BC} is quite remarkable
and highlights how the nNNPDF2.0 determination manages to take into account the extensive
information provided by the global analysis of free-proton
structure.

\begin{figure}[t]
\begin{center}
  \includegraphics[width=0.93\textwidth]{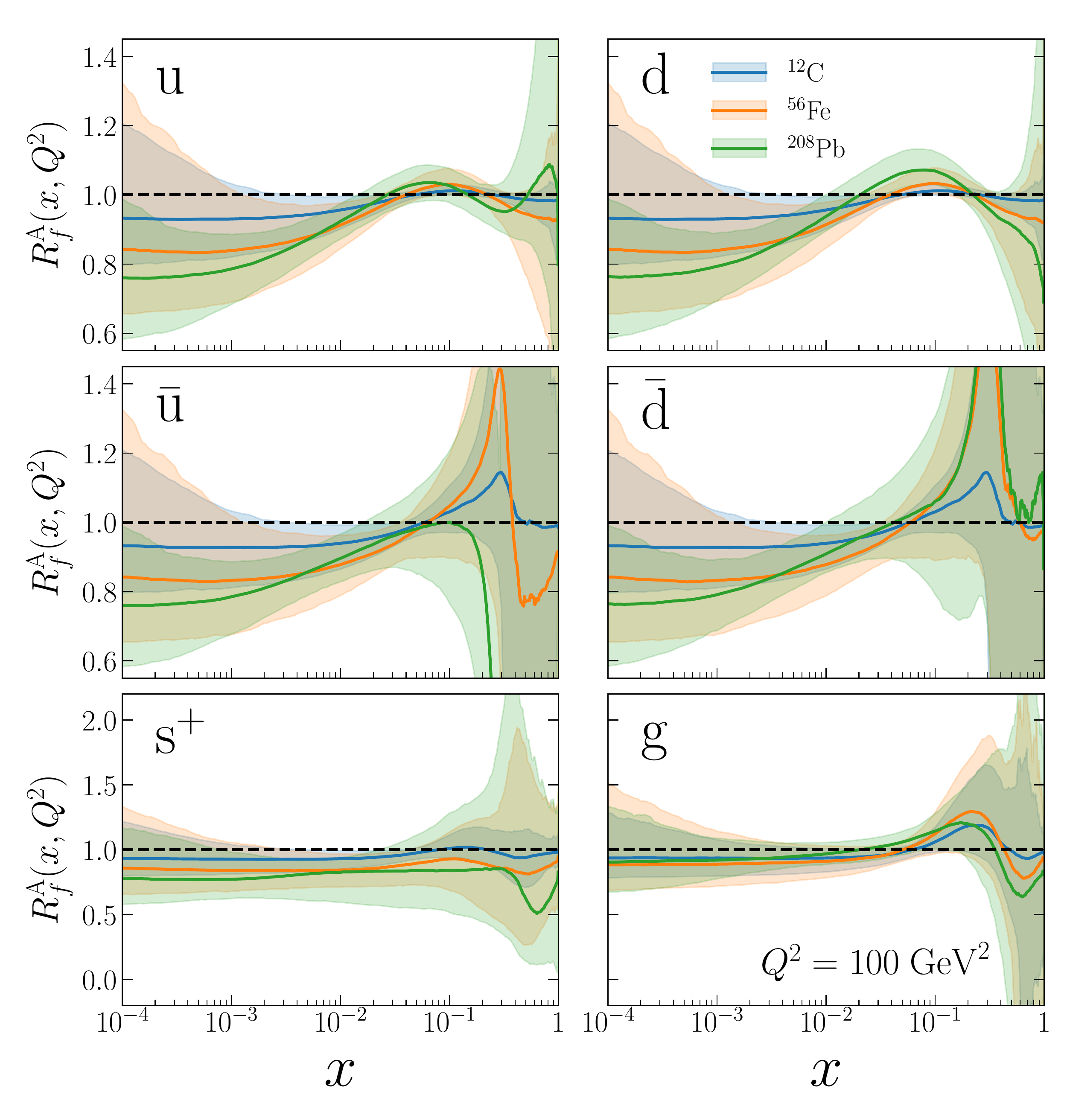}
 \end{center}
\vspace{-0.8cm}
\caption{Comparison of the nuclear PDF ratios,
  Eq.~(\ref{eq:nuclearratioRv3}), for three different nuclei, $^{12}$C,
  $^{56}$Fe, and $^{208}$Pb, at $Q=10$ GeV.
  From top to bottom we show the up and down quarks, the corresponding
  antiquarks, the total strangeness, and the gluon.
  The bands indicate the 90\% confidence level intervals and take into
  account the correlations with the proton baseline used for the
  normalisation.
  \label{R_A}
}
\end{figure}

In Fig.~\ref{R_A} we display the nuclear PDF ratios, defined 
by Eq.~(\ref{eq:nuclearratioRv3}), for the same parton flavors
as in Fig.~\ref{fig:nNNPDF20_vs_BC}.
Here the ratios for $^{12}$C, $^{56}$Fe, and $^{208}$Pb nuclei 
are compared at $Q=10$ GeV. 
The shaded bands indicating the 90\% confidence level intervals 
take into account also the correlations with the proton baseline.
  Overall, the comparison highlights the dependence on the central value
  and uncertainties of the nuclear ratios $R_f^{A}$ as the value of $A$
  is varied from lighter to heavier nuclei.

 For the up and down quark nPDFs in Fig.~\ref{R_A}, we can see that the
 shadowing effects become more prominent at small-$x$ as $A$ increases,
 with the central value reaching $R_f^{A}\simeq 0.75$ at $x=10^{-4}$ for
 the lead ratios.
Interestingly, the nPDF uncertainties on the quarks for $x\lsim 10^{-2}$
are reduced in lead as compared to the lighter nuclei.
This is a consequence of the constraints provided by the LHC data, as
will be shown in Sect.~\ref{sec:comparisonNNPDF10}.
In the large-$x$ region, deviations from the $R_f^{A}=1$ scenario
(no nuclear corrections)
 appear more prominent for the quarks and antiquarks of heavier
nuclei.

Turning now to the valence quarks, one can distinguish the shadowing and
anti-shadowing regions for all values of $A$, though nuclear effects in
carbon are quite small.
While one generally finds a suppression $R_f^{A} < 1$ at large $x$ that
is consistent with the EMC-effect expectation, the position of the
so-called "EMC minimum" is not universal or even guaranteed at the nPDF
level.
For the anti-quarks, the only region where a well-defined qualitative
behavior is observed is the small-$x$ shadowing region, while at
large-$x$ the nPDF uncertainties are too large to draw any solid
conclusion.
Finally, we observe that the nuclear modifications on the gluon nPDF are
rather stable as $A$ is varied.

\begin{figure}[t]
\begin{center}
  \includegraphics[width=0.95\textwidth]{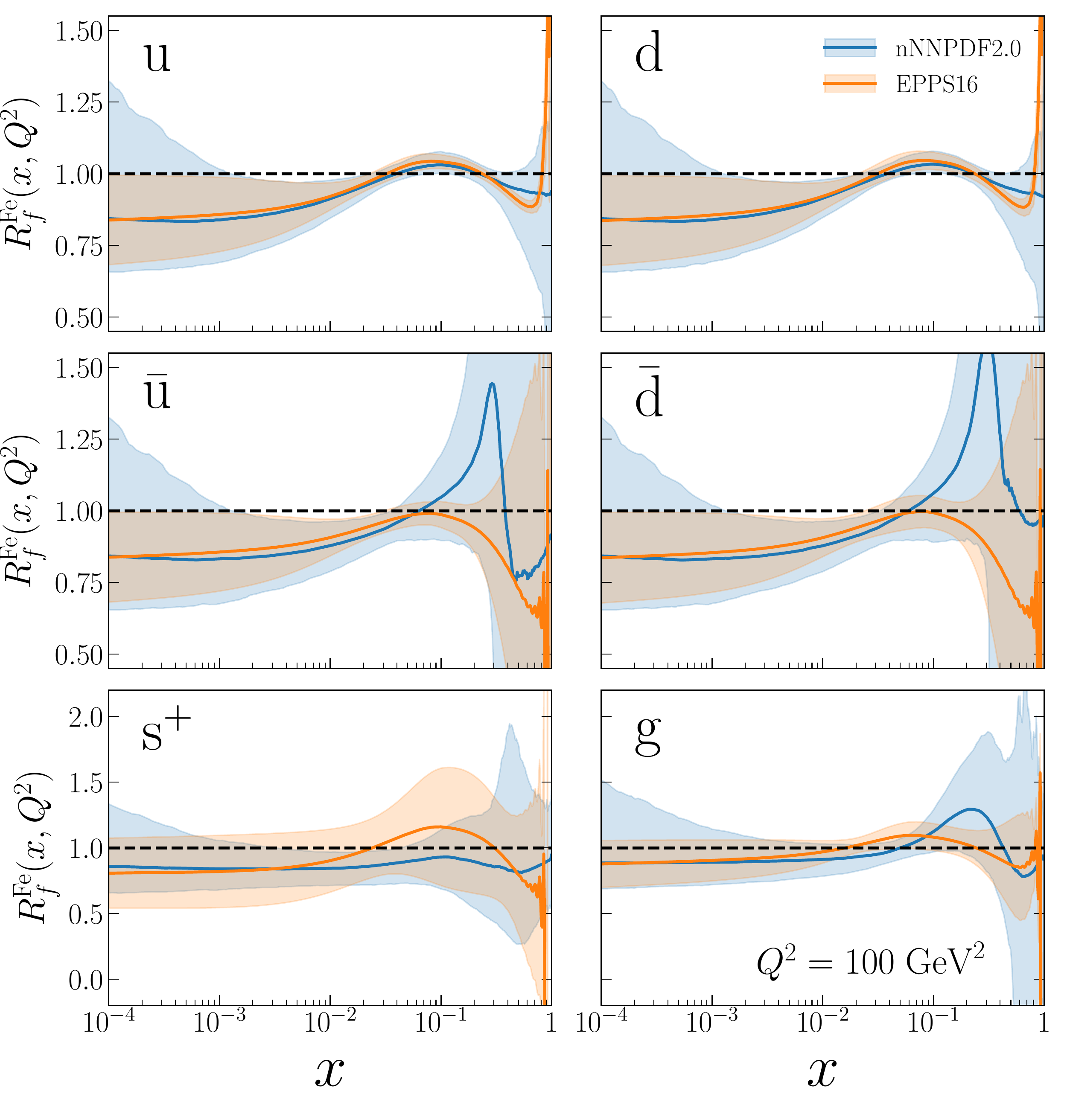}
 \end{center}
\vspace{-0.9cm}
\caption{The nuclear PDF modification ratios for iron, $R_{\rm
  Fe}(x,Q^2)$, as a function of $x$ for $Q=10$ GeV for both nNNPDF2.0
  and EPPS16.
  From top to bottom we show the up and down quarks, the corresponding
  antiquarks, the total strangeness, and the gluon.
  The bands correspond to the 90\% CL uncertainties, and each nPDF set
  is normalised to its corresponding free-proton baseline as indicated
  by Eq.~(\ref{eq:nuclearratioRv3}).
  \label{fig:R_Fe}
}
\end{figure}

\begin{figure}[t]
\begin{center}
  \includegraphics[width=0.95\textwidth]{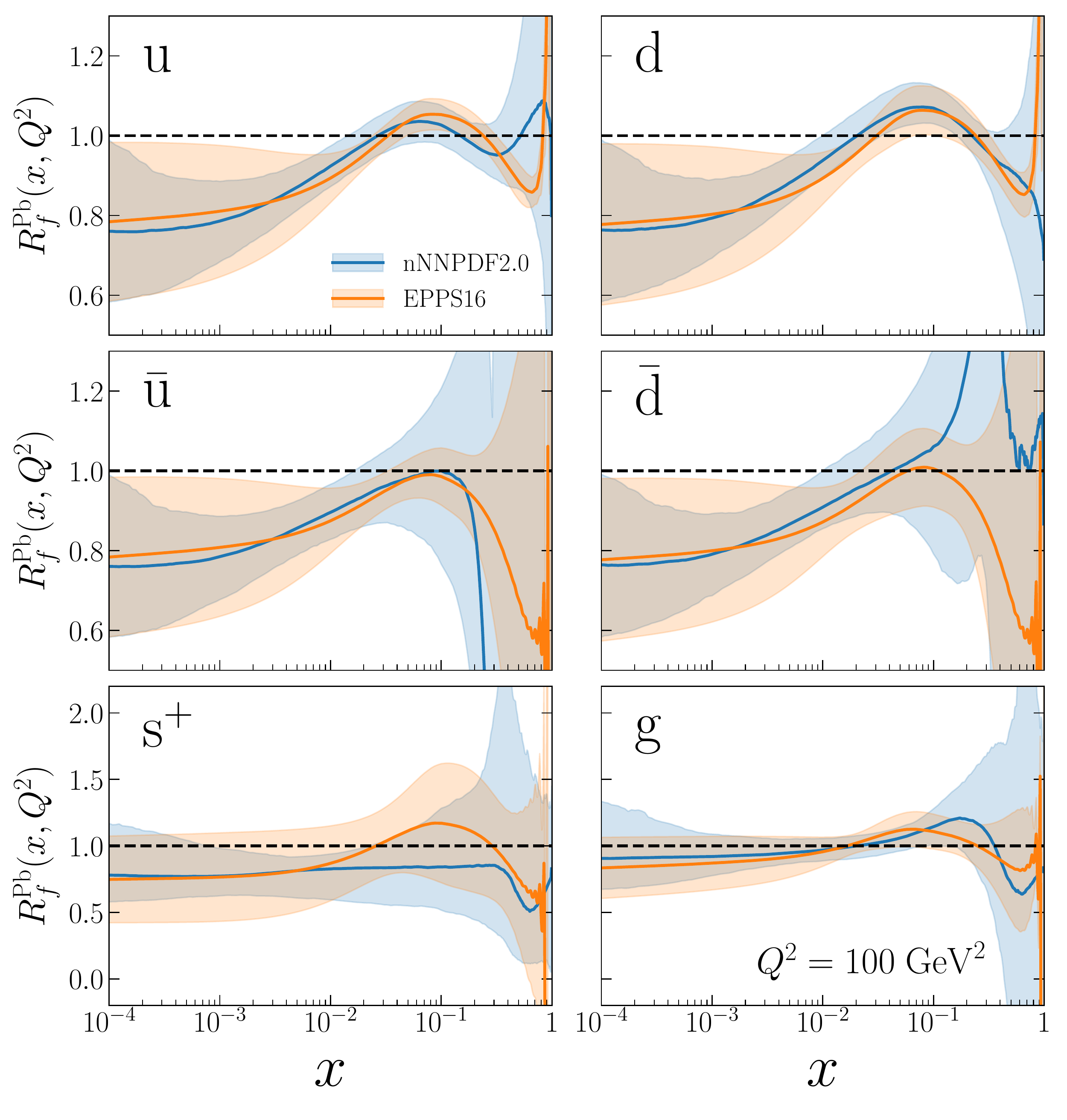}
 \end{center}
\vspace{-0.9cm}
\caption{Same as Fig.~\ref{fig:R_Fe} in the case of lead nuclei, $R_{\rm
Pb}(x,Q^2)$.
  \label{fig:R_Pb}
}
\end{figure}

\paragraph{Comparison with EPPS16.}
In Figs.~\ref{fig:R_Fe} and~\ref{fig:R_Pb}, we display the nuclear PDF
modification ratios  at $Q=10$
GeV for iron, $R_f^{\rm Fe}(x,Q^2)$, and lead, $R_f^{\rm
Pb}(x,Q^2)$, for nNNPDF2.0 and EPPS16, each normalized to the
corresponding free-proton baseline.
  As in Fig.~\ref{R_A}, we show the up and down quarks, the
  corresponding antiquarks, the total strangeness, and the gluon.
  The bands again correspond to the 90\% CL uncertainties constructed
  using the appropriate prescription for each set.
  This means that for EPPS16, the error is computed by adding in
  quadrature the differences in value between the $N_{\rm eg}$
  eigenvectors of the Hessian set and the best fit result.\footnote{When
  computing PDF ratios with EPPS16 we neglect proton PDF uncertainties,
  since adding the EPPS16 and CT14 errors in quadrature is likely to
  represent an overestimate given the missing mutual correlations.}

Beginning with the nPDF comparison for iron nuclei, we find that there
is good agreement between the results of nNNPDF2.0 and EPPS16 both in
terms of central values and of the nPDF uncertainties in the range of
$x$ for which experimental data is available.
In the small- and large-$x$ extrapolation regions, the uncertainties are
larger in the nNNPDF2.0 case.
We also note that the Fermi-motion-like growth of $R_u^{\rm Fe}$ and
$R_d^{\rm Fe}$ at very large $x$, which is built into the EPPS16
parameterization, is absent in the nNNPDF2.0 results.
There instead one finds a suppression compared to the free-proton case,
especially in the case of $R_d^{\rm Fe}$.
As expected, the observed pattern of nuclear modifications is very
similar between up and down quarks and between the corresponding
antiquarks due to iron being nearly isoscalar.

Considering the behavior of the sea quarks, nNNPDF2.0 and EPPS16 agree
well in terms of central values and uncertainties in the shadowing
region, $x\lsim 0.05$.
However, there are more significant differences at large-$x$, where the
qualitative behavior between the two nPDF sets is the opposite:
nNNPDF2.0 favors an enhancement compared to the free-proton baseline,
while EPPS16 prefers instead a suppression.
In any case, the differences are well within the large uncertainty
bands, and additional data is needed to be able to ascertain the correct
behavior in this region.
Note that at large-$x$ the free-proton baseline antiquarks are also
affected by large errors, complicating the interpretation of
$R_{\bar{u}}^{\rm Fe}$ and $R_{\bar{d}}^{\rm Fe}$.

Turning to the nuclear modification of the total strangeness, in
nNNPDF2.0 we find a suppression of $\sim 20$\% compared to the proton
baseline in the relevant range of $x$.
This is consistent with studies of the interplay between the NuTeV
dimuon and the ATLAS W,Z 2011 data in proton global analyses, where
the latter data set largely suppress strangeness in contrast to the
former.
Such behavior is not reported by EPPS16, which exhibits much larger nPDF
uncertainties that are likely due to the absence of the strange-sensitive
NuTeV cross-sections in their analysis. Furthermore, the ATLAS W,Z
2011 distributions are missing from the CT14 proton baseline used by
EPPS16 (although these data have been accounted for in the recent CT18
release~\cite{Hou:2019efy}).

Finally, concerning the gluon PDF we find from this comparison that in
the nNNPDF2.0 fit there is little evidence for nuclear shadowing, with
$R_g^{\rm Fe}\simeq 1$ in the region $x\le 0.05$.
We also find that the nPDF uncertainties on the gluon are larger
compared to EPPS16 by roughly a factor of two.
At larger values of $x$, the uncertainties increase significantly and
nNNPDF2.0 prefers a suppressed central value, unlike EPPS16.
We note that neither of the two analyses include direct constraints on the
large-$x$ nuclear gluons, hence the sizeable nPDF uncertainties,
although available data on dijet and photon production could improve
this situation.

In the corresponding comparison for lead nuclei, displayed in
Fig.~\ref{fig:R_Pb}, one observes a number of similarities and
differences with respect to the nPDFs of iron.
Concerning the up and down quarks, we find our nNNPDF2.0 result provides
significant evidence for shadowing at small-$x$.
For instance, at $x\simeq 5\times 10^{-3}$ we obtain $R_u^{\rm Pb} \ne 1$ at the
four-sigma level or higher.
Interestingly, nPDF uncertainties in the shadowing region are up to a
factor of two smaller in nNNPDF2.0 than in EPPS16.
While in both cases anti-shadowing at $x\simeq 0.1$ is observed, the
larger $x$ qualitative behavior is different between the two analyses,
with EPPS16 finding the (built-in) EMC suppression followed by
Fermi-motion rise while in nNNPDF2.0 the pattern of nuclear
modifications is different.
In any case, the agreement between the central values of nNNPDF2.0 and
EPPS16 for $R_u^{\rm Pb}$ and $R_d^{\rm Pb}$ in the region of $x\lsim
0.3$ is quite remarkable given the very different methodologies employed in
each study.

Concerning the nuclear modifications of the sea quarks in lead nuclei,
one finds a similar qualitative behavior as in the case of iron.
For $x\lsim 0.1$ there is good agreement between the central values of
$R_{\bar{u}}^{\rm Pb}$ and $R_{\bar{d}}^{\rm Pb}$ between EPPS16 and
nNNPDF2.0, with the latter exhibiting smaller uncertainties.
The two sets are more notably different for $x\gsim 0.1$ instead, where
EPPS16 predicts a EMC-like suppression common to $\bar{u}$ and $\bar{d}$
while nNNPDF2.0 favours a large suppression for $\bar{u}$ but an
enhancement for $\bar{d}$.
However, the large nPDF uncertainties in this region prevent any
definitive conclusions, though the two fits are fully consistent
within the 90\% CL bands.
One possible source for the differences could be in the respective
free-proton counterparts, where large-$x$ antiquarks are poorly known.
For the total strangeness, the behavior of $R_{s^+}^{\rm Pb}$ is similar
to that of iron, where nNNPDF2.0 predicts a suppression more or
less independent of $x$, with rather larger uncertainties for EPPS16
compared to our nNNPDF2.0 result due to the missing constraints from the
NuTeV dimuon cross-sections.

Finally, regarding the nuclear modifications of the lead gluon PDF
illustrated in the bottom right panel of Fig.~\ref{fig:R_Pb}, we again
find that $R_g^{\rm Pb}$ agrees with unity across all relevant $x$.
Here the initial-scale differences are washed out partially by DGLAP
evolution, but clearly the shadowing in the nuclear gluons is less
apparent than for the quarks.
Although the nPDF uncertainties increase at large $x$ due to the lack of
direct constraints, the qualitative behavior of $R_g^{\rm Pb}$ differs
between the two PDF determinations.

\paragraph{Nuclear strangeness.}
The strangeness content of the proton in unpolarized PDF fits has
attracted a lot of attention recently.
Traditionally, the determination of $s(x,Q^2)$ in global proton PDF fits
has been dominated by the constraints provided by charm production in
neutrino
DIS~\cite{Bazarko:1994tt,chorus-dimuon,MasonPhD,Mason:2007zz,Samoylov:2013xoa}.
These measurements suggest that the strange sea is suppressed compared
to its up and down quark counterparts, favoring values of around $r_s
\simeq 0.5$ when expressed in terms of the strangeness ratio defined as
\be
\label{eq:strangenessratio}
r_s(x,Q^2) \equiv
\frac{s(x,Q^2)+\bar{s}(x,Q^2)}{\bar{u}(x,Q^2)+\bar{d}(x,Q^2)} \, .
\ee
Other strange-sensitive processes agree qualitatively with the
constraints on $r_s$ provided by the neutrino DIS data, such as W
production in association with charm quarks~\cite{Stirling:2012vh} from
CMS~\cite{Chatrchyan:2013uja,CMS-PAS-SMP-18-013,Sirunyan:2018hde} and
ATLAS 7 TeV~\cite{Aad:2014xca}, and semi-inclusive deep-inelastic
scattering (SIDIS)~\cite{Airapetian:2013zaw,Borsa:2017vwy,Sato:2019yez}.
However, the  ATLAS measurements of the leptonic rapidity distributions
in inclusive W and Z production at 7
TeV~\cite{Aad:2012sb,Aaboud:2016btc} exhibit instead a strong preference
for a symmetric strange sea with $r_s\simeq 1$.
One should point out that general considerations based on perturbative
DGLAP evolution imply that $r_s \to 1$ at large $Q$ and small-$x$, but
at low $Q$ and medium/large-$x$ the value of $r_s$ is dictated by
non-perturbative dynamics.

\begin{figure}[t]
\begin{center}
  \includegraphics[width=0.99\textwidth]{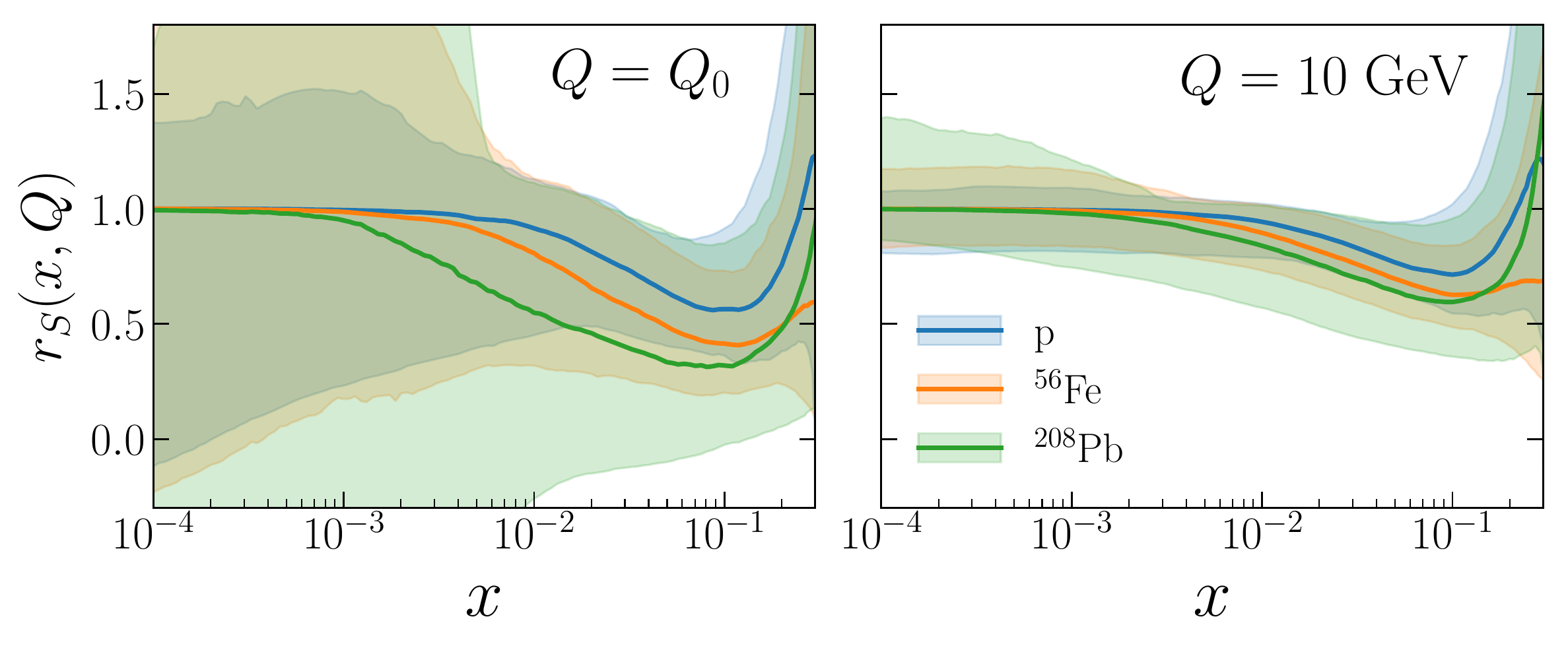}
 \end{center}
\vspace{-0.75cm}
\caption{The strangeness ratio $r_s(x,Q^2)$, defined in
  Eq.~(\ref{eq:strangenessratio}), in nNNPDF2.0 comparing the results
  for the free-proton baseline, $^{56}$Fe, and $^{208}$Pb at both the
  input parametrisation scale $Q_0=1$ GeV (left) and at a higher scale
  $Q=10$ GeV (right plot).
  \label{fig:R_S}
}
\end{figure}

As was motivated in Ref.~\cite{Ball:2018twp}, it is important to
carefully assess the nuclear uncertainties associated to the nuclear
strangeness, given that these will potentially affect the determination
of the proton strangeness from global fits based on neutrino data.
 We display in Fig.~\ref{fig:R_S} the strangeness ratio $r_s(x,Q^2)$
  defined by Eq.~(\ref{eq:strangenessratio}) obtained with our nNNPDF2.0
  result for $^1$p, $^{56}$Fe, and $^{208}$Pb at both the input
  parameterization scale $Q_0=1$ GeV and at a higher scale of $Q=10$
  GeV.

From the comparison in Fig.~\ref{fig:R_S} we find that at the input
parameterization scale $r_s$ is particularly suppressed in the case of
lead, where the central value of nNNPDF2.0 satisfies $r_s < 0.5$ for $5
\cdot 10^{-3} \lsim x \lsim 0.2$.
A similar preference for a suppressed strange sea, albeit less
pronounced, can be seen in iron nuclei.
In any case, the nPDF uncertainties affecting this ratio are rather large,
in particular for the heavier nuclei.
The fact that for $x \lsim 10^{-3}$ one obtains $r_s \simeq 1$ for all
three nuclei is a consequence of the parameterization preprocessing,
whose ranges are chosen to ensure that in the small-$x$ extrapolation
region all quark and antiquark flavors behave in the same way (see
Figs.~\ref{fig:beta_exp} and~\ref{fig:alpha_exp}).
Once DGLAP evolution takes place, $r_s$ tends to become closer to unity
across a wider range in $x$, but even at the higher scale a suppressed
strangeness for $x\gsim 0.01$ is preferred for both iron and lead.

The results in Fig.~\ref{fig:R_S} suggest that including neutrino CC
structure functions such as CHORUS and NuTeV in proton PDF fits without
accounting for nuclear uncertainties might not be a justified
approximation, given the current precision that modern fits achieve.
It will be interesting nonetheless to determine the impact on the global NNPDF
proton PDF fits when nNNPDF2.0  is used to
account for nuclear uncertainties using the procedure outlined in
Ref.~\cite{Ball:2018twp}. 

\subsection{Comparison with nNNPDF1.0}
\label{sec:comparisonNNPDF10}

We now turn to study the differences between the nNNPDF1.0 and nNNPDF2.0
determinations by tracing back the impact of the various improvements in the latter
with a series of comparisons.
The goal of this exercise is to assess which of these differences can be
identified with specific methodological improvements, such as the
cross-section positivity constraint, and which ones are related to the
impact of the new experimental information, either the DIS charged
current structure functions or the LHC gauge boson production
measurements.

The starting point for this study will be a fit denoted nNNPDF1.0r,
which has been obtained with the code used to produce nNNPDF2.0 but
using the same theory, methodology settings, and input dataset as in the
nNNPDF1.0 analysis.
The only differences at this level are related to optimizations and
improvements implemented in the code to speed up its performance.
We have verified that nNNPDF1.0 and nNNPDF1.0r are statistically
indistinguishable, thus we can safely adopt the latter as baseline for
the comparisons in what follows.

\begin{figure}[t]
\begin{center}
  \includegraphics[width=0.95\textwidth]{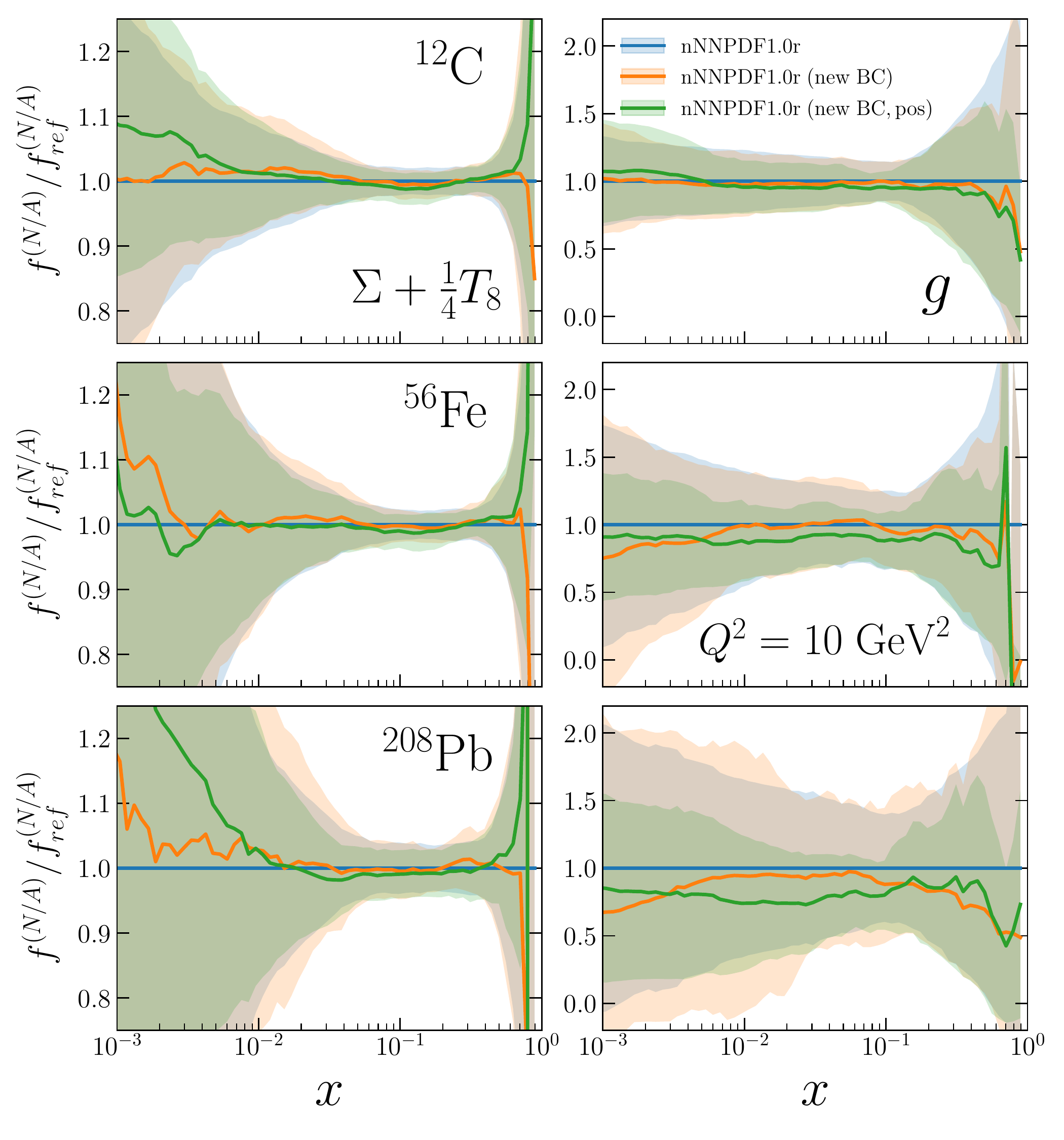}
 \end{center}
\vspace{-0.75cm}
\caption{Comparison between the nNNPDF1.0r baseline and two fit variants
  based on the same input dataset, one where the proton boundary
  condition has been updated and another where in addition the
  positivity of physical cross-sections has been imposed.
  We show the $\Sigma+T_8/4$ quark combination (left) and gluon (right)
  at $Q^2=10$ GeV$^2$ for three values of $A$.
  \label{nNNPDF10Ref}
}
\end{figure}

We have then produced several variants of this nNNPDF1.0r baseline, each
time adding one extra feature or dataset.
The first of these two variants is a fit where the proton boundary
condition has been updated to the no-nuclear NNPDF3.1 fit shown in
Fig.~\ref{fig:NNPDF31_BC_comp}.
The second is a fit where, in addition to the updated boundary
condition, the positivity of cross-sections has been imposed following
the procedure described in Sect.~\ref{sec:positivity}.
We display in Fig.~\ref{nNNPDF10Ref} the comparison between nNNPDF1.0r
and these two fit variants.
Since the isoscalar neutral-current DIS structure functions used
in nNNPDF1.0 are primarily sensitive to the specific quark combination
$\Sigma+T_8/4$, we plot this together with the gluon distribution as a
function of $x$ at $Q^2=10$ GeV$^2$ for carbon, iron, and lead.

First, one can see from Fig.~\ref{nNNPDF10Ref} that the impact of the
new proton boundary condition in the nuclear fit is generally moderate
concerning the size of the uncertainty band.
There are some differences at the central value level for the small-$x$
quarks and for the nuclear gluon PDF of lead, but in both cases the
shifts are much smaller than the associated uncertainties.
This does not imply that using the updated proton boundary condition is
irrelevant for nNNPDF2.0, but rather that this choice is not
particularly impactful for the specific PDF
combinations that can be constrained by
the nNNPDF1.0 dataset.
As shown in Fig.~\ref{fig:NNPDF31_BC_comp}, the differences between the
two variants of the proton boundary conditions are more distinguished
for the total strangeness compared to the other quark flavors.

On the other hand, imposing the positivity of the cross-sections leads
to more important differences.
This is not completely unexpected, since it is well known that in
general a model-independent (n)PDF analysis will lead to some
cross-sections being negative unless their positivity is explicitly
imposed.
In our case, one finds that there is not much difference in the quarks,
but there are clear changes for the nuclear gluons in iron and lead,
especially in the latter.
Here we see that imposing the positivity of cross-sections leads to a
significant reduction of the nPDF uncertainty band, which in the case of
lead can be up to a factor of two.

\begin{figure}[t]
\begin{center}
  \includegraphics[width=0.95\textwidth]{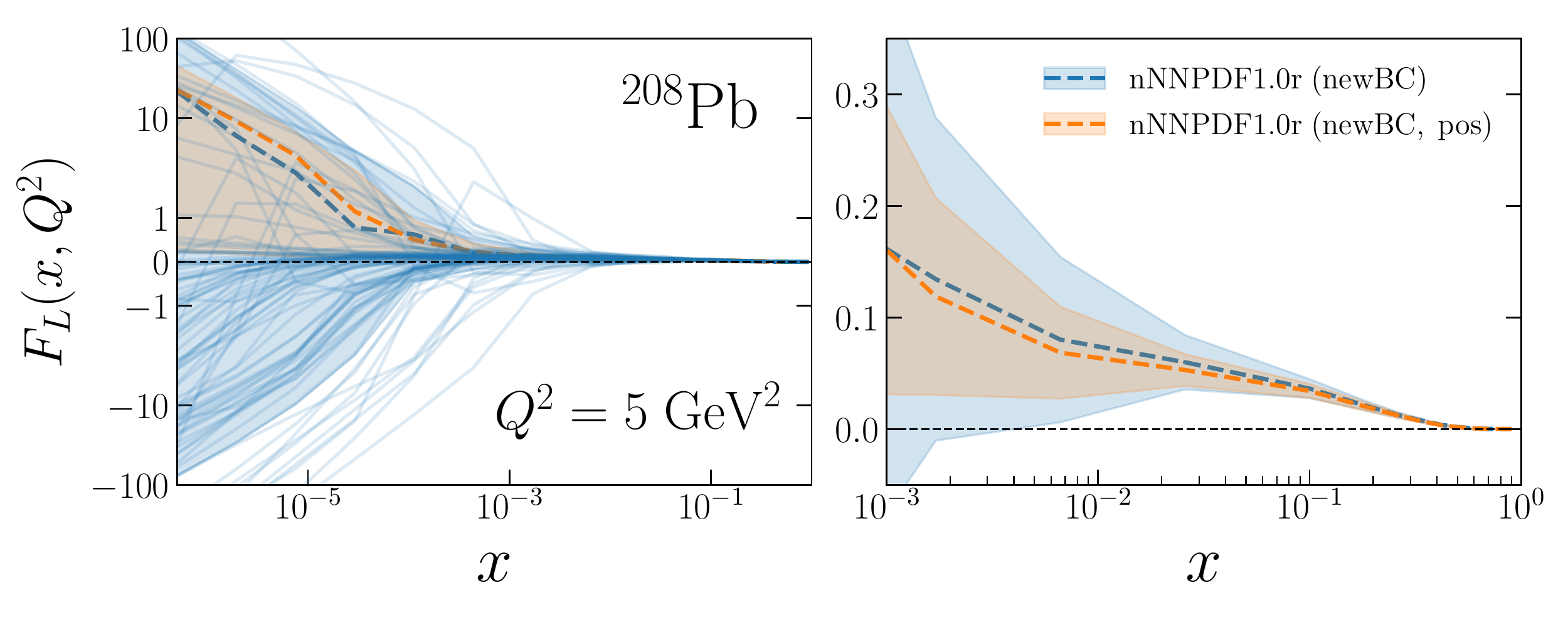}
 \end{center}
\vspace{-0.8cm}
\caption{The longitudinal structure function $F_L(x,Q^2)$ at the
  positivity scale $Q^2=5$ GeV.
  We compare the predictions of the nNNPDF1.0r(newBC) fits with and
  without the cross-section positivity constraint imposed.
  We show the extrapolation (left) and the data (right) regions, in the
  former case displaying also the predictions from the individual
  replicas in the  nNNPDF1.0r(newBC) fit that do not satisfy the
  positivity constraints.
   \label{fig:FLpos}
}
\end{figure}

To illustrate the impact of the cross-section positivity constraint, we
display in Fig.~\ref{fig:FLpos} the longitudinal structure function
$F_L(x,Q^2)$ at the positivity scale $Q^2=5$ GeV.
We compare the predictions of the nNNPDF1.0r fits including the updated
proton boundary condition with and without the cross-section positivity constraint imposed in
both the extrapolation and the data regions.
In the left panel, we display also the predictions from the individual
replicas of the nNNPDF1.0r(newBC) fit that do not
satisfy the cross-section positivity constraints.
Indeed, one can observe that many $F_L$ replicas become negative in some
region of $x$ unless this constraint is explicitly imposed, and that
removing them leads to a significant reduction of the nPDF
uncertainties, particularly in the small-$x$ region.
Interestingly, at medium-$x$ it is largely the upper (rather than the
lower) 90\% CL limit which is reduced by the positivity constraint: this
can be explained by the fact that the very negative $F_L$ replicas at
small-$x$ were actually higher than the median value at medium-$x$ in
order to satisfy the momentum sum rule.

\begin{figure}[t]
\begin{center}
  \includegraphics[width=0.95\textwidth]{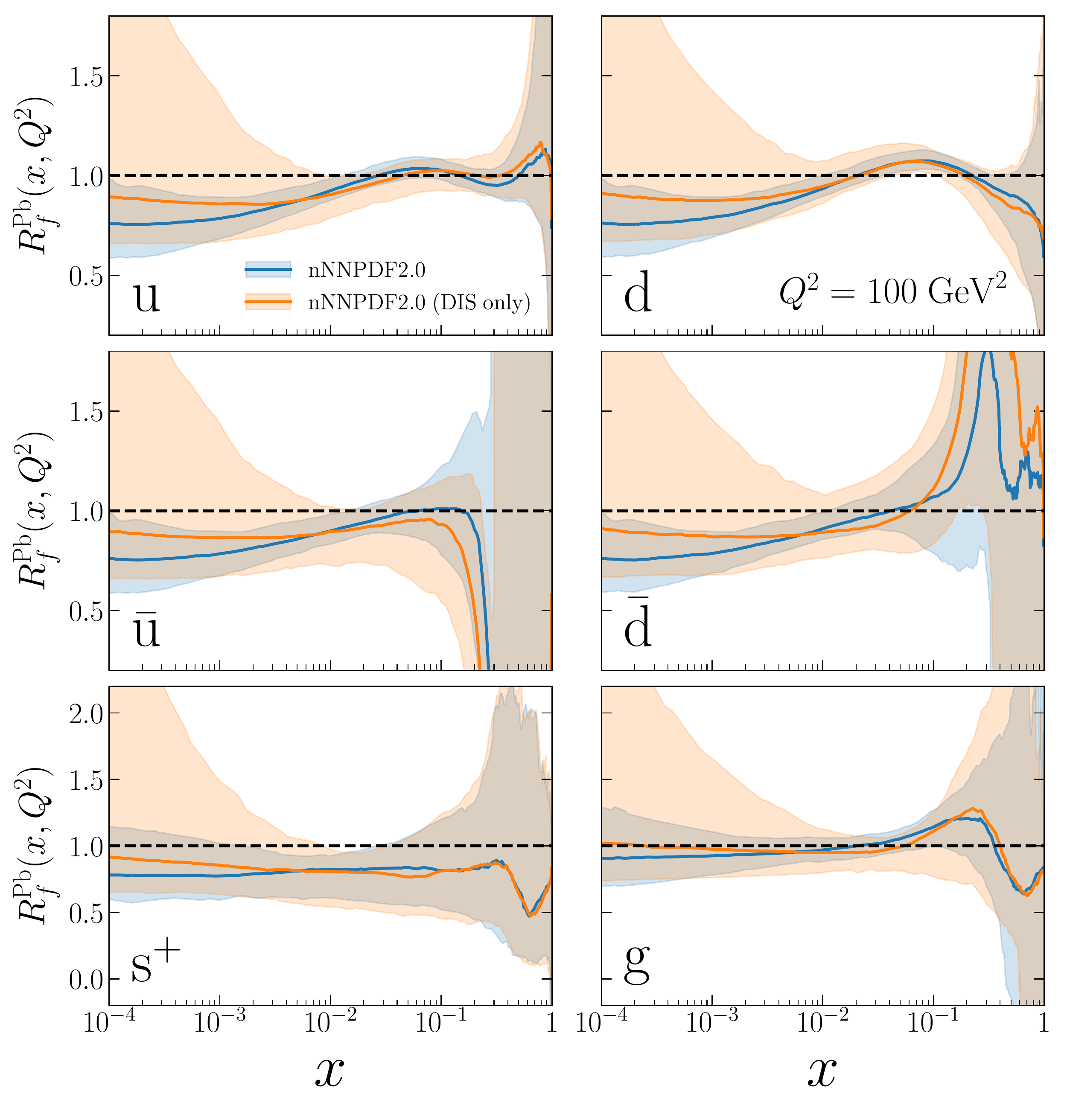}
 \end{center}
\vspace{-0.75cm}
\caption{Same as Fig.~\ref{fig:R_Pb} now comparing the nNNPDF2.0
  baseline results with those of a fit based on identical settings but
  restricted to a DIS-only dataset.
  \label{fig:R_Pb_DIS}
}
\end{figure}

Concerning the impact of the new datasets, a direct comparison of the
nNNPDF1.0r-like fits with those including CC DIS and LHC data is not
possible since as discussed in Sect.~\ref{sec:parametrisation}
the input parameterization basis and the flavor
assumptions are different.
However, we are still able to assess the relative contribution of the CC
structure functions and the LHC gauge boson cross-sections in
determining the nNNPDF2.0 results.
In Fig.~\ref{fig:R_Pb_DIS} we display the nuclear modification ratios
for the nPDFs in lead, as was shown in Fig.~\ref{fig:R_Pb}, but now
comparing the nNNPDF2.0 baseline results with those of a fit that is
restricted to DIS structure functions, including charged-current
scattering, and that uses identical theoretical and methodological
settings.

  One of the most remarkable features of this comparison is the sizeable
  impact that LHC measurements have in reducing the uncertainties of the
  nuclear PDFs.
  This effect is particularly significant for the gluon and for all
  quark flavors at $x\lsim 0.1$.
  On one hand, the LHC data clearly reveals the presence of nuclear
  shadowing at small-$x$ for both the valence and sea quarks, something
  which is not accessible in a DIS-only fit.
  This result is consistent with the nuclear modification ratios for the
  LHC Drell-Yan distributions reported {\it e.g.} in
  Fig.~\ref{nNNPDF20comp_data_3}.
  On the other hand, the impact of the LHC data on the central values
  and uncertainties of nNNPDF2.0 at $x\gsim 0.1$ is less prominent,
  although in that region one also observes a reduction of the
  uncertainties.
  The fact that $R_{\bar{u}}\ll 1$ and $R_{\bar{d}}\gg 1$ for lead
  nuclei at large-$x$ is already present at the level of DIS-only fits
  implies that this trend is favored by the CHORUS and NuTeV
  charged-current structure functions.

\subsection{The momentum and valence integrals in nuclei}
\label{sec:SR}

As was discussed in Sect.~\ref{sec:parametrisation}, we impose three sum
rules in the nNNPDF2.0 determination, namely the momentum sum rule,
Eq.~(\ref{eq:MSR}), and the two valence sum rules,
Eqns.~(\ref{eq:valencesr4}) and~(\ref{eq:valencesr5}).
These constraints are satisfied by adjusting the overall prefactors
$B_f$ in  Eq.~(\ref{eq:param2}) for the gluon $g$, the total valence
$V$, and the valence triplet $V_3$ distributions, respectively.
Furthermore, they are independently imposed for each value of $A$ for
which there is available experimental data.

Here we investigate the role played by these sum rules in the global
nPDF determination.
In particular, we address whether or not the physical requirements of
energy and valence quark number conservation are satisfied by the
phenomenological fit to experimental data (within uncertainties) when
the sum rules are not explicitly imposed.
Recently, theoretical arguments have been put forward that the momentum
sum rule for nucleons in nuclei might not hold~\cite{Brodsky:2019jla}.
Motivated in this respect, we have carried out a similar study to the
one presented in Ref.~\cite{Ball:2011uy}, where global proton PDF fits without
imposing the momentum sum rule were performed.
In the proton case, while the LO prediction for the momentum integral
was to be far from the QCD expectation, both the NLO
and NNLO fits exhibited remarkable agreement at the $\simeq 1\%$ level~\cite{Ball:2011uy}.

We have therefore produced two variants of the nNNPDF2.0 analysis, each
based on $N_{\rm rep}=250$ replicas, where either the momentum sum rule
or the total valence sum rule is not imposed.
Afterwards, we evaluate in each case the corresponding momentum and
total valence integrals, defined as
\be
\label{eq:MSRintegral}
I_{\rm M}(A) \equiv \int_0^1 dx \,x \left(\Sigma^{(p/A)}(x,Q_0) +
g^{(p/A)}(x,Q_0)\right) \, ,
\ee
\be
\label{eq:valencesr4integral}
I_{\rm V}(A) \equiv \int_0^1 dx~ V^{(p/A)}(x,Q_0) \, ,
\ee
and assess whether or not they are in agreement with the QCD
expectations, namely $I_{\rm M}(A)=1$ and $I_{V}(A)=3$
respectively.
One should note that the momentum and valence sum rules are already
satisfied at the level of the proton boundary condition, and thus some
constraints are expected to be propagated to the lighter nuclei.
However, the analysis of Ref.~\cite{Ball:2011uy} demonstrates that
results would be largely unchanged if the momentum and valence sum rules
would have been excluded also from the free-proton baseline.

\begin{figure}[t]
\begin{center}
  \includegraphics[width=0.9\textwidth]{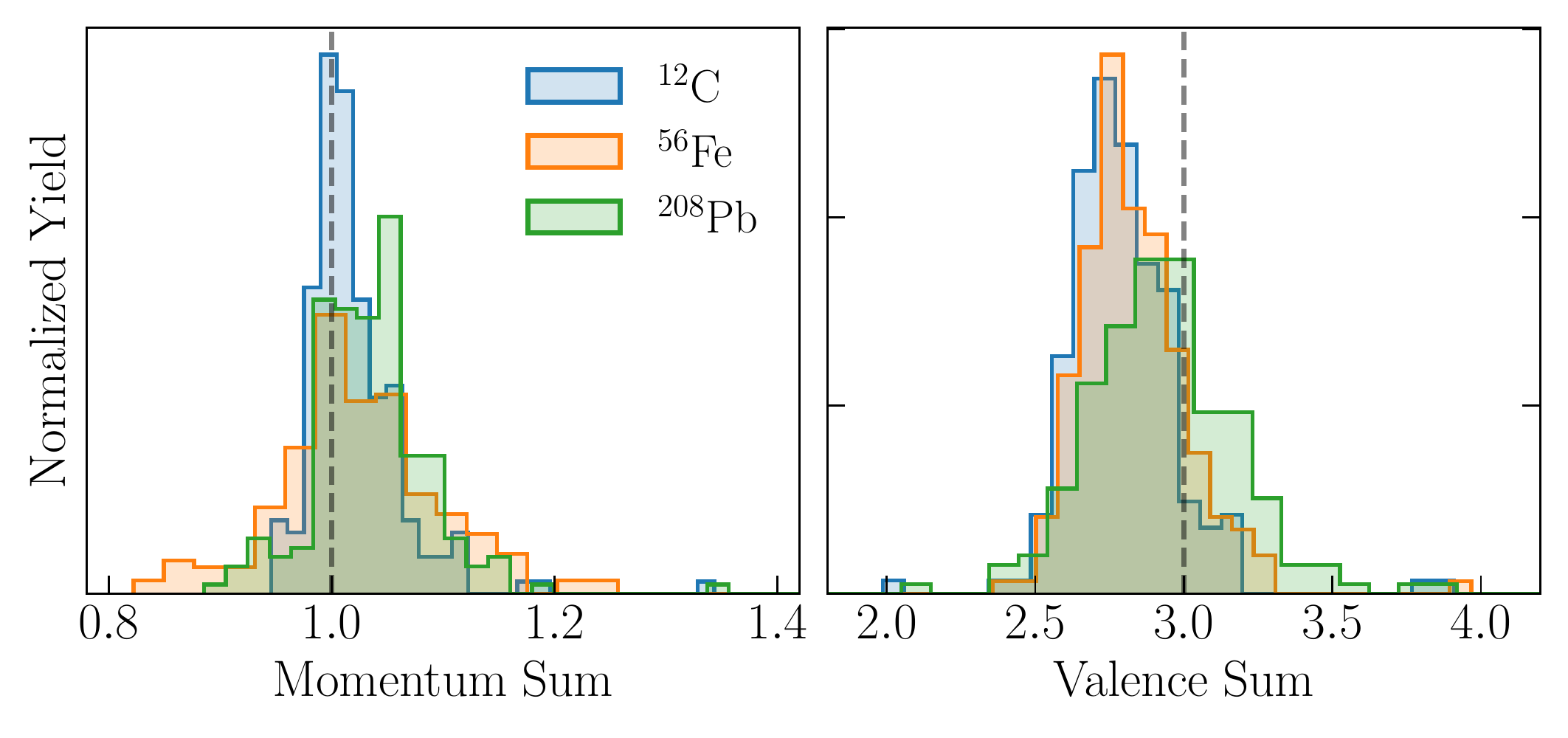}
 \end{center}
\vspace{-0.67cm}
\caption{\small The distribution of the momentum (left) and total
  valence (right panel) integrals, Eqns.~(\ref{eq:MSRintegral})
  and~(\ref{eq:valencesr4integral}) respectively, in the variants of the
  nNNPDF2.0 determination where the corresponding sum rules have not
  been explicitly imposed.
  We show the relative frequency of the momentum and valence integral
 for three representative nuclei: $^{12}$C, $^{56}$Fe, and
 $^{208}$Pb.
 The dashed vertical line indicates the corresponding the QCD expectations,
 $I_{\rm M}(A)=1$ and $I_{\rm V}(A)=3$.
 The associated 90\% CL ranges are reported in
 Table~\ref{tab:momintegral}.}
\label{fig:MomIntegral_noMSR}
\end{figure}

In Fig.~\ref{fig:MomIntegral_noMSR} we display the distribution of the
momentum and valence integrals, Eqns.~(\ref{eq:MSRintegral})
and~(\ref{eq:valencesr4integral}), respectively, in the variants of the
nNNPDF2.0 fit where the corresponding sum rules are not being explicitly
imposed.
  We show the relative frequency of the momentum and valence integral
 for three representative nuclei: $^{12}$C, $^{56}$Fe, and $^{208}$Pb.
 The dashed vertical line in
 Fig.~\ref{fig:MomIntegral_noMSR} indicates the QCD expectations for
 $I_{\rm M}(A)$ and $I_{\rm V}(A)$.
 The corresponding values for the 90\% confidence level intervals for
 each of these two integrals for the relevant values of $A$, as well as
 for the free-proton baseline $A=1$, can be found in
 Table~\ref{tab:momintegral}.
 
\begin{table}[t]
  \renewcommand{\arraystretch}{1.35}
  \centering
  \begin{tabular}{|c|c|c|}
    \toprule
    $\qquad\qquad A\qquad \qquad$  &  $ \qquad\qquad I_{\rm M}(A)
    \qquad\qquad$  & $ \qquad\qquad I_{\rm V}(A) \qquad\qquad$  \\
    \midrule
    1  & $\lc 0.99, 1.06 \rc$ & $\lc 2.53, 3.12 \rc$   \\
    12  & $\lc 0.97, 1.10 \rc$ &  $\lc 2.56, 3.11 \rc$  \\
    56  & $\lc 0.90, 1.16 \rc$ &  $\lc 2.58, 3.16 \rc$  \\
    208  & $\lc 0.94, 1.12 \rc$ & $\lc 2.54, 3.34 \rc$   \\
    \bottomrule
  \end{tabular}
  \vspace{0.3cm}
  \caption{\label{tab:momintegral}
 The 90\% CL ranges for the momentum and valence integrals,
  Eqns.~(\ref{eq:MSRintegral}) and~(\ref{eq:valencesr4integral}),
  in the variants of the nNNPDF2.0 fits whether either one
  or the other sum rule is not imposed. }
\end{table}

From the results presented in Fig.~\ref{fig:MomIntegral_noMSR} and
Table~\ref{tab:momintegral} one finds that the momentum integral is in
agreement with the QCD expectation, $I_{\rm M}(A)=1$, within
uncertainties for all nuclei.
In the case of $^{12}$C for example, one finds that $0.97 \lsim I_{\rm M}
\lsim 1.10$ at the 90\% confidence level, with somewhat larger
uncertainties for the heavier nuclei.
Even for lead, where the proton boundary condition has little effect,
the median of the distribution is reasonably close to the QCD
expectation.
The uncertainties on $I_{\rm M}$ are larger in the nuclear
PDF analysis than the $\simeq 1\%$ error
found in the proton case~\cite{Ball:2011uy}, as expected since the experimental data for
nuclear collisions is far less abundant and further distributed between
different nuclei.
Nevertheless, the overall consistency
with the QCD expectations
is quite compelling.
Note also that here the proton boundary condition is imposed
only for $x\ge 10^{-3}$, and therefore our prediction for $I_{\rm M}(A=1)$
is expected be less accurate as compared to the proton global analysis case.

\begin{figure}[t]
\begin{center}
  \includegraphics[width=0.95\textwidth]{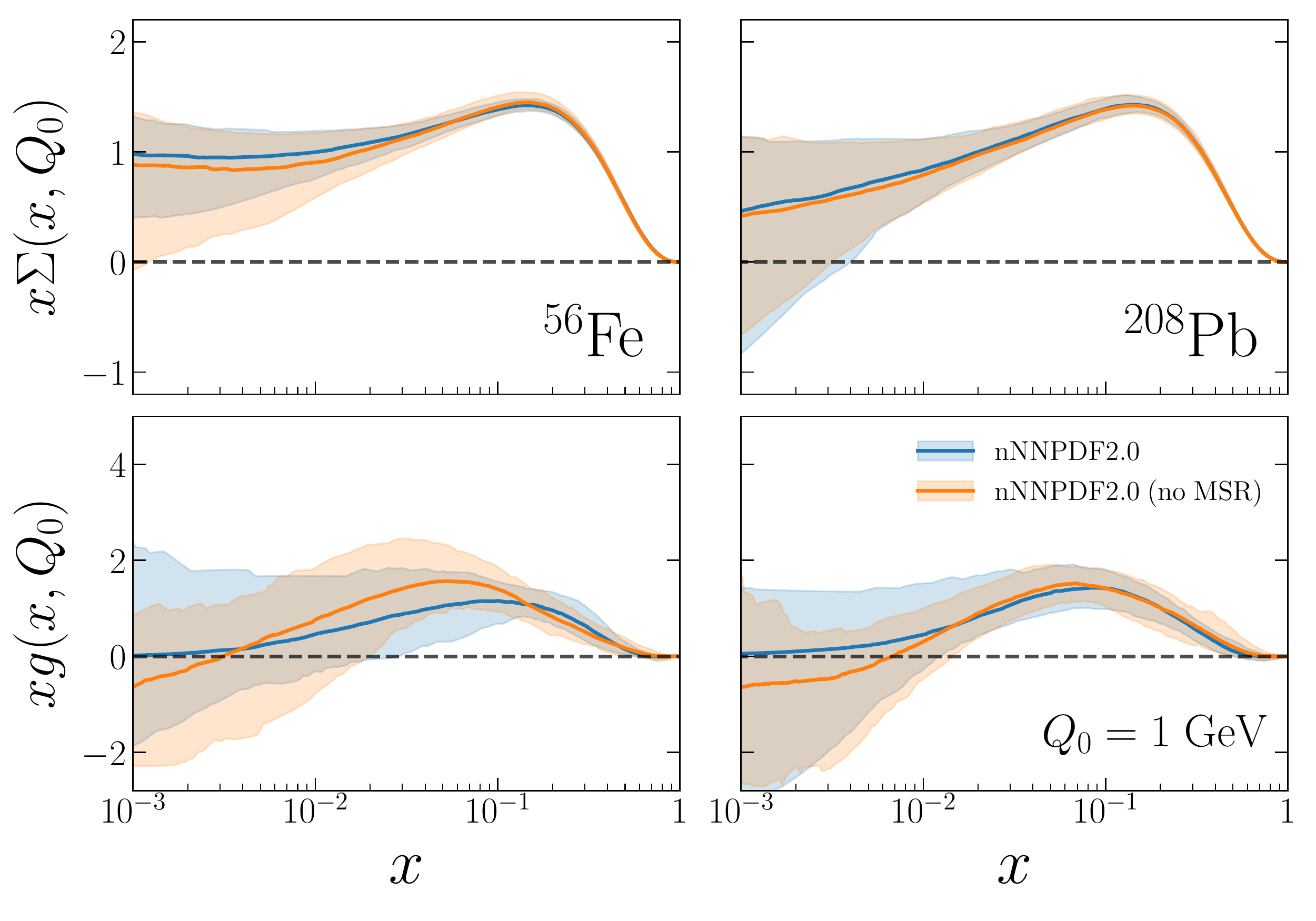}
 \end{center}
\vspace{-0.75cm}
\caption{\small Comparison of the baseline nNNPDF2.0 fit at $Q_0=1$ GeV
  with the variant in which the momentum sum rule is not being imposed.
  We show the total quark singlet (left) and the gluon (right) for
  $^{56}$Fe (upper) and $^{208}$Pb (lower panels).}
\label{fig:PDFcomp_noMSR}
\end{figure}

The result that the momentum integral agrees with the theoretical
predictions for all nuclei is a non-trivial validation of the global
nuclear PDF analysis framework based on the QCD factorization hypothesis.
It further demonstrates the robustness of our fitting methodology, in
that the resulting nPDFs are reasonably stable regardless of whether or
not the momentum sum rule is imposed during the fit.
To illustrate better this latter point, in Fig.~\ref{fig:PDFcomp_noMSR} we
provide a comparison between the baseline nNNPDF2.0 fit at $Q_0=1$ GeV
with the variant in which the momentum sum rule is not being imposed.
  We show the total quark singlet and the gluon for both $^{56}$Fe  and
  $^{208}$Pb.
  Recall that the momentum sum rule is used to fix the overall gluon
  normalization in Eq.~(\ref{eq:param2}).
  In the case of lead, where the experimental constraints are relatively
  abundant, we find that both the singlet and the gluon are reasonably
  similar irrespective of whether or not the momentum sum rule is
  imposed.
  The momentum sum rule plays a larger role in iron, especially in
  reducing the gluon nPDF uncertainties, but interestingly the central
  value of the all distributions is quite stable when comparing the two
  fits.
 This stability is consistent with the results reported in
 Fig.~\ref{fig:MomIntegral_noMSR}.

The main conclusions are qualitatively similar for a fit in which the
total valence sum rule has not been imposed.
Results of this fit are displayed in the right panel of
Fig.~\ref{fig:MomIntegral_noMSR}, where the normalized frequency of the
total valence integral are shown for the same three nuclei discussed
previously.
 The corresponding 90\% CL intervals are also reported in
 Table~\ref{tab:momintegral}.
Similar to the momentum sum results, we find that for the valence
integral the fit results agree with the QCD expectations within
uncertainties.
The preferred value of the valence integral (median) turns out to be
$I_{\rm V} \simeq 2.8$ irrespective of $A$.
This implies that even when Eq.~(\ref{eq:valencesr4integral}) is not
imposed explicitly, the experimental measurements favor the QCD
prediction within 5\% for all values of $A$ relevant for the present
study.
We have also verified that, in a similar way as in
Fig.~\ref{fig:PDFcomp_noMSR}, the resulting nPDFs are reasonably stable
regardless of whether or not the valence sum rule is imposed.

Putting together the results of these two exercises, one can conclude
that the fit results are relatively stable in the nNNPDF framework even
in the absence of the sum rules, consistent with the fact that
experimental data and the QCD expectations based on the factorization
theorem are in agreement with each other for hard-scattering collisions
involving heavy nuclei.

\subsection{The positivity of physical cross-sections}
\label{sec:posresults}

As was discussed in Sect.~\ref{sec:positivity}, we impose the
requirement that the cross-sections of arbitrary physical processes are
positive-definite quantities.
This constraint is implemented by means of an additive penalty term in
the figure of merit, Eq.~(\ref{eq:positivity}).
Moreover, the penalty is constructed from the pseudo-data summarized in
Table~\ref{tab:positivity}, which corresponds to lepton-nuclear
scattering structure functions and Drell-Yan cross-sections in
proton-nucleus collisions.
Recall that the kinematics of the positivity pseudo-data were chosen to
  cover those of the actual data used in the fit, see
  Fig.~\ref{figkinplot}.

\begin{figure}[t]
\begin{center}
  \includegraphics[width=0.9\textwidth]{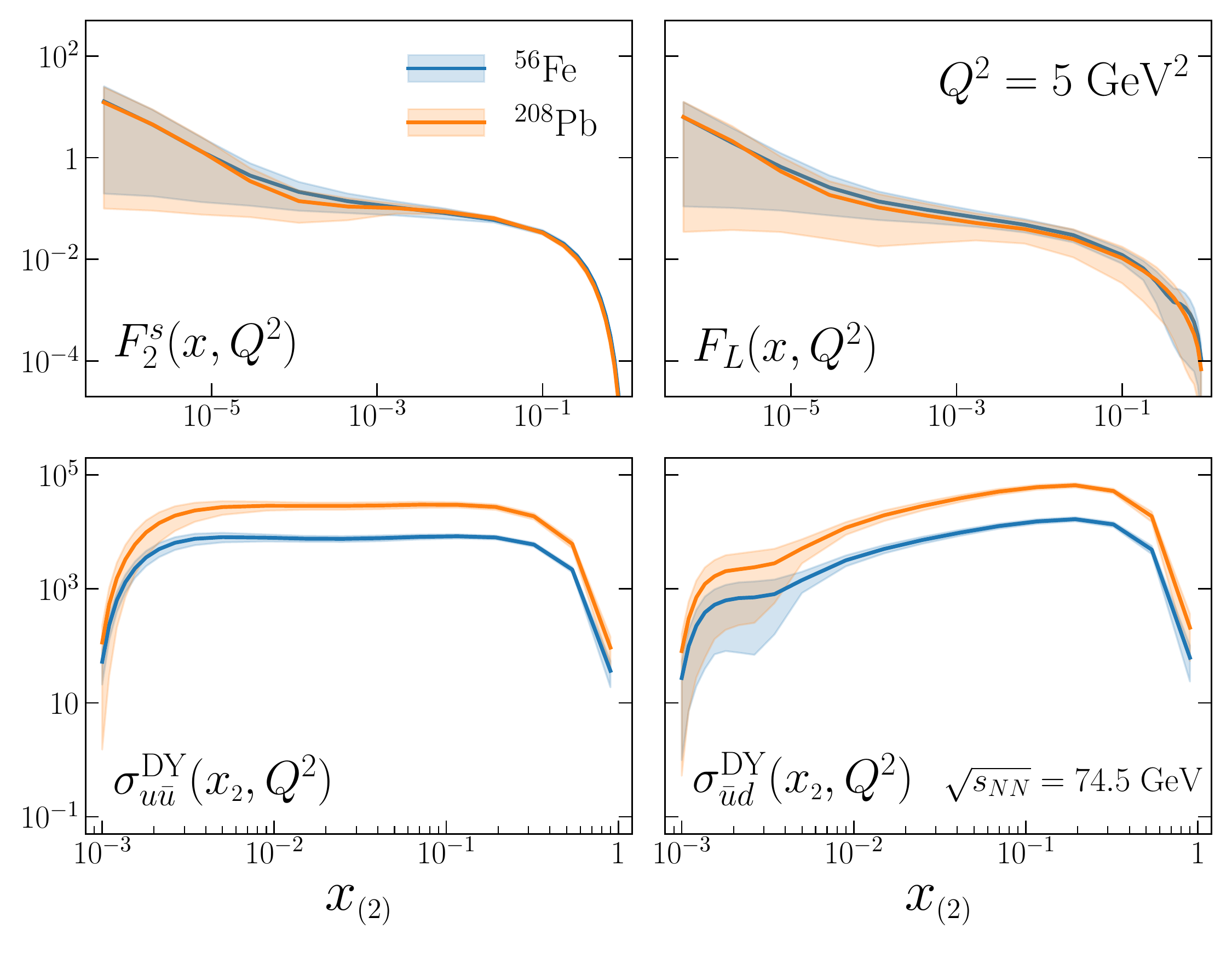}
 \end{center}
\vspace{-0.6cm}
\caption{A representative selection of the positivity observables used
in nNNPDF2.0.
  From top to bottom and from left to right, we show the DIS structure
  functions $F_2^s$ and $F_L$ and the Drell-Yan rapidity distributions
  $\sigma_{u\bar{u}}^{\rm DY}$ and $\sigma_{\bar{u}d}^{\rm DY}$.
  The bands indicate the 90\% confidence level interval.
  \label{fig:positivity}
}
\end{figure}

Here we want to demonstrate that the nNNPDF2.0 determination indeed
satisfies these various positivity constraints.
In Fig.~\ref{fig:positivity} we display a representative selection of
the positivity observables imposed in nNNPDF2.0.
In particular, we show the DIS structure
functions $F_2^s(x,Q^2)$ and $F_L(x,Q^2)$, as well as the Drell-Yan
rapidity distributions $\sigma_{u\bar{u}}^{\rm DY}(y)$ and
$\sigma_{\bar{u}d}^{\rm DY}(y)$, where the bands indicates the 90\%
confidence level uncertainty interval.
  We use a scale of $Q^2=5$ GeV$^2$, which corresponds to the same scale
  in which Eq.~(\ref{eq:positivity}) is imposed.
  Furthermore, we provide the positivity predictions for both iron and
  lead nuclei.
  Note that since the Drell-Yan cross-sections are not normalized by the
  value of $A$, the absolute magnitude of the two nuclei are different.
  Of course, the overall normalization is not relevant for the
  implementation of the positivity constraint.

The selection of positivity observables in Fig.~\ref{fig:positivity} is
representative since it contains one of the quark structure functions
($F_2^s$) constraining a $q^+$ combination, $F_L$ that is sensitive to
the gluon positivity, and a diagonal and off-diagonal DY cross-section
which are relevant for different aspects of quark flavor separation
(confer also the LO expressions in App.~\ref{sec:LOxsecs}).
Here the Drell-Yan cross-sections are represented as a function of
$x_2$, which corresponds to the momentum fraction of the nuclear
projectile obtained using the LO kinematics of
Eq.~(\ref{eq:LOkinematicsDY}).
While the positivity constraint was only implemented for
$x_2\gsim10^{-2}$ with a per-nucleon center-of-mass energy of
$\sqrt{s}=23.5$ GeV, we illustrate instead the positivity for a choice
of kinematics that allow a reach to $x_2\sim10^{-3}$, with a per-nucleon
center-of-mass energy of $\sqrt{s}=74.5$ GeV.
As can be seen from Fig.~\ref{fig:positivity}, the nNNPDF2.0
determination satisfies the positivity of physical cross-sections in the
entire kinematic range.
Here the nPDF uncertainty bands become larger near the kinematic
endpoints ($x=1$ for DIS and $x_2\simeq 10^{-3}$ for Drell-Yan), since
these correspond to regions of the phase space where experimental
constraints are scarce.
Recall that by virtue of DGLAP evolution properties, these results
ensure the cross-sections involving higher momentum transfers, $Q^2 > 5$
GeV$^2$, will also be positive provided one maintains the initial coverage
in $x_2$.
Therefore, we conclude that while we have not explicitly imposed the positivity at
the level of the  nuclear PDFs, physical observables
constructed from nNNPDF2.0  are guaranteed to satisfy the positivity
requirement.

\section{Implications for photon and hadron production in nuclear collisions}
\label{sec:phenomenology}

In this section we discuss some phenomenological
applications of the nNNPDF2.0 determination.
Theoretical predictions for isolated photon
production in proton-lead collisions at the LHC are first
compared with recent 
measurements from the ATLAS collaboration taken at $\sqrt{s}=8.16$ TeV.
We then revisit the potential of the FoCal upgrade to the ALICE
detector in constraining the small-$x$ gluon nuclear PDF using
measurements of direct photon production in the forward region.
Finally, we provide predictions based on nNNPDF2.0 for inclusive hadron
production in proton-nuclear collisions, a process that can constrain both the
quark and gluon nuclear PDFs as well as the corresponding fragmentation
functions in vacuum and in medium.

Several additional applications of our nNNPDF2.0 result
are expected to be of phenomenological interest.
In particular, our nPDFs could be used to
study the constraining power of inclusive and heavy quark
structure function measurements at the recently approved Electron Ion
Collider (EIC)~\cite{Boer:2011fh} and the proposed Large Hadron electron
Collider (LHeC)~\cite{AbelleiraFernandez:2012cc}.
Initial studies based on nNNPDF1.0 and EIC neutral-current structure
function pseudo-data were presented in Ref.~\cite{AbdulKhalek:2019mzd}.
However, updated projections for EIC pseudo-data are now being 
finalized based on more realistic accelerator and detector settings.
We will therefore defer an update to our nNNPDF1.0 
study of EIC pseudo-data to an upcoming Conceptual Design Report 
where the more accurate EIC specifications will be presented 
together with impact studies from various nuclear PDF analysis
groups.

\subsection{Isolated photon production in pA collisions with ATLAS}
\label{sec:photonproduction}

Production of isolated photons in proton-proton collisions is
primarily sensitive to the gluon content of the proton via the QCD Compton
scattering process~\cite{Vogelsang:1995bg,Ichou:2010wc}.
However, several complications associated with
the measurement of photon production 
complicate a clean interpretation in terms of 
hard-scattering cross-sections, such as 
the need for subtracting the fragmentation 
component and the removal of photons coming
from pion decays.
Although early PDF fits used photon production 
in fixed-target scattering to constrain the
gluon data, the data were eventually discarded in favor of
the cleaner and more abundant data on jet production at the
TeVatron~\cite{Gao:2017yyd}.
However, photon production measurements from ATLAS and CMS were
later revisited using NLO QCD theory in Ref.~\cite{d'Enterria:2012yj}
and again at NNLO in Ref.~\cite{Campbell:2018wfu}.
These studies demonstrated the consistency of collider-based 
isolated photon production measurements 
with QCD predictions and with the rest of
the datasets in the global analysis.
Moreover, they help to reduce the uncertainties on the gluon PDF
at $x\simeq 10^{-2}$, a region that is particularly 
relevant for theoretical predictions of Higgs production in gluon fusion.

Photon production is also a highly relevant process in the context of
heavy ion collisions.
Being a QCD-neutral probe, it traverses the quark-gluon plasma without
modifications and thus represents a robust baseline to study the hot and
dense medium properties~\cite{Vitev:2008vk}.
In order to disentangle hot from cold nuclear matter effects, photon
production has been measured in proton-lead collisions at the LHC,
providing a new channel to constrain the nuclear modifications of the
gluon PDF.
Here we focus on the recent ATLAS measurements
from Run II at $\sqrt{s}=8.16$ TeV based on an integrated luminosity of
$\mathcal{L}=165$ nb$^{-1}$~\cite{Aaboud:2019tab}.
This analysis provides the inclusive production rates of isolated prompt
photons in three different rapidity regions as a function of
$E_{T}^{\gamma}$, the photon transverse energy, in the range 20
GeV to 550 GeV.

To compute the corresponding theoretical predictions, 
we use NLO QCD theory with the same settings as
in Ref.~\cite{Campbell:2018wfu} by adopting a modified 
version of {\tt MCFM} v6.8 interfaced to {\tt APPLgrid}.
The settings of the calculation have been adjusted to map the
experimental isolation conditions and thus bypass the need to explicitly
account for the fragmentation component.
We have benchmarked the results of this {\tt MCFM}-based calculation
with the theory predictions presented in Ref.~\cite{Aaboud:2019tab} based on
the {\tt JETPHOX} program~\cite{Catani:2002ny}, finding a reasonable
agreement but also some differences at large rapidities.

\begin{figure}[t]
\begin{center}
  \includegraphics[width=0.90\textwidth]{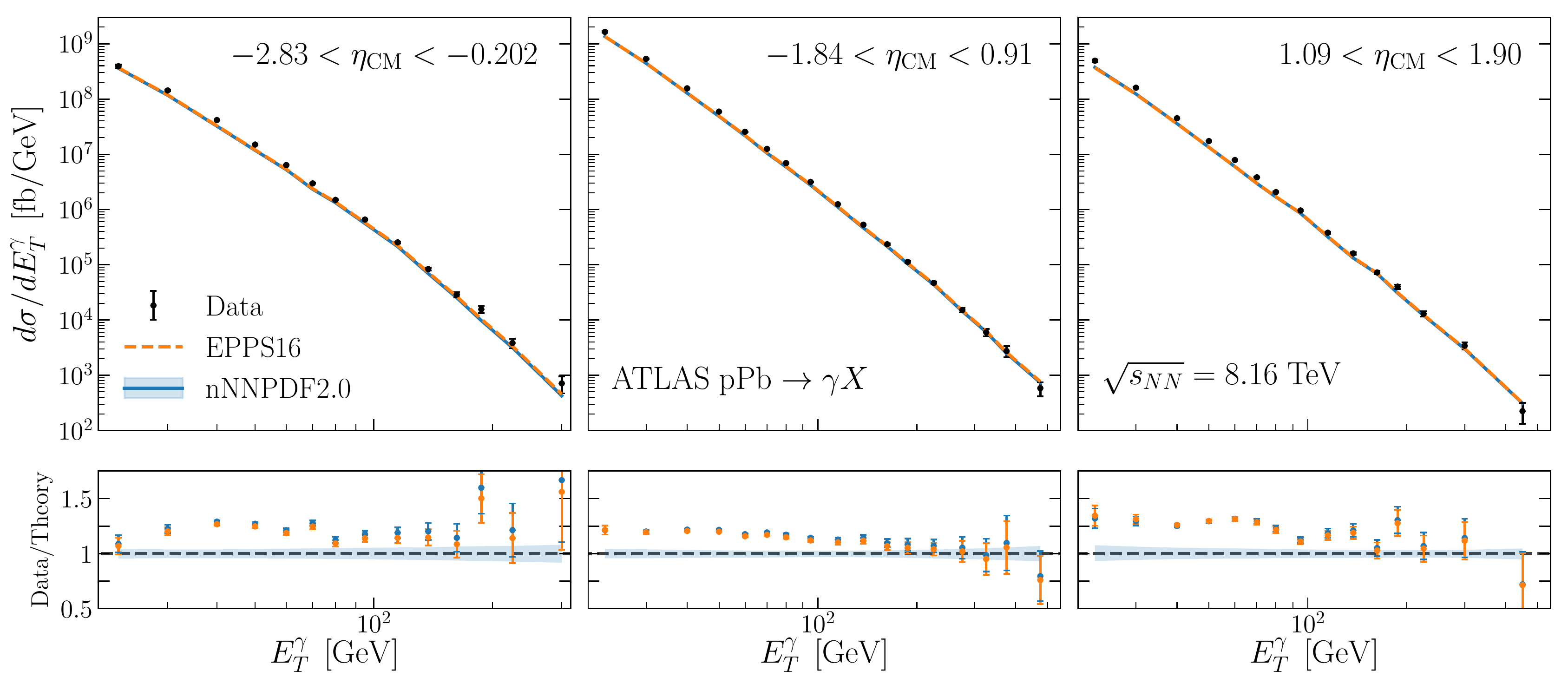}
 \end{center}
\vspace{-0.7cm}
\caption{Comparison between the ATLAS measurements of the photon
  $E_T^\gamma$ distributions in three rapidity bins in the center of
  mass frame with the corresponding  NLO QCD theory calculations based
  on {\tt MCFM} with nNNPDF2.0 and EPPS16 as input.
  For each rapidity bin, the upper panels display the absolute
  distributions and the lower panels the corresponding ratio between the
  theory calculations and the central value of the experimental data.
  \label{fig:ATLASthdat}
}
\end{figure}

In Fig.~\ref{fig:ATLASthdat}, we display the comparison between the ATLAS
measurements of the photon transverse energy ($E_T^\gamma$)
distributions in three rapidity bins with the corresponding NLO QCD
theory calculations based on {\tt MCFM} with nNNPDF2.0 and EPPS16 as
input.
For each rapidity bin, the upper panels display the absolute
distributions and the lower panels the corresponding ratio between the
central value of the experimental data and the theory calculations.
The error bands in the experimental measurements indicate the sum in
quadrature of the statistical and systematic experimental uncertainties, while in
the theory calculations the error bands correspond to the 90\% CL ranges.

From this comparison one finds that the theory calculations appears to
undershoot the experimental data by roughly $25\%$ 
in the three rapidity bins for most of
the $E_T^\gamma$ range, both for nNNPDF2.0 and EPPS16.
This discrepancy cannot be accommodated within the stated experimental
and theoretical uncertainties.
A similar qualitative behavior was reported in the
original ATLAS publication based on the {\tt JETPHOX} predictions.
On the other hand, the NNPDF3.1 proton PDFs are known to describe well the
corresponding isolated photon measurements from proton collisions 
at both $\sqrt{s}=8$ and 13 TeV~\cite{Campbell:2018wfu}.

As expected, the disagreement between the experimental data and the
theory calculations found in Fig.~\ref{fig:ATLASthdat} is
translated into poor $\chi^2$ values.
Using nNNPDF2.0, one obtains that $\chi^2/n_{\rm dat}=9.1, 10.5,$ and
$8.5$ in the forward, central, and backwards rapidity bin.
Similar numbers are obtained in the case of the theory predictions based
on EPPS16.
The situation does not improve by much if the ATLAS photon data is added
to the nNNPDF2.0 global analysis.
In such a case the agreement between the theory calculations and the
experimental data improves somewhat, with $\chi^2/n_{\rm dat}=6.1, 7.5,$
and $5.7$ for the three rapidity bins, but it remains far from
satisfactory.

Until the origin of this disagreement between theory and data is better
understood, it will not be possible to include the ATLAS prompt photon
production measurements in a global nPDF fit.
Alternatively, one could instead consider fitting related observables
that are presented in the same ATLAS publication.
The first of these is the nuclear modifications ratio $R_{p{\rm Pb}}$,
where the absolute $E_{T}^\gamma$ distributions in pPb collisions are
normalized to their pp counterparts, the latter being derived from a
simulation-derived extrapolation from data in proton-proton collisions
at $\sqrt{s}=8$ TeV.
The second is the ratio between different rapidity bins, such as between
forward and backward rapidities, as a function of $E_T^\gamma$.
The advantages of such ratios is that many experimental systematic
uncertainties partially cancel out, thus facilitating the comparison
with theoretical predictions.
On the other hand, these observables might also exhibit a reduced nPDF
sensitivity, in particular for the ratio between different rapidity
bins.
Future studies should shed more light on the usefulness of the prompt
photon measurements to constrain nuclear PDFs within a global analysis.

\subsection{Isolated photon production in pA collisions with FoCal}

Current measurements of direct photon production at the LHC, such as
those discussed above from the ATLAS collaboration~\cite{Aaboud:2019tab}
as well as related measurements from CMS and
ALICE~\cite{Acharya:2020sxs}, are restricted to the central rapidity
region.
The reason is that this is the only region instrumented with
electromagnetic calorimeters and thus suitable to identify photons.
A measurement of isolated photon production in the forward region,
however, is also highly interesting for nPDF studies.
Not only would such measurements provide direct access to the
poorly-known gluon nuclear modifications at small-$x$, but 
it would also allow testing for the possible onset of QCD non-linear
dynamics~\cite{Benic:2016uku}.

With this motivation, a new forward calorimeter extension of the ALICE
detector, dubbed
FoCal~\cite{vanderKolk:2020fqo,ALICECollaboration:2719928}, has been
proposed.
Both the acceptance and instrumentation of this detector have been optimized
to provide access to the nuclear PDFs at low scales and small momentum
fractions via the measurement of isolated photon production at low
transverse momenta and forward rapidities in proton-ion collisions.
The FoCal is proposed for installation during the Long Shutdown 3
(2025-2026) phase of the LHC.

The impact of future FoCal measurements on the small-$x$ nuclear PDFs
was first studied in Ref.~\cite{vanLeeuwen:2019zpz}.
In that analysis, pseudo-data based on the expected kinematical reach
and experimental uncertainties for FoCal was generated and used to
constrain the nNNPDF1.0 determination by means of the Bayesian
reweighting method~\cite{Ball:2010gb,Ball:2011gg}.
It was found that the FoCal measurement would constrain the nuclear
gluon modifications down to $x\simeq 10^{-5}$, leading to an uncertainty
reduction by up to an order of magnitude as compared to the baseline
fit.
These results indicated a comparable or superior constraining power on
the small-$x$ nPDFs when compared to related projections from future
facilities, such as the Electron Ion Collider~\cite{Accardi:2012qut}.

Motivated by the new and improved projections for the FoCal 
pseudodata that have recently became available, 
we revisit their impact on nuclear PDFs
using the present nPDF determination. 
In this case, the nNNPDF2.0 PDFs represent a more realistic 
baseline since they provide a robust quark flavor
separation with a better handle on the gluon.
Moreover, the positivity of physical cross-sections is guaranteed, 
a constraint that helps to reduce the small-$x$ nuclear PDF uncertainties.

\begin{figure}[t]
\begin{center}
\includegraphics[width=0.49\textwidth]{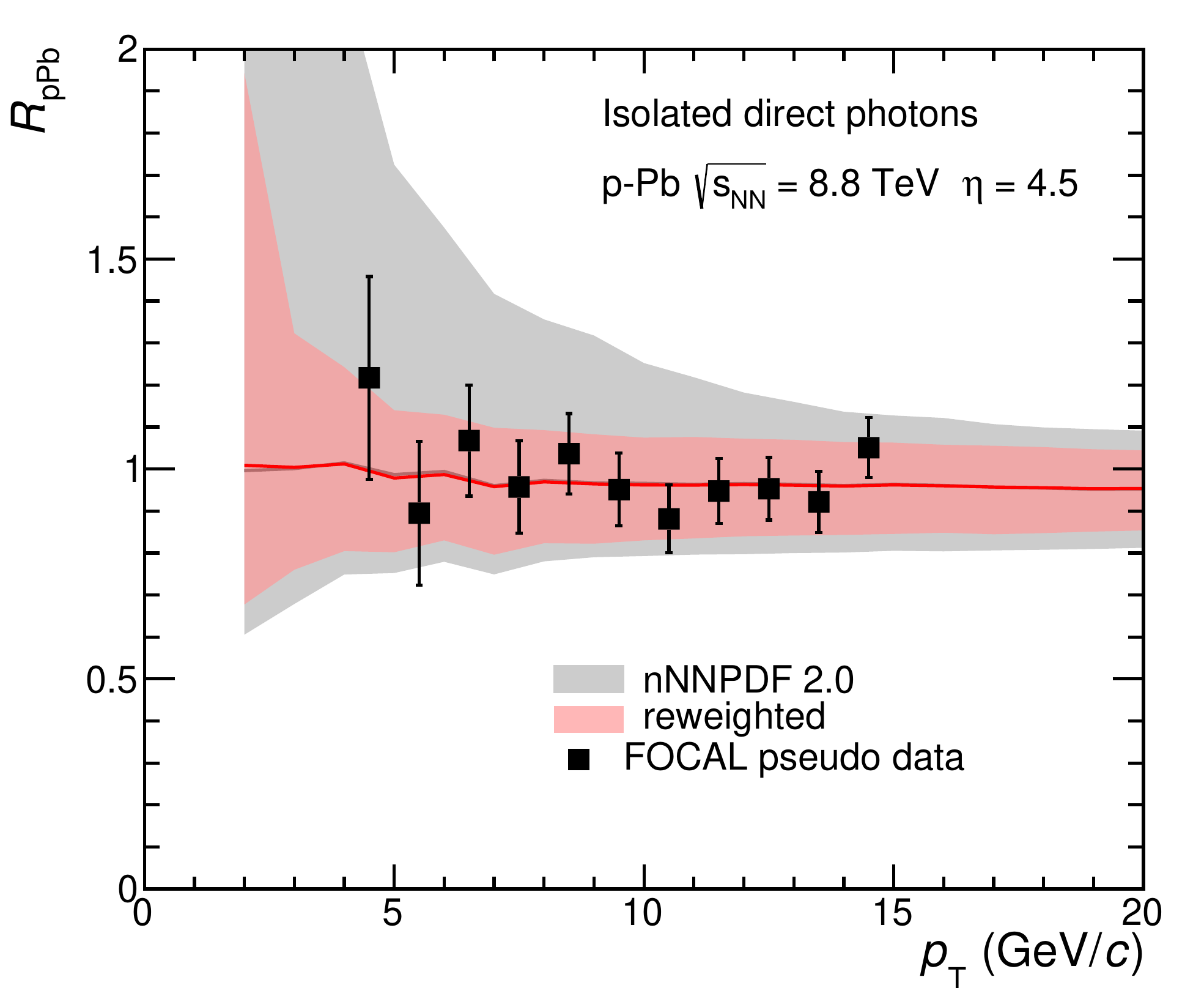}
\includegraphics[width=0.49\textwidth]{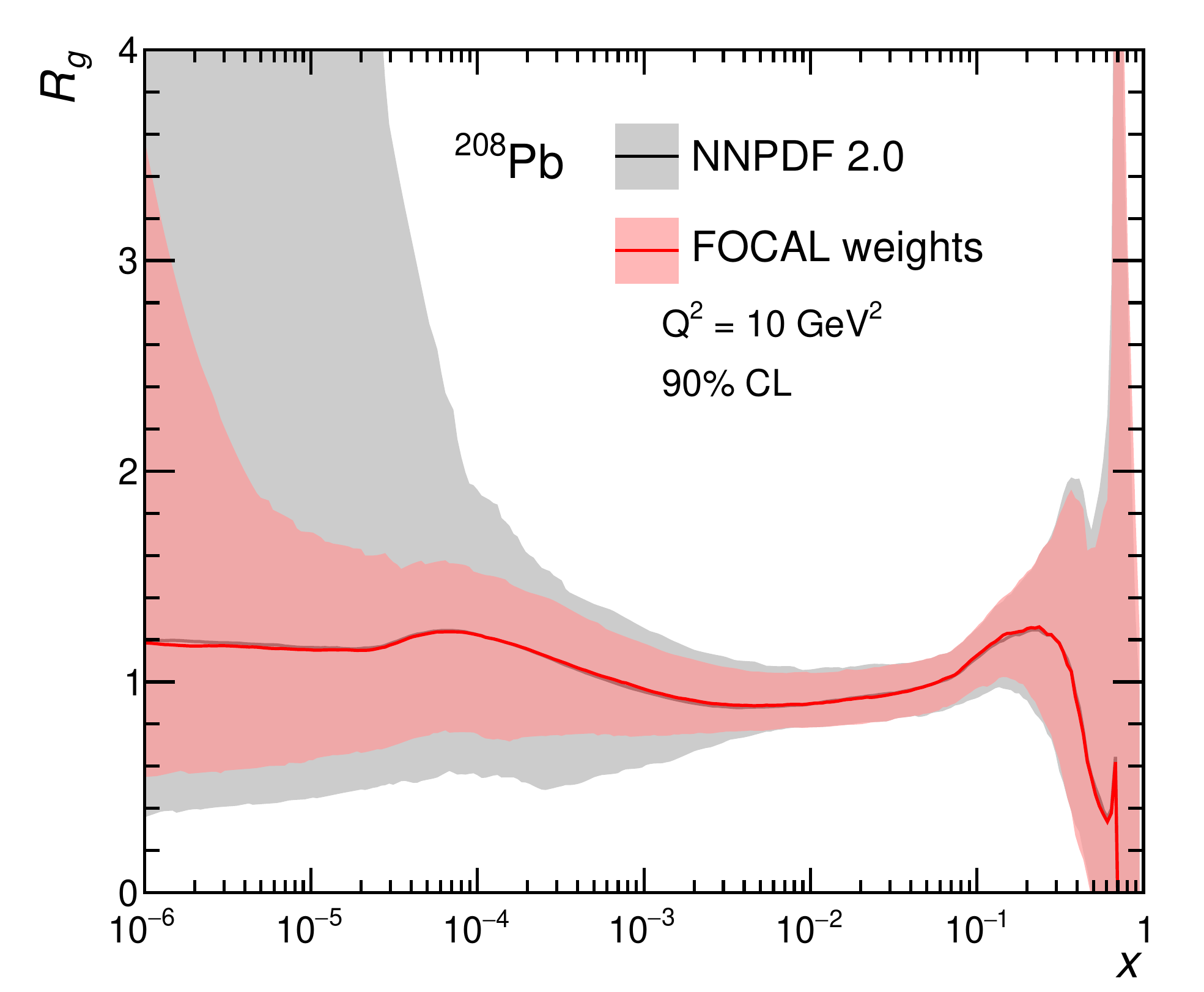}
  \end{center}
\vspace{-0.4cm}
\caption{Left: the nuclear modification factor $R_{\rm pPb}(p_T^\gamma)$
  for direct photon production in pPb collisions at $\sqrt{s}=8.8$ TeV
  for a rapidity of $\eta_\gamma=4.5$ as a function of the photon
  transverse momentum $p_T^\gamma$.
  The theoretical predictions are compared with the FoCal pseudo-data
  for two sets of input nPDFs: the original nNNPDF2.0 set, and the
  variant that has been reweighted with with FoCal projections.
  Here the FoCal pseudo data assumes the central value of the nNNPDF2.0
  prediction.
  Right: the gluon nuclear modification factor $R_g(x,Q)$ for $Q^2=10$
  GeV$^2$ for both the original and the reweighted nNNPDF2.0 fits.
  The nPDF uncertainties correspond in both cases to the 90\% confidence
  level intervals.
  \label{fig:focal}
}
\end{figure}

For this study we have adopted the same settings as
in Ref.~\cite{vanLeeuwen:2019zpz} and computed NLO QCD predictions with a
modified version of {\tt INCNLO} that benefits from improved numerical
stability at forward rapidities~\cite{Helenius:2014qla}.
Theoretical predictions for FoCal cross-sections have been computed with
$N_{\rm rep}=400$ replicas of nNNPDF2.0, which are subsequently used to
account for the impact of the FoCal pseudo-data by means of Bayesian
reweighting.\footnote{We are grateful to Marco van Leeuwen for providing
us with the results presented here.}
Fig.~\ref{fig:focal} displays the nuclear modification factor $R_{\rm
  pPb}(p_T^\gamma)$ for direct photon production in pPb collisions at
  $\sqrt{s}=8.8$ TeV for a rapidity of $\eta_\gamma=4.5$ as a function
  of the photon's transverse momentum $p_T^\gamma$.
  The theoretical predictions based on NLO QCD theory are compared with
  the FoCal pseudo-data for two sets of input nPDFs: the original
  nNNPDF2.0 set, and the variant that has been reweighted with the FoCal
  projections.
  Here the central value of the FoCal pseudo-data has been
  chosen to be the same as that of the nNNPDF2.0 prediction.
  In the right panel of Fig.~\ref{fig:focal} we show the gluon nuclear
  modification factor $R_g(x,Q)$ for $Q^2=10$ GeV$^2$ for both the
  original and the reweighted nNNPDF2.0 fits.
  In all cases, the nPDF uncertainty bands correspond to the 90\%
  confidence level intervals.

From the results of Fig.~\ref{fig:focal}, one finds that the FoCal
measurements would still impact the uncertainties of the nuclear gluon
modifications at small-$x$, especially in the upper limit of the
uncertainty band.
The effective number of replicas in this case is $N_{\rm
eff}=345$.
Note that nNNPDF2.0 exhibits a preference for $R_{\rm pPb}\simeq 1$,
and thus shadowing is not favored in the gluon sector, consistent
with the results reported in Fig.~\ref{R_A}.
On the other hand, nNNPDF2.0 does not contain any dataset with
particular sensitivity to the nuclear gluon modifications, implying that
the projections for the impact of FoCal in the global nPDF analysis
could be somewhat over-optimistic (see also the discussion in
Sect.~\ref{sec:summary}).

\begin{figure}[t]
  \begin{center}
    \includegraphics[width=0.49\textwidth]{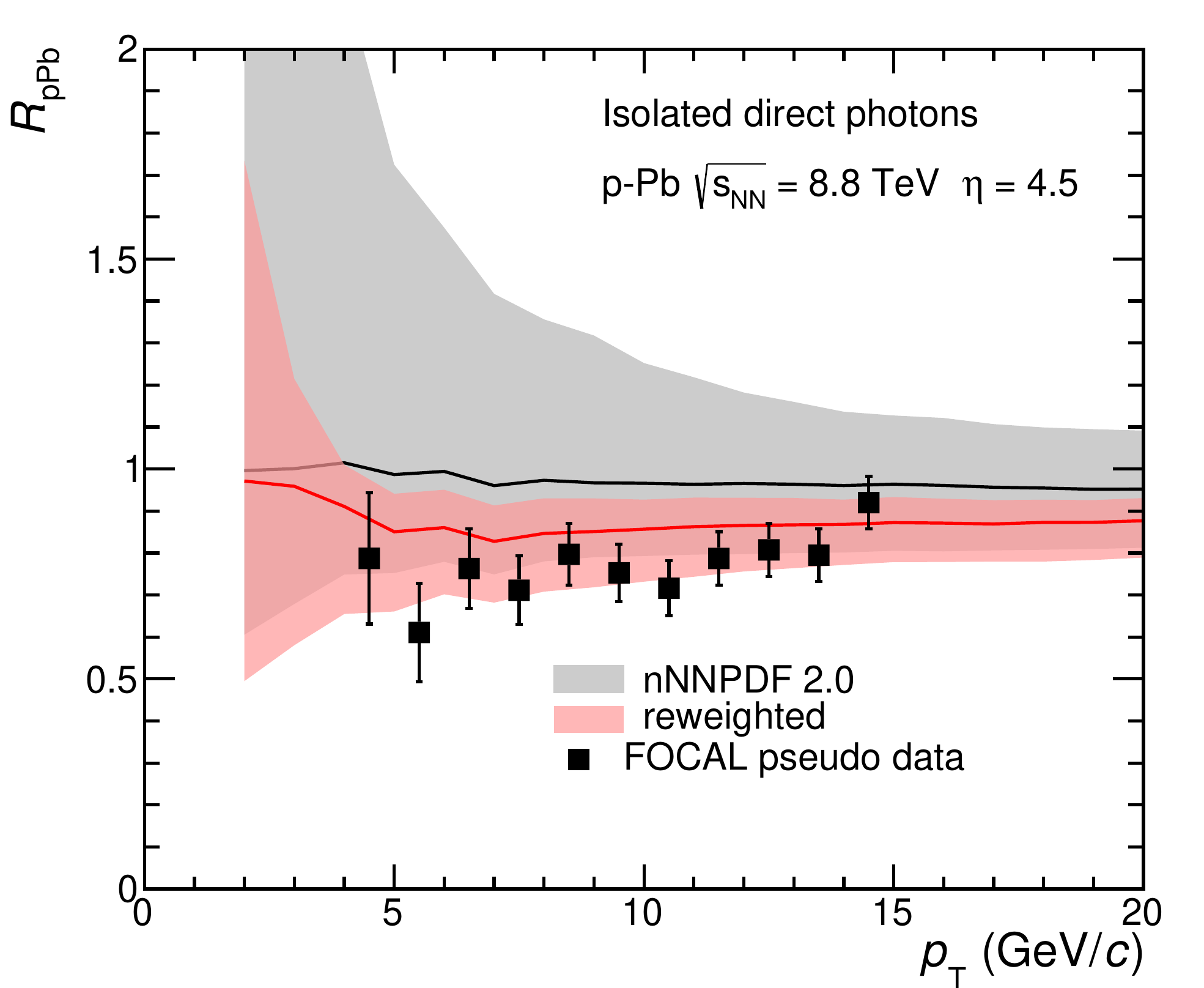}
    \includegraphics[width=0.49\textwidth]{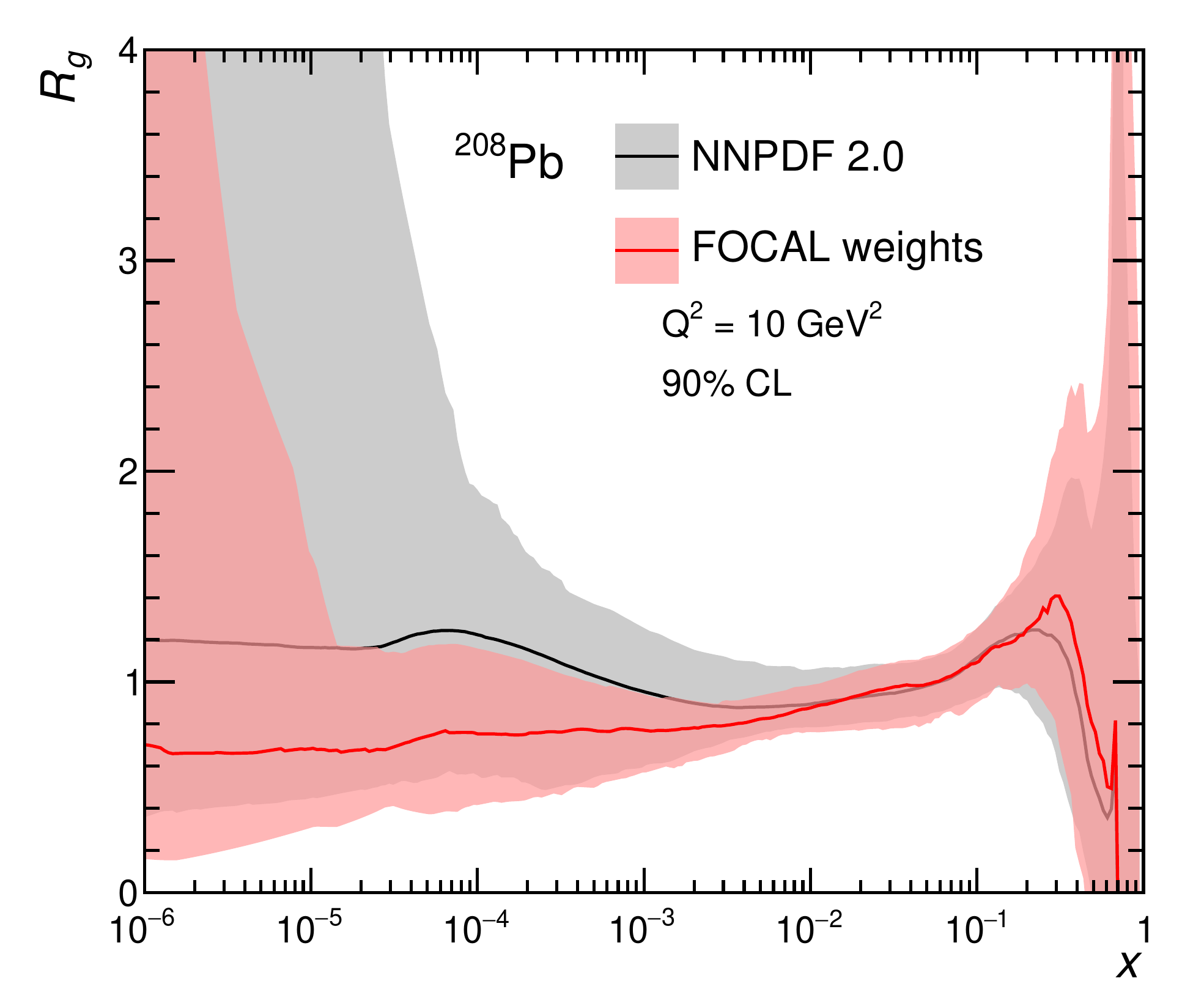}
    \end{center}
\vspace{-0.4cm}
\caption{Same as Fig.~\ref{fig:focal} now for the case where the FoCal
  pseudo data has been generated under the assumption that $R_{\rm pPb}
  \simeq 0.6$ rather than based on the nNNPDF2.0 central value.
  \label{fig:focal2}
}
\end{figure}

Crucially, however, we have assumed in this exercise that the central value of the
FoCal measurement would be unchanged compared to the initial baseline
prediction.
In Fig.~\ref{fig:focal2} we display instead the results of the
reweighting for a scenario in which the FoCal pseudodata 
have a value of $R_{\rm pPb} \simeq 0.6$.
  In this case, the effective number of replicas is much smaller, $N_{\rm
  eff}=117$, indicating that the FoCal data are adding a significant
  amount of new information to the global fit.
  Here the resulting value for the gluon nuclear modification
  ratio at small-$x$ would be $R_g\simeq 0.7$.\footnote{
  Note that the reweighting technique may lead to unreliable 
  uncertainty bands when using data values that fall 
  outside the predictions produced by the prior.}
  Therefore, this analysis indicates that FoCal measurements could be
  sensitive either to the gluon shadowing effects or to possible
  non-linear QCD dynamics.
  To disentangle one from the other, a dedicated analysis of the
  $\chi^2$ and nPDF behavior in the small-$x$ region would
  be required, following the approach developed in Ref.~\cite{Ball:2017otu}. 

\subsection{Inclusive hadron production in pA collisions}

The inclusive production of pions and kaons in hadronic collisions
provides information not only on the initial state (parton
distribution functions) but also on the final-state hadronization
mechanism of partons into hadrons.
The latter is described by the fragmentation functions (FFs), which are
extracted from experimental data by means of a global analysis akin to
that of the PDFs~\cite{deFlorian:2014xna,Sato:2019yez,d'Enterria:2013vba,Bertone:2018ecm,Bertone:2017tyb,Albino:2005me,Hirai:2007cx}.
Likewise, in proton-nuclear collisions the production of identified
hadrons can provide information on the initial state nuclear PDFs
as well the parton-to-hadron hadronization in the
presence of cold nuclear matter effects.

\begin{figure}[t]
\begin{center}
  \includegraphics[width=0.99\textwidth]{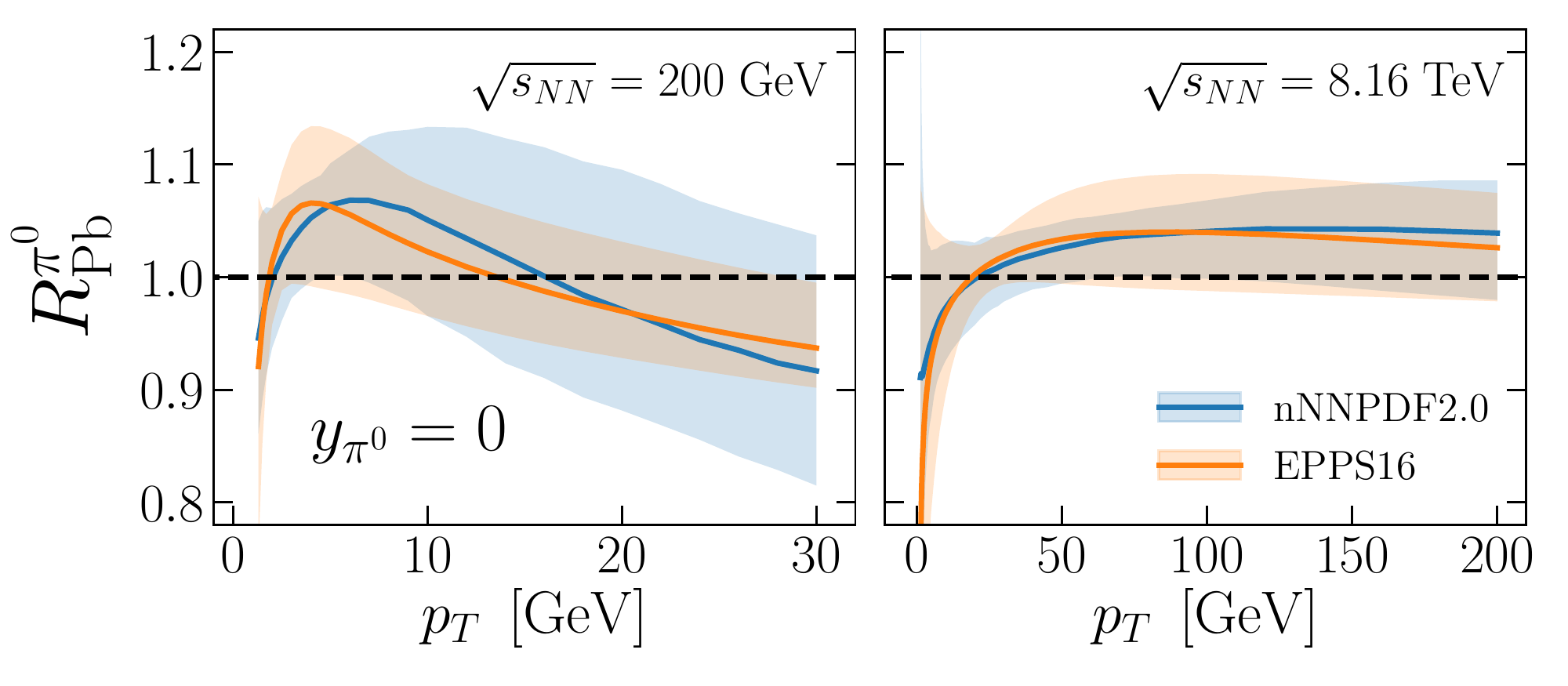}
 \end{center}
\vspace{-0.3cm}
\caption{\small The nuclear modification ratio $R_{\rm Pb}^{\pi^0}$ for
  the production of neutral pions in proton-lead collisions as a
  function of the pion transverse momentum $p_T$.
  We provide theoretical predictions based on NLO QCD and the DSS14
  hadron fragmentation functions both for nNNPDF2.0 and EPPS16, with the
  corresponding nPDF uncertainties in each case.
  Results are provided for the RHIC kinematics (left), corresponding to
  $\sqrt{s}=200$ GeV, and for the LHC kinematics (right), where
  $\sqrt{s}=8.16$ TeV, and in both cases pions are produced centrally,
  $|y_{\pi^0}|=0$.
  \label{fig:plot_R_all}
}
\end{figure}

In Fig.~\ref{fig:plot_R_all}, we display the nuclear modification ratio
  $R_{\rm Pb}^{\pi^0}$ for the production of neutral pions in
  proton-lead collisions as a function of the pion transverse momentum
  $p_T$.
  The theoretical calculations
  are based on NLO QCD and use the DSS14
  hadron fragmentation functions~\cite{deFlorian:2014xna}
  for both the nNNPDF2.0 and EPPS16 predictions.\footnote{We are grateful to Ilkka Helenius for providing us with the results of this calculation.}
  Moreover, the central values and 90\% CL uncertainties 
  are provided for RHIC kinematics, corresponding to
  $\sqrt{s}=200$ GeV, and for LHC kinematics, where $\sqrt{s}=8.16$
  TeV.
  In both cases, the pions are assumed to be measured at central
  rapidities, $y_{\pi^0}= 0$. 
  See Refs.~\cite{d'Enterria:2013vba,Albacete:2017qng} for additional details
  regarding the theoretical calculation of inclusive pion production in
  hadronic collisions.

From Fig.~\ref{fig:plot_R_all} we can see that the nNNPDF2.0 prediction
for $R_{\rm Pb}^{\pi^0}$ is consistent with unity within uncertainties
for all values of $p_T$ both at RHIC and LHC kinematics.
At RHIC kinematics, we find that the ratio is less than one at the
smallest $p_T$ values, becomes $R > 1$ between $p_T=3$ and 17 GeV, and
then goes back to $R<1$.
Since inclusive hadron production is dominated by quark-gluon
scattering, in particular the scattering of valence quarks for
neutral pion production, this behavior is consistent with the results shown in
Fig.~\ref{fig:R_Pb}. 
From low to high $p_T$, one moves from the
shadowing region to the anti-shadowing enhancement, and ends
in the region sensitive to EMC suppression.
A similar explanation can be made for the trends in $R_{\rm Pb}^{\pi^0}$
at the LHC kinematics. 
However, here the ratio
$p_T/\sqrt{s}$ does not become large enough to reach the EMC region, and
thus the ratio remains larger than one for most of the $p_T$ range as a result 
of anti-shadowing effects.
Lastly, the EPPS16 predictions agree with the nNNPDF2.0
result well within uncertainties, reflecting the underlying consistency 
at the nPDF level.

Overall, the results of Fig.~\ref{fig:plot_R_all} confirm that inclusive hadron
production in proton-nucleus collisions can provide 
a handle on the nuclear PDF modifications at medium and large-$x$, 
although an optimal interpretation of the experimental data can only be 
achieved by the simultaneous determination of the nPDFs together 
with the hadron fragmentation functions.

\section{Summary and outlook}
\label{sec:summary}

In this work we have presented a model-independent global determination
of nuclear parton distributions by incorporating the
constraints from nuclear DIS
structure functions and gauge boson production in
proton-lead collisions.
We have demonstrated that a satisfactory description of all the fitted
data sets can be achieved, highlighting the reliability of the QCD
factorization paradigm in the heavy nuclear sector.
Our results demonstrate significant nuclear effects
among the quark flavors in
nuclei, in particular a shadowing of the up and down
quark distributions in heavy nuclei such as lead.
Nuclear modifications are 
found also in the strangeness of 
heavier nuclei, displaying a suppression with respect to the free proton
across a large region of $x$.
In addition, we have shown that upon releasing the momentum
and valence sum rule constraints, 
the data prefer integral values that agree with QCD expectations
for all values of $A$.

We have also explored some phenomenological
implications of the nNNPDF2.0 determination.
We first compared nNNPDF2.0 theoretical predictions with ATLAS
measurements of isolated photon production at the LHC,
an important hard probe in proton-lead collisions.
We then studied the impact on the small-$x$ nuclear gluon PDF from
the forward isolated photons production at the FoCal upgrade of the
ALICE detector.
Lastly, we analyzed our theory predictions for
neutral pion production in proton-lead collisions at RHIC and LHC
center-of-mass energies.
Apart from these applications, the nNNPDF2.0 PDF set can 
be used as input to theoretical predictions for a range of other hard processes
in pPb and PbPb collisions, in particular 
for heavy ion collisions involving lighter nuclei,
in comparisons with non-perturbative nuclear models, and with
QCD calculations at small-$x$ involving dense nuclear matter.
We also 
expect the nNNPDF2.0 release to be used in future proton global PDF fits to
estimate the theory uncertainties associated with neutrino scattering
data~\cite{Ball:2018twp}, and also in high-energy
astroparticle physics processes that involve hard scattering on
nuclei~\cite{Garcia:2020jwr,Bertone:2018dse}.
 
While the input data set used in this work allowed for a 
state-of-the-art determination 
of the nuclear quarks and anti-quarks, it 
only provided loose constraints on
the nuclear gluon PDF, especially for heavier
nuclei where uncertainties are relatively large.
To bypass this limitation, the next step in the nNNPDF family of nuclear
PDF fits will be to include additional datasets that provide direct
information on the nuclear gluon modifications.
In addition to the isolated photon production measurements discussed in
Sect.~\ref{sec:photonproduction}, perhaps the most attractive candidate
in this respect is dijet production in pPb collisions.
Measurements of dijet production from Run I in pp collisions have been
recently analyzed in the framework of NNLO QCD theory
in Ref.~\cite{AbdulKhalek:2020jut}, demonstrating a good compatibility with
the global dataset and a marked constraining power on the large-$x$
gluon.
In the corresponding pPb case, an EPPS16-based profiling
analysis~\cite{Eskola:2019dui} of CMS dijet data at $\sqrt{s}=5.02$
TeV~\cite{Sirunyan:2018qel} revealed a significant pull of this
measurement on the nuclear gluon modifications.

Another process that is known to provide important information on the
nuclear gluon PDF is charmed meson production, in particular from
the LHCb measurements in the forward region~\cite{Aaij:2017gcy}.
This process offers unique sensitivity to the small-$x$ (n)PDFs down to
$x \simeq 10^{-6}$, as was demonstrated by proton~\cite{Zenaiev:2015rfa,Bertone:2018dse,Gauld:2016kpd} and nuclear studies~\cite{Kusina:2017gkz,Eskola:2019bgf}.
Fully exploiting the constraints provided by these measurements
requires, as for the rest of hard probes in nuclear collisions, a
consistent theoretical and methodological treatment of charm production
in both proton and nuclear global QCD analyses.

On a longer timescale, one might aim to achieve a
determination of the proton and nuclear PDFs simultaneously 
from a universal analysis, thus bypassing the need to 
include proton information by means of the proton boundary condition penalty. 
In the same spirit of the QCD analyses of proton PDFs and fragmentation functions
presented in Refs.~\cite{Sato:2019yez,Ethier:2017zbq}, such an integrated fit
of proton and nuclear PDFs would ensure the ultimate theoretical and
methodological consistency of the determination of the nuclear
modifications of the free-nucleon quark and gluon structure.
\\
\hrule

\noindent
\\
\\
The nNNPDF2.0 determination is available in the {\tt LHAPDF6}
library~\cite{Buckley:2014ana} for all relevant nuclei from $A=1$ to
$A=208$.
The nNNPDF2.0 sets are available both for the nPDFs of bound protons,
$f^{(p/A)}(x,Q^2)$, and those of bound nucleons, $f^{(N/A)}(x,Q^2)$,
following the conventions in Sect.~\ref{sec:defs}.
Each of these sets is composed by $N_{\rm rep}=250$ correlated replicas,
see Sect.~5 of~\cite{AbdulKhalek:2019mzd} for their usage prescriptions.
The naming convention used for the sets is the following:\\

\begin{center}
  \renewcommand{\arraystretch}{1.15}
\begin{tabular}{ll}
   $\qquad\qquad f^{(N/A)}(x,Q^2)$ &  $\qquad\qquad f^{(p/A)}(x,Q^2)$ \\
{\tt nNNPDF20\_nlo\_as\_0118\_N1}     &  {\tt nNNPDF20\_nlo\_as\_0118\_N1} \\
{\tt nNNPDF20\_nlo\_as\_0118\_D2}     &  {\tt nNNPDF20\_nlo\_as\_0118\_p\_A2\_Z1}\\
{\tt nNNPDF20\_nlo\_as\_0118\_He4}    &  {\tt nNNPDF20\_nlo\_as\_0118\_p\_A4\_Z2}\\
{\tt nNNPDF20\_nlo\_as\_0118\_Li6}    &  {\tt nNNPDF20\_nlo\_as\_0118\_p\_A6\_Z3}\\
{\tt nNNPDF20\_nlo\_as\_0118\_Be9}    &  {\tt nNNPDF20\_nlo\_as\_0118\_p\_A9\_Z4}\\
{\tt nNNPDF20\_nlo\_as\_0118\_C12}    &  {\tt nNNPDF20\_nlo\_as\_0118\_p\_A12\_Z6}\\
{\tt nNNPDF20\_nlo\_as\_0118\_N14}    &  {\tt nNNPDF20\_nlo\_as\_0118\_p\_A14\_Z7}\\
{\tt nNNPDF20\_nlo\_as\_0118\_Al27}   &  {\tt nNNPDF20\_nlo\_as\_0118\_p\_A27\_Z13}\\
{\tt nNNPDF20\_nlo\_as\_0118\_Ca40}   &  {\tt nNNPDF20\_nlo\_as\_0118\_p\_A40\_Z20}\\
{\tt nNNPDF20\_nlo\_as\_0118\_Fe56}   &  {\tt nNNPDF20\_nlo\_as\_0118\_p\_A56\_Z26}\\
{\tt nNNPDF20\_nlo\_as\_0118\_Cu64}   &  {\tt nNNPDF20\_nlo\_as\_0118\_p\_A64\_Z29}\\
{\tt nNNPDF20\_nlo\_as\_0118\_Ag108}  &  {\tt nNNPDF20\_nlo\_as\_0118\_p\_A108\_Z47}\\
{\tt nNNPDF20\_nlo\_as\_0118\_Sn119}  &  {\tt nNNPDF20\_nlo\_as\_0118\_p\_A119\_Z50}\\
{\tt nNNPDF20\_nlo\_as\_0118\_Xe131}  &  {\tt nNNPDF20\_nlo\_as\_0118\_p\_A131\_Z54}\\
{\tt nNNPDF20\_nlo\_as\_0118\_Au197}  &  {\tt nNNPDF20\_nlo\_as\_0118\_p\_A197\_Z79}\\
{\tt nNNPDF20\_nlo\_as\_0118\_Pb208}  &  {\tt nNNPDF20\_nlo\_as\_0118\_p\_A208\_Z82}\\
\end{tabular}
\end{center}

\noindent
Additional variants of the nNNPDF2.0 NLO fit present in this work, such as the fits
without the momentum and valence sum rules and a $N_{\rm rep}=1000$ replica set
for lead, are available on the NNPDF collaboration website:
\begin{center}
\url{http://nnpdf.mi.infn.it/for-users/nuclear-pdf-sets/}
\end{center}  

\subsection*{Acknowledgments}

We are grateful to our colleagues of the NNPDF collaboration for support
and stimulating discussions, and in particular to Emanuele R. Nocera for his
continuous assistance during this work.
We thank Kari Eskola, Petja Paakkinen, and Hannu Paukkunen for discussions about
the EPPS16 analysis.
We thank Marco van Leeuwen for providing the predictions for FoCal and
Ilkka Helenius for those of pion production in heavy ion collisions.
R.~A.~K., J.~E., and J.~R. are
supported by the Netherlands Organization for Scientific
Research (NWO).
 We acknowledge grants of computer capacity from the Finnish Grid and
   Cloud Infrastructure (persistent identifier
   urn:nbn:fi:research-infras-2016072533).

\appendix
\section{PDF sensitivity of input cross-sections}
\label{sec:LOxsecs}

In this appendix we indicate the PDF sensitivity of all processes used
as input in the nNNPDF2.0 determination by providing their explicit cross-section expressions at leading order (LO) in perturbative QCD. 
We provide these expressions for a general nuclear target with $Z$ 
protons and $(A-Z)$ neutrons in terms of bound proton distributions 
of the physical and evolution bases, and in terms of the average 
bound nucleon PDFs. 
In what follows, we further simplify the conventions adopted in Sect.~\ref{sec:defs} to
\begin{align}
  f &\equiv f^{(p/A)}\,,\\
  f_{A} &\equiv f^{(N/A)} \,.
\end{align}

\paragraph{1. Lepton-nucleus scattering.}
The double differential cross-section for the DIS of a charged lepton
off a nucleus with mass number $A$ is given by
\begin{equation}
  \frac{d^2\sigma^{i,l^\pm}}{dxdQ^2}(x,Q^2,A)=\frac{2\pi\alpha^2}{xQ^4}\eta^i\left[Y_{+}F_2^i(x,Q^2,A) \mp Y_{-}xF_3^i(x,Q^2,A)-y^2F_L^i(x,Q^2,A)\right]\, ,
\end{equation}
with $i = \rm NC,\,CC$, $\eta^{\rm NC} = 1$, $\eta^{\rm CC} =
(1\pm\lambda)^2\eta_W$, where $\eta_W$ denotes the 
squared ratio of the W-boson couplings and propagator
with respect to those of the photon, and $\lambda$ 
is the helicity of the incoming lepton.
In the case of NC DIS we restrict ourselves to photon-mediated
processes, for which $F^{\rm NC}_3 = 0$. 
The usual DIS kinematic variables are defined as
$Y_{\pm} = 1 \pm (1-y)^2$ and
\begin{equation}
  x = \frac{Q^2}{2P\cdot q},\quad Q^2 = -q^2, \quad y=\frac{q\cdot P}{k \cdot P} \,.
\end{equation}

\paragraph{A1. Neutral-current DIS.}
For NC DIS, the underlying process is lepton-nucleus scattering
mediated by a virtual photon exchange, 
$l^{\pm}+A\xrightarrow[]{\gamma}l^{\pm}+X$.
Since the available data is at $Q^2<<M_Z$, 
the contributions from $Z$ boson exchange can be neglected.
The double-differential cross-section for the
scattering of a charged lepton off a nucleus with atomic and mass numbers $Z$ and $A$
is proportional at leading order to the $F_2$ structure function which can be expressed as
\begin{align}
  \begin{split}
    \label{eq:F2NC}
    F_{2}^{\rm NC} & = \frac{x}{9A}\bigg[(4A-3Z) d^{+} + A s^+ + (A+3Z) u^+ \bigg]=
    \frac{x}{9}\bigg[4u^+_A + d^+_A + s^+_A\bigg] \\ &=  \frac{x}{18A}\bigg[ 4 A \Sigma + A T_{8} + 3(2Z- A) T_{3} \bigg] \, ,
  \end{split}
\end{align}
Here we define the structure function per nucleon, rather than the structure
function of the nucleus as a whole.
We can see from the above expression that for an isoscalar nucleus $A=2Z$ the contribution
proportional to $T_3$ vanishes and the nuclear structure functions
depend only on $\Sigma$ and $T_8$ via the $\Sigma+T_8$ combination.
At LO only Eq.~(\ref{eq:F2NC}) is relevant for the description of neutral-current
DIS since the longitudinal structure function vanishes due to the Callan-Gross relation,
 $F_L^{\rm NC} = F_2^{\rm NC} - 2xF_1^{\rm NC} = 0$.

\paragraph{B1. Charged-current DIS.}
For CC DIS, the underlying process is neutrino scattering off of a nucleus via the
exchange of a W boson, $\nu+
A\xrightarrow[]{W^{-}}l^{-}+X$ and $\bar{\nu}+ A\xrightarrow[]{W^{+}}+l^{+}+X$.
Due to the fact that W$^+$ and W$^-$ bosons couple to different quark
flavors, the difference between neutrino and anti-neutrino structure
functions provides a handle on quark flavor separation.
Here we provide the expressions for $\nu(\bar{nu})A$ scattering, the expressions
for the conjugate process involving the CC scattering of charged leptons
is the same.
The expressions for the inclusive
CC structure functions via $W^{-}$-boson exchange, $\bar{\nu} +A\xrightarrow[]{W^{-}}+l^{+}X$, 
are the following:

  \begin{align}
    \begin{split}
  F_{2}^{\bar{\nu} A\xrightarrow[]{W^{-}}l^{+}X}
  =  \frac{2x}{A} \bigg[&
  |V_{ud}|^2  \bigg(
    Z (u + \bar{d})
    + (A - Z) (d + \bar{u})
    \bigg)+
  |V_{us}|^2 \bigg(
    Z u 
    + (A-Z) d
    + A \bar{s}
    \bigg)\bigg]\\
  = 2x \bigg[&
  |V_{ud}|^2  \bigg(
    u_A + \bar{d}_A
    \bigg)+
  |V_{us}|^2 \bigg(
    u_A
    + \bar{s_A}
    \bigg)
  \bigg]\\
  = \frac{x}{6A}\bigg[&
  |V_{ud}|^2 \bigg(
    2 A( 2\Sigma+ T_{8})
    +6(2Z-A) V_{3}
    \bigg)\\
 +&|V_{us}|^2 \bigg(
    3(2Z-A) T_{3} 
    + 3(2Z-A) V_{3}
    + A (4\Sigma - T_{8} + 3  V_{8})
    \bigg)\bigg]
    \end{split}
\end{align}

\begin{align}
\begin{split}
F_{3}^{\bar{\nu} A\xrightarrow[]{W^{-}}l^{+}X}
= \frac{2}{A} \bigg[&
|V_{ud}|^2\bigg(
Z(u-\bar{d})
+(A-Z)(d-\bar{u})
\bigg)
+|V_{us}|^2\bigg(
Z u
+ (A-Z) d
- A \bar{s}
\bigg)\bigg] \\
= 2 \bigg[&
|V_{ud}|^2  \bigg(
  u_A - \bar{d}_A
  \bigg)+
|V_{us}|^2 \bigg(
  u_A
  - \bar{s_A}
  \bigg)
\bigg]\\
= \frac{-1}{6A} \bigg[&
|V_{ud}|^2\bigg(
  6 (A-2Z) T_{3}
  - 2 A(2V + V_{8})
  \bigg)\\
+&|V_{us}|^2\bigg(
  + 3 (A-2Z) T_{3}
  + 3 (A-2Z) V_{3}
  + A (-3T_{8} -4V +  V_{8})
  \bigg)\bigg]
\end{split}
\end{align}

Here we consider only the contribution from up-, down-, and strange-initiated
processes for simplicity. 
The generalization to heavy-quark initiated processes is straightforward.
An interesting observation from these expressions is that 
$F_2$ and $F_3$ provide complementary information on quark flavor separation.

The corresponding expressions for the inclusive charged-current structure functions
in the case of neutrino scattering via $W^{+}$-boson exchange, $\nu+
A\xrightarrow[]{W^{+}}+l^{-}X$, are given by
\begin{align}
  \begin{split}
F_{2}^{\nu
A\xrightarrow[]{W^{+}}l^{-}X}
  = \frac{2x}{A}\bigg[
    &|V_{ud}|^2\bigg(
    Z(\bar{u} + d)
    + (A-Z)(u + \bar{d})
    \bigg)
    +|V_{us}|^2\bigg(
    Z \bar{u}
    + (A-Z) \bar{d}
    + A s
    \bigg)
    \bigg]\\
    = 2x\bigg[
      &|V_{ud}|^2\bigg(
      \bar{u}_A + d_A
      \bigg)
      +|V_{us}|^2\bigg(\bar{u}_A
      + s_A
      \bigg)
      \bigg]\\
  = \frac{x}{6A}\bigg[
    &|V_{ud}|^2\bigg(
      2 A(2 \Sigma + T_{8})
      + 6 (A-2Z) V_{3}
      \bigg)\\
    +&|V_{us}|^2\bigg(
      A (4\Sigma-T_{8}-3V_{8})
      +3(2Z- A) T_{3}
      + 3( A-2Z) V_{3}
      \bigg)
  \bigg]
\end{split}
\end{align}
\begin{align}
  \begin{split}
  F_{3}^{\nu
  A\xrightarrow[]{W^{+}}l^{-}X}
  = \frac{2}{A} \bigg[
  &|V_{ud}|^2\bigg(
  Z (d-\bar{u})
  +(A-Z)( u-\bar{d})
  \bigg)
  +|V_{us}|^2\bigg(
  - Z \bar{u}
  -(A-Z)\bar{d}
  + A s
  \bigg)\bigg]\\
  = 2\bigg[
    &|V_{ud}|^2\bigg(
    -\bar{u}_A + d_A
    \bigg)
    +|V_{us}|^2\bigg(-\bar{u}_A
    + s_A
    \bigg)
    \bigg]\\
  = \frac{1}{6A}\bigg[
    &|V_{ud}|^2\bigg(
      2A(2V+V_{8})
      +6(A - 2Z) T_{3}
      \bigg)\\
      +&|V_{us}|^2\bigg(
      A(- 3 T_{8} + 4V - V_{8})
      + 3(A-2Z) T_{3}
      + 3(2Z-A) V_{3}
      \bigg)
  \bigg]
\end{split}
\end{align}

We now turn to the exclusive CC charm production structure functions, required
for the description of the NuTeV cross-sections.
Following the FONLL treatment of heavy quark structure functions,
the charm contribution to the inclusive structure function is defined by all terms
where the charm quark couples to the W boson.
Note that this definition is somewhat different from the experimental definition,
where charm structure functions are identified by requesting the presence
of charm in the final state, while the theory definition includes charm-initiated
contributions as well.
The charm production structure functions in the case of antineutrino- and
neutrino-initiated scattering reads for $W^-$ as
\begin{align}
  \begin{split}
F_{2}^{\bar{\nu} A\xrightarrow[]{W^{-}}l^{+}cX} = \frac{2x}{A}\bigg[
&|V_{cd}|^2 \bigg(
Z\bar{d}
+(A-Z) \bar{u}
\bigg)
+|V_{cs}|^2 
 A \bar{s}
\bigg]\\
=2x\bigg[&|V_{cd}|^2
\bar{d}_A
+|V_{cs}|^2
\bar{s}_A
\bigg]\\
=\frac{x}{6A}\bigg[&|V_{cd}|^2 \bigg(
A (2\Sigma + T_{8} - 2 V - V_{8})
+ 3 (A-2Z) T_{3}
+ 3(2Z- A) V_{3}
\bigg)\\
+&|V_{cs}|^2 2A\bigg(
\Sigma
- T_{8}
- V
+ V_{8}
\bigg)\bigg]
  \end{split}
\end{align}
\begin{align}
  \begin{split}
F_{3}^{\bar{\nu} A\xrightarrow[]{W^{-}}l^{+}cX} = \frac{2}{A}\bigg[&|V_{cd}|^2 \bigg(
  - Z \bar{d}
  - (A- Z) \bar{u}
  \bigg)
  - |V_{cs}|^2A
  \bar{s}
  \bigg]\\
  = - 2\bigg[&|V_{cd}|^2
  \bar{d}_A
  + |V_{cs}|^2
   \bar{s}_A
  \bigg]\\
  =\frac{-1}{6A}\bigg[&|V_{cd}|^2 \bigg(
    3(A-2Z)T_{3}
    +3(2Z-A)V_{3}
    +A( 2\Sigma+ A T_{8}- 2 A V- A V_{8})
    \bigg)\\
    + &|V_{cs}|^2 2A\bigg(
    + \Sigma
    - T_{8}
    - V
    + V_{8}
    \bigg)\bigg]    
  \end{split}
\end{align}
and for $W^+$:
\begin{align}
  \begin{split}
F_{2}^{\nu A\xrightarrow[]{W^{+}}l^{-}cX} =
\frac{2x}{A}\bigg[&|V_{cd}|^2 \bigg(
Z d
+ (A-Z) u
\bigg)
+ |V_{cs}|^2A
s
\bigg]\\
= 2x\bigg[&|V_{cd}|^2
d_A
+ |V_{cs}|^2
s_A
\bigg]\\
=
\frac{x}{6A}\bigg[&|V_{cd}|^2 \bigg(
A (2\Sigma+  T_{8} + 2 V + V_{8})
+ 3(A-2Z) T_{3}
+ 3 (A-2Z) V_{3}
\bigg)\\
+ &|V_{cs}|^22A\bigg(
+ \Sigma
- T_{8}
+ V
- V_{8} 
\bigg)\bigg]
  \end{split}
\end{align}
\begin{align}
  \begin{split}
F_{3}^{\nu A\xrightarrow[]{W^{+}}l^{-}cX} =\frac{- 2}{A}\bigg[&|V_{cd}|^2 \bigg(
  + Z d
  + (A-Z) u
  \bigg)
  + |V_{cs}|^2A
  s
  \bigg]\\
  =
  - 2\bigg[&|V_{cd}|^2
   d_A
  + |V_{cs}|^2
  s_A
  \bigg]\\
  =\frac{-1}{6A}\bigg[&|V_{cd}|^2 \bigg(
    A( 2\Sigma+ T_{8}+ 2 V+ V_{8})
    + 3(A-2Z) T_{3}
    + 3 (A-2Z) V_{3}
    \bigg)\\
    + &|V_{cs}|^22A\bigg(
    + \Sigma
    - T_{8}
    + V
    - V_{8}
    \bigg)\bigg]
  \end{split}
\end{align}
%

\paragraph{2. Weak boson production.}
Here we provide the corresponding leading order expressions
for weak gauge boson production in proton-nucleus collisions.
Experimental measurements for NC observables are provided in terms 
of the invariant mass $M$ and rapidity of the dilepton final state, while 
CC measurements are binned in terms of the charged lepton rapidity.
Here we show the LO expressions without accounting for the gauge boson decay,
so in the case of W$^\pm$ production the connection with the experimental data is 
somewhat less direct than for $Z$ boson production.

\paragraph{A2. Neutral current DY}
We start by discussing NC Drell-Yan (DY), namely
$Z$-boson production in proton-nucleus collisions, $p+A \xrightarrow[]{Z} l^+l^-$.
The leading order expressions are given by
\begin{align}
  \begin{split}
\frac{d\sigma_Z^{pA}}{dM^2dy} & \propto A\bigg[a_u\bigg(u_1\bar{u}_2 + \bar{u}_1u_2 \bigg) + a_d\bigg(d_1\bar{d}_2 + \bar{d}_1d_2 + s_1\bar{s}_2 + \bar{s}_1s_2\bigg)\bigg]
  \end{split}
\end{align}
where $q_1 = q(x_1,M^2)$ and $q_2 = q(x_2,M^2)$ and the Drell-Yan kinematics at LO are:
\begin{equation}
  \label{eq:LOkinematicsDY}
x_{1,2} = \frac{M}{\sqrt{s}}e^{\pm y}\,, \quad\quad M^2 = x_1x_2s \, ,
\end{equation}
with
$\sqrt{s}$ being the center-of-mass energy of the collision, $M$ being the
invariant mass of the final state, and $y$ the rapidity of the final state system.
The effective weak couplings are given by
\begin{equation*}
a_q = (\bar{g}^q_V + \bar{g}^q_A), \quad\quad \bar{g}^f_V=(t^{(f)}_{3}-2Q_f \sin^2\theta_W), \quad\quad 
\bar{g}^f_A=t^{(f)}_{3}
\end{equation*}
where $t^{(f)}_{3}$ is the weak isospin of fermion $f$, $Q_f$ is the electric charge and $\theta_W$ is Weinberg's angle.

\paragraph{B2. Charged current DY}
The corresponding LO expressions in the case of W$^-$ and W$^+$ boson production
in pA collisions are given by
  \begin{align}
    \begin{split}
  \frac{d\sigma_{W^-}^{pA,}}{dM^2dy} \propto\,\, & |V_{ud}|^2\bigg(Z(d_1\bar{u}_2 + \bar{u}_1d_2) +(A-Z)(\bar{d}_1u_2 + u_1\bar{d}_2)\bigg) \\
  + &|V_{us}|^2 \bigg(Z( d_1\bar{s}_2 + \bar{s}_1d_2) +(A-Z)(u_1\bar{s}_2 + \bar{s}_1u_2) \bigg)
    \end{split}
\end{align}
and:
\begin{align}
  \begin{split}
  \frac{d\sigma_{W^+}^{pA,}}{dM^2dy} \propto\,\,  &|V_{ud}|^2\bigg(Z(u_1 \bar{d}_2+ \bar{d}_1 u_2) +(A-Z)(d_1 \bar{u}_2 + \bar{u}_1d_2)\bigg)\\
  +&|V_{us}|^2\bigg(Z(u_1\bar{s}_2+\bar{s}_1u_2) + (A-Z)(d_1\bar{s}_2+ \bar{s}_1 d_2)\bigg)
\end{split}
\end{align}
where in this case $y$ stands for the $W$ boson rapidity, which in
general is different from the pseudo-rapidity of the charged lepton
that is measured by experiments.
However, in either case the 
PDF combinations that enter at LO remain the same.
From the comparison between the LO expressions of the DIS structure
functions and the $W$ and $Z$ production cross-sections in pA
collisions, it is clear that each group of process is sensitive to a
different combination of quark and antiquark PDFs.
Therefore, their combination (due to the PDF universality in QCD) into a
global QCD analysis provides a unique handle for a robust separation
between quark and antiquark flavors in nuclei.


\providecommand{\href}[2]{#2}\begingroup\raggedright\endgroup

\end{document}